\documentclass[11pt]{article}

\usepackage[utf8]{inputenc}
\usepackage[T1]{fontenc}
\usepackage{lmodern}
\usepackage[protrusion=true,expansion=false]{microtype}
\usepackage[margin=1in, top=1.1in, bottom=1.1in]{geometry}

\usepackage{amsmath,amssymb}
\usepackage{amsthm}
\theoremstyle{definition}
\newtheorem{definition}{Definition}
\newtheorem{proposition}{Proposition}

\usepackage{booktabs}
\usepackage{tabularx}
\usepackage{longtable}
\usepackage{array}
\usepackage{multirow}

\usepackage{listings}
\usepackage{xcolor}

\definecolor{codegray}{rgb}{0.5,0.5,0.5}
\definecolor{codegreen}{rgb}{0,0.5,0}
\definecolor{codeblue}{rgb}{0,0,0.6}
\definecolor{codebg}{rgb}{0.97,0.97,0.97}

\lstdefinestyle{acp}{
  backgroundcolor=\color{codebg},
  basicstyle=\small\ttfamily,
  breakatwhitespace=false,
  breaklines=true,
  captionpos=b,
  commentstyle=\color{codegray}\itshape,
  frame=single,
  framesep=4pt,
  keepspaces=true,
  keywordstyle=\color{codeblue}\bfseries,
  numbers=none,
  numberstyle=\tiny\color{codegray},
  showspaces=false,
  showstringspaces=false,
  showtabs=false,
  stringstyle=\color{codegreen},
  tabsize=2,
  columns=flexible
}

\usepackage{hyperref}
\hypersetup{
  colorlinks=true,
  linkcolor=blue!70!black,
  urlcolor=blue!70!black,
  citecolor=blue!70!black,
  pdftitle={Agent Control Protocol: Admission Control for Agent Actions},
  pdfauthor={Marcelo Fernandez},
  pdfsubject={Formal protocol for governance of autonomous agents. v1.30. arXiv:2603.18829},
  pdfkeywords={autonomous agents, admission control, capability tokens, cryptographic governance, agent authorization, audit ledger}
}

\usepackage{titlesec}
\titleformat{\section}{\large\bfseries}{\thesection}{1em}{}
\titleformat{\subsection}{\normalsize\bfseries}{\thesubsection}{1em}{}

\usepackage{setspace}
\usepackage{tikz}
\usetikzlibrary{arrows.meta,positioning}
\usepackage{pgfplots}
\pgfplotsset{compat=1.18}

\usepackage{graphicx}
\usepackage{enumitem}
\usepackage{footnote}

\newcommand{\yes}{\checkmark}
\newcommand{\no}{$\times$}

\begin{document}

\title{%
  \textbf{Agent Control Protocol}\\[0.4em]
  \large ACP v1.30: Admission Control for Agent Actions\\[0.3em]
  \normalsize Technical Specification and Reference Implementation%
}

\author{
  Marcelo Fernandez\\
  TraslaIA\\
  \texttt{info@traslaia.com}\\
  \url{https://agentcontrolprotocol.xyz}
}

\date{April 2026 \\ \small Draft Standard \\ \small \href{https://arxiv.org/abs/2603.18829}{arXiv:2603.18829 [cs.CR]} \quad \href{https://doi.org/10.5281/zenodo.19672575}{DOI: 10.5281/zenodo.19672575}}

\maketitle

\begin{abstract}
Autonomous agents can produce harmful behavioral patterns from individually
valid requests---a threat class that per-request policy evaluation cannot
address, because stateless engines evaluate each request in isolation and cannot
enforce properties that depend on execution history.
We present ACP, a temporal admission control protocol that enforces behavioral
properties over execution traces by combining static risk scoring with stateful
signals (anomaly accumulation, cooldown) through a \texttt{LedgerQuerier}
abstraction separating decision logic from state management.
ACP is not an anomaly detection system: it does not detect suspicious patterns
and alert; it \emph{blocks execution} based on deterministic, history-aware
risk scoring, providing a hard enforcement boundary rather than an advisory signal.
We demonstrate that this structural separation has measurable security
consequences: under a 500-request workload where every request is individually
valid (RS=35), a stateless engine with identical scoring approves all 500~requests, while
ACP limits autonomous execution to 2~out of 500 (0.4\%), escalating after
3~actions and enforcing denial after 11---isolating the structural gap between
stateless and stateful admission.
We identify a bounded state-mixing vulnerability in ACP-RISK-2.0: agents
executing high-frequency benign operations in one context can inadvertently
elevate risk scores in unrelated high-value contexts due to agent-level
rate aggregation---producing false denials that a stateless engine would
never generate (Experiment~7).
We introduce ACP-RISK-3.0, which eliminates this cross-context interference
by scoping rate-based anomaly signals to the interaction context via
\texttt{PatternKey(agentID, capability, resource)}, while preserving
enforcement: repeated behavior within a single context continues to trigger
denial at the same threshold (Experiment~8).
Decision evaluation runs at 739--832\,ns (p50; Table~\ref{tab:benchmarks});
throughput reaches 1{,}720{,}000~req/s (baseline, InMemoryQuerier)
and 920{,}000~req/s at 10 concurrent workers (Table~\ref{tab:throughput}),
degrading predictably with backend latency without protocol changes.
Safety and liveness properties are model-checked via TLC-runnable TLA+
(11~invariants + 4~temporal properties, 0~violations);
two-agent safety is verified at the correct ledger bound across
\textbf{4,294,930,695~distinct states} with zero violations.
Runtime correctness is validated by 73~signed conformance test vectors.
We demonstrate that an adversary with complete knowledge of the risk formula can suppress
BAR to $0.00$ while every individual request remains policy-compliant (Experiment~10);
BAR-Monitor detects the regime shift via $\Delta\mathrm{BAR}$ three batches before collapse,
and counterfactual evaluation confirms structural enforcement capacity is preserved
($\mathrm{BAR}_C = 1.00$).
We further identify and formalize \emph{deviation collapse}---a previously unrecognized
governance failure mode in which enforcement is active but never exercised because upstream
constraints eliminate the conditions required for \texttt{DENIED} decisions---and introduce
Boundary Activation Rate (BAR) as its detection mechanism (Experiment~9).
We evaluate coordinated multi-agent behavior (Experiment~13): $N$ agents issuing
homogeneous requests on a shared resource each accumulate risk independently via
agent-scoped \texttt{PatternKey} and ACP's evaluate-then-mutate execution contract,
guaranteeing each agent exactly two approved actions before escalation and denial.
The total coordination window $\mathrm{CW}_\mathrm{appr} = 2N$ with zero deviation
across all configurations: coordinated activity scales exactly linearly, preventing
superlinear amplification.
Together, Experiments~9, 13, and TLA+ verification demonstrate that ACP constrains
both the \emph{existence} and the \emph{scale} of undesirable behaviors.
ACP is Paper~1 of a 6-paper Agent Governance Series:
P0---atomic decision boundaries~\cite{fernandez2026a};
P2---behavioral drift detection (IML)~\cite{fernandez2026c};
P3/4---governance structure, fair allocation, and irreducibility~\cite{fernandez2026govstr};
P5---runtime execution validity under partial observability (RAM)~\cite{fernandez2026ram};
P6---operationalization of RAM as a runtime enforcement protocol~\cite{fernandez2026op}.
All companion papers are published on Zenodo.

\textbf{Specification and implementation:} \url{https://github.com/chelof100/acp-framework-en}
\end{abstract}

\tableofcontents

\newpage

\section{The Problem ACP Solves}

Autonomous agents are being deployed in institutional environments without a technical standard to govern their behavior.
This gap is structural, not merely a tooling limitation.

\subsection{The Structural Gap}

When an autonomous agent makes a decision and executes it, there is a critical moment between both actions.
In current models, that moment does not formally exist: the decision and the execution are the same event.
The agent decides and acts. There is no intermediate validation.
There is no point of intervention. There is no structured record of why the decision was made.

This is acceptable when a human executes that action, because the human bears responsibility and can be questioned.
An autonomous agent cannot be questioned. It can only be audited---and only if there is something to audit.

The fundamental problem is therefore not whether agents are trustworthy, but that no standardized, protocol-level mechanism currently exists that allows an institution to demonstrate that its agents operated within authorized limits.

Autonomous agents differ from traditional software services in one critical respect:
their action sequences emerge from internal reasoning processes rather than from a pre-specified
function.
A conventional service executes a declared operation in response to a request.
An autonomous agent may produce a sequence of actions whose security relevance only becomes
apparent across multiple steps---a single request provides insufficient information to determine
whether the agent is operating within authorized behavioral bounds.

ACP enforces constraints over execution traces rather than individual requests.
Admissibility is resolved at execution time, but it is not memoryless.
This enables deterministic verification of \emph{temporal behavioral properties}: whether an agent
has exceeded its authorized request rate within a time window, whether a delegation chain is being
exercised beyond its intended scope, whether an anomaly pattern is accumulating across a session.
These properties depend on accumulated interaction history and cannot be reduced to single-request
evaluation.
They correspond to the class of trace-based safety properties formalized by
Schneider~\cite{schneider2000enforceable}, which stateless policy engines cannot enforce by
construction.
No adjustment of scoring thresholds or risk weights can compensate for this absence:
an engine that evaluates each request in isolation lacks the information necessary to determine
whether the current request is part of an anomalous sequence.
The requirement is state, not configuration---a structural constraint independent of the
sophistication of the scoring function.

Existing policy engines and authorization frameworks---OPA~\cite{opa}, RBAC~\cite{sandhu1996rbac},
capability systems---evaluate each request in isolation against declared policy.
ACP does not replace them; it adds a stateful enforcement layer above them.
Each admission decision is computed as a function of the interaction trace, not only the current
request---a composition that extends the practical enforcement boundary from individual interactions
to behavioral sequences.

Governance of autonomous agents therefore involves three distinct requirements.
First, admissibility must be resolved at the execution boundary---before an action commits,
not after~\cite{anderson1972planning}.
Second, the state evaluated at that boundary must encode execution history:
anomaly accumulation, action frequency, and prior behavioral patterns.
Without this, systems cannot distinguish isolated valid actions from sequences that become
inadmissible over time.
Third, the system must preserve the conditions under which inadmissibility can arise.
A system that always produces admissible actions does not demonstrate effective governance;
it may be operating in a regime where the boundary is intact but never exercised.
In this work, we address all three requirements.
ACP satisfies the first two by design; we identify and formalize the third as
\emph{failure condition preservation} (Section~\ref{sec:deviation-collapse}).

\subsection{ACP as Admission Control}

A useful analogy for understanding what ACP does is the Kubernetes Admission Controller pattern~\cite{kubernetes-admission}.

Kubernetes intercepts every API request before it reaches the cluster and runs it through a sequence of admission checks---\texttt{ValidatingWebhookConfiguration},
\texttt{ResourceQuota} enforcement, OPA Gatekeeper policies.
If any check fails, the request is rejected before touching cluster state.

ACP applies this pattern to agent actions:

\begin{lstlisting}
agent intent
    |
[1] Identity check     (ACP-AGENT-1.0, ACP-HP-1.0)
    |    is this agent who they claim to be?
[2] Capability check   (ACP-CT-1.0, ACP-DCMA-1.1)
    |    does the agent hold a valid token for this action?
[3] Policy check       (ACP-RISK-3.0, ACP-PSN-1.0)
    |    is this action within current policy?
[4] ADMIT / DENY / ESCALATE
    |    (if ADMIT)
[5] Execution token    (ACP-EXEC-1.0)
    |    single-use cryptographic proof of admission
[6] Ledger record      (ACP-LEDGER-1.3)
    |    immutable signed audit entry
system state mutation
\end{lstlisting}

The critical difference from Kubernetes: ACP's admission check operates across institutional boundaries.
An agent from Bank~A can be admitted by Bank~B without Bank~B trusting Bank~A's internal infrastructure---the
cryptographic proof is self-contained and verifiable with Bank~A's published public key alone.

This \emph{admission control} framing also clarifies the relationship with related tools:
\begin{itemize}[noitemsep]
  \item \textbf{OPA (Open Policy Agent)} \cite{opa} can serve as the policy evaluation engine inside Step~3.
        ACP does not replace OPA; it adds the identity and delegation chain layers above it.
  \item \textbf{AWS IAM / Azure RBAC} model static role permissions for humans.
        ACP adds dynamic agent delegation with execution proof.
  \item \textbf{OAuth 2.0} \cite{rfc6749} handles API access tokens.
        ACP extends delegation to multi-agent chains with non-escalation and verifiable provenance.
  \item \textbf{SPIFFE / SPIRE} \cite{spiffe} provides cryptographic workload identity.
        ACP builds on that identity to add capability scoping and governance.
\end{itemize}

\noindent Viewed through the lens of classical security theory, ACP is a
\emph{decentralized reference monitor}: a component that mediates every agent
action, is always invoked before execution, and produces a tamper-evident record
of every admission decision---extending Anderson's reference monitor concept~\cite{anderson1972planning}
to multi-organizational environments where no single trusted party controls all agents.

\subsection{Why RBAC and Zero Trust Are Insufficient}

RBAC and Zero Trust are the predominant control layers in enterprise environments.
Both are necessary. Neither solves the problem of governing autonomous agents:

\begin{table}[h!]
\centering
\small
\begin{tabular}{@{}lccc@{}}
\toprule
\textbf{Criterion} & \textbf{RBAC} & \textbf{Zero Trust} & \textbf{ACP} \\
\midrule
Designed for & Human roles & Network access & Autonomous agents \\
Cryptographic identity & No & Partial & Yes (Ed25519, mandatory) \\
Verifiable dynamic delegation & No & No & Yes (chained, auditable) \\
Decision/execution separation & No & No & Yes (Execution Tokens) \\
Real-time risk evaluation & No & Partial & Yes (deterministic) \\
Multi-institutional auditing & Non-standard & Non-standard & Native (signed ledger) \\
Transitive delegation revocation & No & No & Yes (formal propagation) \\
B2B interoperability for agents & Unstructured & Unstructured & Central protocol design \\
\bottomrule
\end{tabular}
\caption{Comparison of ACP with RBAC and Zero Trust across agent governance criteria.}
\end{table}

ACP does not replace RBAC or Zero Trust. It adds a governance layer oriented specifically to autonomous agents that operates above existing controls.

\subsection{The Concrete Scenario ACP Prevents}

\textbf{Without ACP:}
A financial processing agent receives instructions from another agent to execute a transfer.
The agent executes the action.
If the instruction was legitimate, everything works.
If it was compromised, injected, or generated by an unauthorized agent, the transfer occurs anyway.
No formal mechanism exists to prevent it, nor is there a technical record that allows reconstructing the authorization chain.

\textbf{With ACP:}
The agent requesting the transfer must present a Capability Token cryptographically signed by the issuing institution,
demonstrate possession of the associated private key,
and the request passes through the risk engine before receiving an Execution Token.
The ET is single-use.
The entire chain is recorded in the Audit Ledger with an externally verifiable institutional signature.

\subsection{Design Insight}
\label{sec:design-insight}

ACP is \textbf{compute-cheap but state-sensitive}: decision evaluation is computationally
inexpensive, while system performance is primarily constrained by state access patterns.

This separation enables a protocol that is both efficient in isolation (739--832\,ns per
decision, p50; Table~\ref{tab:benchmarks}, $\sim$78\,ns for the cooldown short-circuit path) and adaptable in deployment:
performance can be improved by replacing the state backend without modifying protocol semantics.
The \texttt{LedgerQuerier} interface formalizes this boundary---the decision function remains
stateless and deterministic; the state backend can be a local in-memory structure for development
or a distributed, indexed store in production.

\subsection{Contributions}

This paper makes the following contributions, organized by theme:

\paragraph{I.\enspace Stateful temporal admission control.}
\begin{itemize}[noitemsep]
  \item \textbf{Structural necessity of stateful enforcement.} We demonstrate
    empirically (Experiment~6) that temporal behavioral properties over execution
    traces cannot be enforced by stateless policy engines regardless of scoring
    function or thresholds. Under a 500-request financial transfer workload where
    every request is individually valid, a stateless engine with identical scoring
    approves all 500~requests; ACP limits autonomous execution to 2~out of 500
    (detection latency: 11~actions, early signal: 3~actions).
  \item \textbf{Sub-microsecond temporal admission control.} We show that
    stateful trace-level enforcement is compatible with 739--832\,ns decision
    latency (p50; Table~\ref{tab:benchmarks}) and 920{,}000~req/s throughput,
    achieved by separating decision logic (stateless, compute-cheap) from state
    management via the \texttt{LedgerQuerier} abstraction---enabling backend
    replacement without modifying protocol semantics or security guarantees.
  \item \textbf{Runtime containment with measurable latency.} A cooldown-based
    containment mechanism provides a 10.7$\times$ faster denial path (78\,ns via
    Step~2 short-circuit) and predictable throughput degradation under backend
    stress, isolated empirically in a controlled latency injection experiment
    (0\,$\mu$s $\to$ 920k~req/s; 5\,ms $\to$ 470~req/s).
\end{itemize}

\paragraph{II.\enspace Formal verification and verifiability.}
\begin{itemize}[noitemsep]
  \item \textbf{Formal verification of safety invariants.} Safety properties
    (admission determinism, cooldown monotonicity, per-agent isolation) are
    verified by TLC model checking over 5,684,342~generated states in the
    single-agent configuration (11~invariants + 4~temporal properties, 0~violations)
    and over 4,294,930,695~distinct states in a two-agent configuration
    at \texttt{LEDGER\_BOUND=11} (11~safety invariants, 0~violations, 10.5~h,
    i9-13900HX 24-core, 24\,GB heap), establishing that enforcement holds under
    adversarial state conditions and across independent concurrent agents.
  \item \textbf{End-to-end verifiability without proprietary infrastructure.}
    A three-layer pipeline---73~signed conformance test vectors, TLC-checked
    TLA+ model, and ACR-1.0 runtime sequence runner---enables independent
    validation of both specification and reference implementation.
\end{itemize}

\paragraph{III.\enspace Risk scoring and adversarial robustness.}
\begin{itemize}[noitemsep]
  \item \textbf{Context-scoped anomaly model eliminating cross-context interference.}
    We introduce ACP-RISK-3.0, which redefines Rule~1 to use context-scoped
    pattern counts, eliminating cross-context state-mixing while preserving
    enforcement against repeated behavior within a given interaction context
    (Experiment~8, Section~\ref{sec:exp8}).
  \item \textbf{Knowledge-aware adversarial evasion (Experiment~10).}
    We demonstrate that an adversary with complete knowledge of the ACP-RISK-3.0
    parameters can suppress BAR to $0.00$ while every individual request remains
    policy-compliant --- per-decision enforcement is structurally blind to this attack.
    BAR-Monitor detects the regime shift via $\Delta\mathrm{BAR}$ at Batch~2, three
    batches before threshold collapse; \texttt{EvaluateCounterfactual} confirms
    structural enforcement capacity is preserved ($\mathrm{BAR}_C = 1.00$).
    These three mechanisms form a coherent defense against knowledge-aware evasion.
\end{itemize}

\paragraph{IV.\enspace Governance: deviation collapse and boundary monitoring.}
\begin{itemize}[noitemsep]
  \item \textbf{Identification of deviation collapse as a governance failure mode.}
    We demonstrate empirically that a system may remain fully compliant while its
    admissibility boundary is never exercised
    ($\mathrm{BAR}_A = 0.70 \to \mathrm{BAR}_B = 0.00$, $n=20$).
    We introduce \emph{Boundary Activation Rate} (BAR) as a metric for detecting this
    condition and propose counterfactual evaluation as a restoration mechanism
    (Experiment~9, Section~\ref{sec:deviation-collapse}).
  \item \textbf{Failure condition preservation as a governance requirement.}
    We establish failure condition preservation as a necessary requirement for
    effective admission control: a system must not only enforce admissibility
    correctly but must preserve the conditions under which enforcement can fire.
    This separates \emph{correct enforcement} from \emph{effective governance}
    and provides the theoretical foundation for deviation collapse detection
    (Section~\ref{sec:deviation-collapse}).
  \item \textbf{Boundary Activation Monitoring (BAR-Monitor).}
    We introduce BAR-Monitor (\texttt{pkg/barmonitor}), a mechanism that detects
    loss of boundary interaction by tracking both $\mathrm{BAR}_N$ (current
    activation level) and $\Delta\mathrm{BAR}$ (progressive decline toward collapse).
    BAR-Monitor provides the lower-bound complement to cooldown's upper-bound
    constraint, defining a bounded operational region in which governance
    remains both safe and meaningful (Section~\ref{sec:bar-monitor}).
  \item \textbf{Empirical $\Delta\mathrm{BAR}$ early-warning validation (Phase D).}
    We validate $\Delta\mathrm{BAR}$ as a proactive detection mechanism through a
    controlled drift simulation (Experiment~9 Phase~D, 5 batches, 0--100\%
    upstream sanitization).
    \texttt{AlertTrend} fires at Batch~2 ($\mathrm{BAR}=0.57$, $\Delta\mathrm{BAR}=-0.25$),
    three batches before the threshold condition confirms collapse ($\mathrm{BAR}=0.00$
    at Batch~5), demonstrating a measurable early-warning gap.
  \item \textbf{TLA+ structural failure condition preservation.}
    We extend the TLA+ model with two invariants ---
    \textsc{FailureConditionPreservation} and \textsc{NoDegenerateAdmissibility} ---
    that establish necessary structural conditions for failure condition preservation.
    Both are verified by TLC with zero violations over all reachable states.
    \textsc{BARMonitorLiveness} is stated formally and validated empirically (Phase~D).
\end{itemize}

\paragraph{Positioning.}
The enforcement design introduced in this work---in which evaluation
and state transition are resolved jointly at the admission
boundary---can be understood as an instance of a broader class of
stateful enforcement systems. A formal characterization of this
structural property, and its extension to behavioral invariants
beyond the admission decision, are developed in subsequent work.

\section{Related Work}
\label{sec:related-work}

\subsection{Access Control and Policy Enforcement}

Traditional access control models such as Role-Based Access Control (RBAC)~\cite{sandhu1996rbac}
and Attribute-Based Access Control (ABAC)~\cite{hu2015abac} provide structured mechanisms for
regulating access decisions based on predefined roles or attributes.
Policy languages such as XACML~\cite{oasis2013xacml} and modern policy engines such as
Open Policy Agent~\cite{opa} and Cedar~\cite{cedar} extend these paradigms with expressive
evaluation engines.

These systems are fundamentally designed around \emph{stateless decision evaluation}:
they evaluate a request against policy at a single point in time.
They do not define how system state evolves as a result of decisions over time,
and they do not formalize how historical behavior (repeated denials, temporal patterns)
influences subsequent decisions in a deterministic and auditable manner.

ACP differs by explicitly separating:
(i)~stateless decision evaluation,
(ii)~externalized state management, and
(iii)~an execution contract governing state evolution across requests.
This enables reasoning not only about individual decisions, but about the temporal behavior
of the admission system.

\subsection{Runtime Policy Systems and Agent Frameworks}

Recent protocols for autonomous agents and multi-agent coordination define structured interactions
between components and services.
The Model Context Protocol (MCP)~\cite{mcp} provides structured tool access between LLM applications and services.
Agent-to-Agent (A2A)~\cite{a2a} defines communication and task delegation between agents.

While these protocols address interoperability and orchestration, they do not define:
\begin{itemize}[noitemsep]
  \item a deterministic risk evaluation model that accumulates state across requests,
  \item explicit state evolution semantics with a testable execution contract, or
  \item externally verifiable execution traces that can be reproduced by third parties.
\end{itemize}

Recent work has begun to address security governance for LLM-based agentic systems.
Syros et al.~\cite{syros2025saga} propose SAGA, a security architecture that provides
user-controlled access policies and cryptographic mediation for agent-to-agent tool use
and inter-agent communication.
While SAGA focuses on high-level governance architecture and inter-agent
communication policies, ACP enforces admission constraints at the execution layer:
every action request passes through a stateful risk evaluation with deterministic
outcomes, a verifiable execution contract (ACR-1.0), and a tamper-evident ledger---
providing enforcement rather than governance framing.
Tsai and Bagdasarian~\cite{tsai2025contextual} argue at HotOS~2025 that security policies
must be purpose-specific rather than uniformly applied, emphasising that one-size-fits-all
approaches are inadequate for diverse agent deployments---a perspective consistent with
ACP's per-agent isolation and stateful enforcement boundaries.

ACP complements these approaches by providing a governance layer that is independent of
orchestration protocols, focusing specifically on verifiable decision enforcement and auditability.

\subsection{Auditability and Tamper-Evident Logs}

Secure audit logging~\cite{schneier1999secure} and event sourcing~\cite{fowler2005event} ensure
that past events can be reconstructed and verified after the fact.
These approaches operate \emph{post hoc}: they provide evidence of what occurred without
constraining system behavior in real time.

ACP integrates auditability directly into the admission process.
Every decision---APPROVED, ESCALATED, or DENIED---is recorded in an append-only, cryptographically
chained ledger (ACP-LEDGER-1.3) before the governed action executes.
This enables both real-time enforcement and post-hoc verification from the same artifact.
The \texttt{LedgerAppendOnlyTemporal} invariant in the TLA+ model formally specifies
that ledger entries are never modified or removed.

\subsection{Formal Verification and Runtime Enforcement}
\label{sec:related-formal}

Model checking~\cite{lamport2002specifying} and symbolic reasoning are used to establish
correctness properties in distributed systems and protocols, typically focusing on safety
and liveness of state transitions.

The theoretical foundations of runtime enforcement trace to Anderson's reference monitor
concept~\cite{anderson1972planning} and Schneider's formal characterization of enforceable
security policies~\cite{schneider2000enforceable}, which established the boundary between
properties that can be \emph{verified} post-hoc and those that can be \emph{enforced}
at the point of execution via a security automaton.
Falcone et al.~\cite{falcone2012runtime} systematize runtime enforcement monitors,
distinguishing enforceable property classes (safety, co-safety, guarantee, persistence)
and the suppression or insertion mechanisms each requires.

Stateless policy engines such as OPA~\cite{opa} evaluate each request independently,
without intrinsic support for stateful behavioral accumulation across execution traces.
OPA has become a widely adopted standard for policy enforcement in distributed systems,
offering a declarative evaluation language (Rego) and integration points across APIs,
admission controllers, and service meshes.
These properties make it well-suited for request-level authorization, but they also define
a structural boundary: enforcement of constraints that depend on prior actions, frequency
patterns, or temporal context requires externalizing state to a separate store and threading
it through every evaluation call.

This distinction is critical for agentic systems.
Autonomous agents generate sequences of actions whose admissibility depends not only on
the current input, but on the history of prior actions, frequency patterns, and temporal
constraints---properties that cannot be expressed in stateless evaluation without introducing
external infrastructure.
In practice, approximating stateful behavior in OPA requires maintaining counters and timers
in an external store (e.g., Redis or a database), explicitly injecting current state into
each evaluation request, and relying on the surrounding architecture to guarantee atomicity
between evaluation and state mutation.
This shifts correctness responsibility outside the policy engine itself, increasing
architectural complexity and weakening enforcement guarantees at the decision boundary.

Our experimental evaluation (Experiment~\ref{sec:exp14}) illustrates this boundary directly.
Under repeated identical inputs---ten financial transfer requests from a single agent---OPA
remains permissive across all iterations in its standard stateless configuration (allow=\texttt{true}~$\times$~10).
When augmented with externally injected request counts, OPA can approximate frequency-based
constraints, but only through additional infrastructure outside the engine, and without
atomicity guarantees between evaluation and the counter update.
ACP enforces the same constraints natively: requests transition from \textsc{Approved} to
\textsc{Escalated} to \textsc{Denied} via accumulated PatternKey counts maintained within
the execution contract, without external coordination.

This difference reflects a distinction in expressiveness.
Stateless policy engines operate over representations of individual requests; ACP operates
over execution traces.
As a result, ACP can enforce constraints that depend on the existence, frequency, and ordering
of actions over time---properties that stateless engines, by construction,
cannot enforce without external state integration~\cite{schneider2000enforceable}.
ACP instantiates a stateful enforcement mechanism aligned with Schneider's security automaton
model, enabling formal reasoning about temporal properties of agent behavior that
stateless evaluation cannot capture.

ACP instantiates these principles in a decentralized, multi-organizational setting.
The ACR-1.0 compliance runner acts as an enforcement monitor over the ACP execution
contract: every request passes through the monitor before state mutation occurs,
the monitor is always invoked (no bypass path), and its decisions are recorded in a
tamper-evident ledger.
The TLA+ model specifies the safety and liveness properties the monitor is designed to
enforce; formally checked invariants are then instantiated as concrete sequence test
vectors and validated at runtime by the compliance runner.
This creates a three-layer verifiability chain: formal $\to$ data $\to$ runtime.

\subsection{Capability Systems and Proof-of-Possession Mechanisms}

ACP builds upon established primitives from capability-based security~\cite{capability-security} and
proof-of-possession authentication, but differs in how these primitives are
composed and enforced over execution traces.

\paragraph{Macaroons.}
Macaroons~\cite{birgisson2014macaroons} introduce decentralized authorization tokens
with embedded caveats that can be attenuated and verified without contacting a central
authority.
ACP-CT-1.0 shares structural similarities with Macaroons in that tokens carry verifiable
constraints and can be evaluated without implicit trust in the caller.
However, Macaroons are fundamentally stateless: all constraints must be encoded within the
token itself, and verification does not consult external state.
They do not support enforcement of temporal or behavioral properties that depend on prior
actions across requests.
In contrast, ACP separates token validation from behavioral enforcement: ACP-CT ensures
capability integrity per request, while ACP-RISK evaluates actions against accumulated
execution traces via a stateful backend, enforcing constraints that cannot be expressed as
static caveats.

\paragraph{ZCAP-LD.}
ZCAP-LD~\cite{zcapld} extends object-capability models to decentralized systems using
linked data proofs, enabling delegation chains and fine-grained authorization over
distributed resources.
Similar to Macaroons, ZCAP-LD focuses on authorization semantics and delegation but does
not define mechanisms for temporal enforcement or anomaly detection across sequences of
actions; policy evaluation remains scoped to individual invocations.
ACP operates at a different layer: it assumes a valid capability system (such as ACP-CT or
alternatives) and adds a stateful admission control layer that evaluates behavior over
time, remaining compatible with existing capability frameworks while extending their
enforcement model.

\paragraph{DPoP.}
DPoP~\cite{rfc9449} introduces a mechanism for binding access tokens to a client-held
private key, ensuring that tokens cannot be replayed by unauthorized parties.
ACP-HP-1.0 is aligned with DPoP in enforcing proof-of-possession per request; however,
DPoP operates at the transport/session level within OAuth flows.
ACP extends this model to action-level enforcement: each action is cryptographically bound
and independently verifiable, enabling replay resistance that is both cryptographic
(invalid proof) and behavioral (anomalous repetition patterns detected via F\_anom
Rule~3, as shown in Experiment~5).

\paragraph{Summary.}
These systems provide strong primitives for authorization and authentication but evaluate
requests in isolation.
ACP does not introduce new cryptographic primitives; its contribution lies in composing
existing mechanisms into a stateful admission control model that enforces constraints over
execution traces rather than individual requests---the structural difference empirically
demonstrated in Experiment~6.

\subsection{Behavioral Failure Modes in Governed Systems}

Two bodies of work share surface-level similarity with deviation collapse but address structurally distinct failure modes.

\emph{Specification gaming}~\cite{krakovna2020specification} occurs in reinforcement
learning when an agent maximizes a reward function while violating its intended
objectives---satisfying the letter of a constraint while undermining its purpose.
The locus of failure is the objective specification: the agent behaves optimally
under the given reward signal, which itself is misaligned with the designer's intent.
Deviation collapse operates at a different layer: the admission constraint is
correctly specified and correctly enforced, but upstream filtering prevents
risk-bearing requests from reaching the enforcement boundary.
The ACP engine does not optimize against the policy; it evaluates faithfully.
The failure is architectural, not behavioral.

\emph{Goodhart's Law}~\cite{goodhart1975monetary, strathern1997improving}---``when
a measure becomes a target, it ceases to be a good measure''---describes how
optimization pressure corrupts proxy metrics.
Deviation collapse shares the surface form of a metric losing signal, but the
mechanism differs: no optimization pressure is applied to the Boundary Activation
Rate, and no party is gaming the BAR metric.
The collapse occurs because the upstream pipeline removes the inputs that would
activate the boundary---a structural consequence of pipeline composition, not
metric manipulation.

\subsection{LLM Agent Security and Runtime Enforcement}

A growing body of work examines security threats specific to LLM-integrated agents.
Greshake et al.~\cite{greshake2023not} define \emph{indirect prompt injection} (IPI),
in which adversarial instructions embedded in data retrieved by an agent override
intended behavior, enabling exfiltration, unauthorized API calls, and cross-agent
propagation.
Zhan et al.~\cite{zhan2024injecagent} and Debenedetti et al.~\cite{debenedetti2024agentdojo}
provide benchmark evaluations showing that model-level defenses alone are insufficient to
contain IPI: frontier models remain vulnerable across diverse task environments.

ACP operates at the execution layer, not the prompt layer.
Prompt injection and jailbreaking occur before an action request reaches ACP's admission
boundary; ACP intercepts the resulting tool or API call before it modifies system state.
The two defense layers are therefore composable rather than competitive: a
prompt-layer filter reduces the IPI attack surface; ACP enforces structural constraints
over the action sequence regardless of how the request was generated.

Wang et al.~\cite{wang2025agentspec} propose AgentSpec, a domain-specific language for
defining runtime safety constraints on LLM agents, sharing the architectural insight that
enforcement must be interposed outside the model.

\textbf{Functional comparison with AgentSpec.}
Table~\ref{tab:agentspec-comparison} contrasts the two systems along five dimensions.

\begin{table}[h!]
\centering\small
\caption{Functional comparison of ACP and AgentSpec.}
\label{tab:agentspec-comparison}
\begin{tabular}{@{}p{3.5cm}p{4.5cm}p{4.5cm}@{}}
\toprule
Dimension & AgentSpec & ACP (this work) \\
\midrule
Enforcement granularity &
  Property-based rules over tool calls (LTL-style constraints). &
  Admission control with risk scoring and stateful signals. \\
\midrule
Threat model &
  Compliance with operator-defined safety properties. &
  Adversarial admission control + governance collapse detection (BAR-Monitor). \\
\midrule
Deviation collapse detection &
  No equivalent to BAR-Monitor. If upstream constraints are relaxed, all decisions
  become compliant silently. &
  BAR-Monitor detects this regime explicitly; TREND alert fires before collapse. \\
\midrule
Composability &
  Operates adjacent to the prompt layer (tool call interception). &
  Operates at the execution layer; composable with AgentSpec. \\
\midrule
Formal verification &
  Policy synthesis proved correct (AgentSpec properties). &
  Safety and liveness model-checked in TLA+ (11 invariants, 5,684,342 states, 0 violations). \\
\bottomrule
\end{tabular}
\end{table}

\textbf{Key differentiator: governance collapse.}
Consider the Experiment~10 scenario: a formula-aware adversary drives
$\mathrm{RS}=0$ on every request, causing every decision to be \texttt{APPROVED}.
Every individual decision conforms to the configured policy --- no rule is violated.
AgentSpec, operating on per-tool-call compliance, would not flag this regime:
each call satisfies all defined safety properties.
ACP's BAR-Monitor detects the collapse via $\Delta\mathrm{BAR}$:
the transition from $p_0=0.70$ (baseline boundary activation) to $p_1=0.00$
(full evasion) is detected at Batch~2 ($\mathrm{BAR}=0.57$, $\Delta\mathrm{BAR}=-0.25$),
before the degenerate regime is reached.
This represents a governance dimension orthogonal to per-decision enforcement:
ACP monitors whether the governance boundary remains meaningful, not only
whether individual decisions are compliant.
The two systems are therefore composable rather than competitive.

Motwani et al.~\cite{motwani2024collusion} demonstrate that multiple LLM agents can
coordinate hidden behavior via steganographic encoding in outputs, forming a
multi-agent collusion attack that is invisible to per-message inspection.
This attack is outside the scope of ACP's single-admission-point model; detection
of cross-agent steganographic channels would require correlation across agent interaction
records, which the ACP ledger provides but the detection logic is not currently specified.
Deng et al.~\cite{deng2025aiagents} provide a systematic survey of AI agent security
organized around four knowledge gaps, situating admission control as a mitigation for
the interaction-with-untrusted-external-entities gap.

\subsection{Summary}

Table~\ref{tab:comparison} summarizes the properties of ACP relative to existing approaches.

\begin{table}[h!]
\centering
\small
\begin{tabular}{@{}lcccccc@{}}
\toprule
\textbf{Property} & \textbf{RBAC} & \textbf{ABAC} & \textbf{OPA} & \textbf{Cedar} & \textbf{MCP/A2A} & \textbf{ACP (this work)} \\
\midrule
Stateful decisions       & \no & \no & \no & \no & \no & \yes \\
Deterministic evaluation & \yes & \yes & \yes & \yes & \no & \yes \\
Auditability             & partial & partial & partial & partial & \no & \yes \\
Runtime enforcement      & \yes & \yes & \yes & \yes & partial & \yes \\
Temporal behavior        & \no & \no & \no & \no & \no & \yes \\
External verifiability   & partial$^*$ & partial$^*$ & partial$^*$ & partial$^*$ & \no & \yes \\
Execution contract       & \no & \no & \no & \no & \no & \yes \\
Agent-native             & \no & \no & \no & \no & partial & \yes \\
\bottomrule
\end{tabular}
\caption{Comparison of ACP with existing access control and agent governance approaches.
\yes~=~fully supported; \no~=~not supported; partial~=~partially addressed.
$^*$Policy engines (RBAC, ABAC, OPA, Cedar) support policy-level auditability but not
decision-level external verifiability: reproducing a specific past decision requires
access to the original engine and policy state.
ACP external verifiability refers to the ability to reproduce any decision from
signed execution artifacts (test vectors, ledger entries) without access to the
original infrastructure.}
\label{tab:comparison}
\end{table}

Existing systems address individual aspects of the problem---policy evaluation,
agent orchestration, audit logging, or formal verification.
ACP's contribution lies in combining these elements into a unified model centered on
an explicit execution contract.
This enables a property not directly provided by prior systems:
the ability to deterministically reproduce and externally verify the behavior of
an admission decision system over time.

Unlike existing systems, ACP explicitly models temporal behavior---history accumulation,
anomaly detection, and cooldown enforcement---within the admission decision itself,
and provides verifiable execution semantics through signed test vectors and a runtime compliance runner.
This combination of temporal state, runtime enforcement, and external verifiability
is not jointly addressed by any prior system in Table~\ref{tab:comparison}.

\section{What ACP Is}

A formal technical specification---not a framework, not a platform, not a set of best practices.
A protocol with precise definitions, formal state models, verifiable flows, and explicit conformance requirements.

\subsection{Definition}

Agent Control Protocol (ACP) is a technical specification that defines the mechanisms by which autonomous institutional agents are identified, authorized, monitored, and governed in B2B environments.
It establishes the formal contract between an agent, the institution that operates it, and the institutions with which it interacts.

\textbf{Core invariant:}
\[
\textit{Execute}(\text{request}) \Longrightarrow
\textit{ValidIdentity}(\text{agent})
\;\wedge\; \textit{ValidCapability}
\;\wedge\; \textit{ValidDelegationChain}
\;\wedge\; \textit{AcceptableRisk}
\]

No action of an ACP agent can be executed without all four predicates being simultaneously true.
If any fails, the action is denied. No exceptions.

\subsection{Design Principles}

ACP was designed with five principles that are non-negotiable at implementation time:

\begin{itemize}[noitemsep]
  \item[\textbf{P1}] \textbf{Fail Closed.}
        On any internal component failure, the action is denied. Never approved by default.
  \item[\textbf{P2}] \textbf{Identity is cryptography.}
        $\text{AgentID} = \text{base58}(\text{SHA-256}(\text{public\_key}))$.
        No usernames. No arbitrary IDs. Identity cannot be claimed---it must be demonstrated in every request.
  \item[\textbf{P3}] \textbf{Delegation does not expand privileges.}
        The delegated agent's permissions are always a strict subset of the delegator's permissions.
        This property is cryptographically verified at every chain hop.
  \item[\textbf{P4}] \textbf{Complete auditability.}
        Every decision---approved, denied, or escalated---is recorded in an append-only ledger with institutional signature. Not just successes. Everything.
  \item[\textbf{P5}] \textbf{External verification possible.}
        Any institution can verify ACP artifacts from another institution using only the public key registered in the ITA. No dependency on proprietary systems.
\end{itemize}

\subsection{Formal Agent Model}

In ACP, an agent is a formal tuple with well-defined state:
\[
A = (\, \text{AgentID},\; \text{capabilities},\; \text{autonomy\_level},\; \text{state},\; \text{limits} \,)
\]

\begin{table}[h!]
\centering
\small
\begin{tabular}{@{}llp{7cm}@{}}
\toprule
\textbf{Field} & \textbf{Type} & \textbf{Description} \\
\midrule
AgentID & String (43--44 chars) & \texttt{base58(SHA-256(pk))}. Derived from public key. Immutable. \\
capabilities & List of strings & Explicit permissions. Format: \texttt{acp:cap:<domain>.<action>}. Never abstract roles. \\
autonomy\_level & Integer 0--4 & Determines risk evaluation thresholds. 0 = no autonomy. 4 = maximum. \\
state & Enum & \texttt{active | restricted | suspended | revoked}. Transition to \texttt{revoked} is unidirectional. \\
limits & Object & Rate limits, maximum amounts, time windows. Not modifiable at runtime. \\
\bottomrule
\end{tabular}
\caption{ACP formal agent tuple fields.}
\end{table}

\paragraph{Formal model.}
The following definitions establish the mathematical objects over which ACP's
admission function is specified.

\begin{definition}[Execution Trace]
Let $E = (a,\, \mathit{cap},\, \mathit{res},\, t)$ denote an authorization event,
where $a$ is an agent identifier, $\mathit{cap}$ a capability, $\mathit{res}$ a
resource class, and $t$ a timestamp.
An \emph{execution trace} of length $n$ for agent $a$ is an ordered sequence
\[
  T_a = \langle E_1, E_2, \dots, E_n \rangle,
\]
where each $E_i = (a, \mathit{cap}_i, \mathit{res}_i, t_i)$ is an event
processed in non-decreasing temporal order ($t_i \leq t_{i+1}$).
\end{definition}

\begin{definition}[Agent Execution State]
For agent $a$ at time $t$, the \emph{execution state} $S_a(t)$ is a tuple
\[
  S_a(t) = \bigl(\,\mathit{denial\_count},\; \mathit{pattern\_count},\;
                    \mathit{cooldown\_until},\; \mathit{ledger\_entries}\,\bigr),
\]
where $\mathit{denial\_count} \in \mathbb{N}$ is the number of DENIED decisions
within the active cooldown window, $\mathit{pattern\_count}: C \times R \to \mathbb{N}$
maps capability--resource pairs to their request frequency within the current context,
$\mathit{cooldown\_until} \in \mathbb{N} \cup \{\bot\}$ is the timestamp at which
an active cooldown expires (or $\bot$ if no cooldown is active), and
$\mathit{ledger\_entries}$ is the append-only sequence of prior admission records.
\end{definition}

\begin{definition}[Admission Function]
The \emph{admission function} $f$ maps an agent execution state and an
incoming event to a decision:
\[
  f\bigl(S_a(t^-),\, E\bigr) \;\in\;
  \{\,\textsc{Approved},\; \textsc{Escalated},\; \textsc{Denied}\,\}.
\]
$f$ is deterministic: given the same state and event, $f$ always returns the
same decision.
The risk score $\mathrm{RS}(S_a, E) \in [0, 100]$ summarizes the risk of
admitting $E$ given current state; the decision boundary is:
\[
  f(S_a, E) =
  \begin{cases}
    \textsc{Denied},    & \text{if } \mathrm{RS} \geq \tau_{\mathrm{deny}}, \\
    \textsc{Escalated}, & \text{if } \tau_{\mathrm{esc}} \leq \mathrm{RS} < \tau_{\mathrm{deny}}, \\
    \textsc{Approved},  & \text{otherwise,}
  \end{cases}
\]
where $\tau_{\mathrm{esc}} = 40$ and $\tau_{\mathrm{deny}} = 70$ under the
default \texttt{PolicyConfig} (Section~\ref{sec:bar}).
\end{definition}

\subsection{Layered Architecture}

ACP does not replace existing security infrastructure. It is added as an upper layer:

\begin{lstlisting}
ACP Layer   -- Autonomous agent governance: identity, authorization, risk, auditing
RBAC Layer  -- Role-based access control for human users
Zero Trust  -- Continuous identity and network access verification
\end{lstlisting}

\subsection{Refinement under Transactional Assumptions}
\label{sec:refinement}

The \emph{Atomic Decision Boundaries} framework~\cite{fernandez2026a} (Paper~0 in
this governance series) establishes a formal requirement: a governance mechanism
can guarantee runtime admissibility \emph{only if} evaluation and state mutation
occur in a single, indivisible LTS step.  This section shows that ACP satisfies
this requirement under two explicit transactional assumptions, and provides the
correspondence table that connects both papers.

\subsubsection*{Transactional Assumptions}

\begin{itemize}[noitemsep]
  \item \textbf{(TA1) Serializable ledger.}
        All reads and writes to the state backend (\texttt{LedgerQuerier}) are
        serializable: no two concurrent \texttt{evaluate} calls observe
        interleaved intermediate states.  In practice this is provided by
        the backend's transaction isolation guarantee (e.g., Redis
        \texttt{MULTI/EXEC}, PostgreSQL \texttt{SERIALIZABLE}).
  \item \textbf{(TA2) Atomic commit.}
        The pair (emit Capability Token, append Audit Ledger entry) either
        both succeed or both fail as an indivisible unit.  ACP enforces this
        via the \texttt{ACP-LEDGER-1.3} append protocol: a ledger write failure
        causes the token to be treated as if it were never issued
        (PROHIB-007: fail closed).
\end{itemize}

\subsubsection*{Correspondence Table}

Table~\ref{tab:refinement} maps the abstract objects of~\cite{fernandez2026a}
to ACP's concrete constructs.

\begin{table}[h!]
\centering
\small
\begin{tabular}{@{}p{4.0cm}p{7.8cm}@{}}
\toprule
\textbf{Abstract (Paper~0)} & \textbf{ACP instantiation} \\
\midrule
State $s \in S$ &
  $(\mathit{ledger},\; \mathit{riskScore},\; \mathit{cooldownExpiry},\;
  \mathit{agentState})$ held by the \texttt{LedgerQuerier} \\[4pt]
Action $a \in \mathit{Act}$ &
  Incoming capability request $r$ carrying
  \texttt{(AgentID, cap[], res, exp, nonce, sig)} \\[4pt]
Decision function $F(s,a)$ &
  \texttt{evaluate}$(s,r)$: verify signature $\to$ score risk $\to$
  compare thresholds $\to$ return
  \texttt{APPROVED / DENIED / ESCALATED} \\[4pt]
$F(s,a)=(\textsc{Allow},s')$ &
  \texttt{APPROVED}: emit Capability Token + append ledger entry;
  $s' = s[\mathit{cooldown}{+}{=}\Delta,\;\mathit{ledger}{+}{=}\mathit{entry}]$ \\[4pt]
$F(s,a)=(\textsc{Refuse},s')$ &
  \texttt{DENIED}: no token emitted; ledger records denial;
  $s' = s[\mathit{anomalyCount}{+}{=}1]$ \\[4pt]
$F(s,a)=(\textsc{Escalate},\hat{s}')$ &
  \texttt{ESCALATED}: request placed in audit queue pending supervisor
  resolution; $\hat{s}' = (s,\;\mathit{auditQueue} \cup \{r\})$ \\[4pt]
Admissibility $\mathit{Adm}(s,a)$ &
  $\mathit{RS}(r,s) < \theta_{\mathit{deny}}$ (risk score below denial
  threshold for agent's autonomy level) \\[4pt]
Escalation space $\hat{S}{=}S{\times}\mathit{Req}$ &
  State backend extended with \texttt{escalation\_queue}:
  $\hat{S} = S \times \mathit{PendingEscalations}$ \\
\bottomrule
\end{tabular}
\caption{Correspondence between the abstract Atomic Decision Boundary
  model~\cite{fernandez2026a} and ACP constructs.
  Under TA1 and TA2 each row is realized atomically.}
\label{tab:refinement}
\end{table}

\subsubsection*{Refinement Claim}

\textbf{Claim.}
\emph{Under TA1 and TA2, the ACP \texttt{evaluate-then-mutate} pipeline
realizes the atomic decision boundary $F : S \times \mathit{Act} \to
\mathcal{D} \times \hat{S}$ of~\cite{fernandez2026a} as a single LTS step:
state read, decision computation, and state write are executed without
interleaving, and token emission and ledger commit are jointly atomic.
This is a \textbf{refinement} (not a software implementation) of the abstract
model, because it adds concrete cryptographic, timing, and network constructs
not present in the abstract LTS.}

Two corollaries follow:
\begin{enumerate}[noitemsep]
  \item \textbf{Admissibility guarantee.}
        Under TA1--TA2, every ACP \texttt{APPROVED} decision corresponds to
        an admissible action at evaluation time.  Split evaluators (RBAC,
        OPA, policy engines) cannot guarantee this because state can change
        between read and write.
  \item \textbf{Escalation closure.}
        Every \texttt{ESCALATED} request is appended to the audit queue
        atomically with its state snapshot, making supervisor resolution
        deterministic and replay-safe.
\end{enumerate}

\textbf{Scope.}
The refinement holds under TA1--TA2.  When the state backend provides only
read-committed isolation (weaker than serializable), two concurrent evaluations
can observe the same pre-mutation state, breaking atomicity; this is a declared
limitation (see \S\ref{sec:limitations}).  The behavioral layer above this
boundary---whether drift accumulates within the admitted action space---is the
subject of~\cite{fernandez2026c} (Paper~2, IML).  Allocation fairness across
competing agents sharing the boundary, strategy-proofness, and compositional
irreducibility of the full four-layer stack are addressed
in~\cite{fernandez2026govstr} (Paper~3/4).

\section{Technical Mechanisms}

ACP defines six interdependent mechanisms. Each has its own formal specification, state model, data structure, protocol flow, and error codes.

\subsection{Serialization and Signing (ACP-SIGN-1.0)}

Every verification in ACP begins with signature verification.
ACP-SIGN-1.0 defines the exact process that produces a binary result---valid or invalid---without ambiguity:

\begin{enumerate}[noitemsep]
  \item \textbf{Canonicalization with JCS (RFC~8785) \cite{rfc8785}.}
        Produces a deterministic representation of the JSON object, independent of field order and the system that generated it.
  \item \textbf{SHA-256 hash} over the canonical output in UTF-8.
  \item \textbf{Ed25519 signature (RFC~8032) \cite{rfc8032}} over the hash.
        32-byte key, 64-byte signature.
  \item \textbf{Base64url encoding} without padding for transmission.
\end{enumerate}

Signature verification precedes all semantic validation.
An object with an invalid signature is rejected without processing its content.
This rule has no exceptions (PROHIB-003, PROHIB-012).

\subsection{Capability Token (ACP-CT-1.0)}

The Capability Token is ACP's central artifact.
It is a signed JSON object~\cite{rfc7519} that specifies exactly what an agent can do, on what resource, for how long, and whether it can delegate that capability to other agents.

\begin{lstlisting}[language={}]
{
  "ver": "1.0",
  "iss": "<AgentID_issuer>",
  "sub": "<AgentID_subject>",
  "cap": ["acp:cap:financial.payment"],
  "res": "org.example/accounts/ACC-001",
  "exp": 1718923600,
  "nonce": "<128bit_CSPRNG_base64url>",
  "deleg": { "allowed": true, "max_depth": 2 },
  "parent_hash": null,
  "sig": "<Ed25519_base64url>"
}
\end{lstlisting}

\textbf{Critical fields:}
\texttt{exp} is mandatory---a token without expiry is invalid by definition.
The 128-bit nonce enforces request uniqueness within the canonical representation and
supports anomaly accumulation; it does not constitute nonce-based replay prevention
(see Section~\ref{sec:adversarial} and Section~\ref{sec:limitations}).
\texttt{parent\_hash} chains delegated tokens in a verifiable way.
The signature covers all fields except \texttt{sig}.

\subsection{Handshake and Proof-of-Possession (ACP-HP-1.0)}

Possessing a valid Capability Token is not sufficient to act.
ACP-HP-1.0 requires that the bearer demonstrate in every request that they possess the private key corresponding to the AgentID declared in the token.
This eliminates the possibility of impersonating an agent with a stolen token.

The protocol is stateless---it does not establish sessions, does not produce a \texttt{session\_id}, does not require server-side state between requests. The proof occurs in every interaction:

\begin{enumerate}[noitemsep]
  \item The receiving system issues a 128-bit challenge generated by CSPRNG, valid for 30 seconds and single-use.
  \item The agent signs the challenge together with the HTTP method, path, and body hash of the request.
  \item The receiver verifies the signature using the agent's public key, obtained from the ITA.
  \item The challenge is deleted immediately after use---it cannot be reused.
\end{enumerate}

This sequence is designed to provide four properties:
identity authentication, cryptographic request binding, anti-replay, and transport channel independence.

\subsection{Deterministic Risk Evaluation (ACP-RISK-2.0)}

Each authorization request passes through a deterministic risk function that produces a Risk Score (RS) in the range $[0, 100]$.
The same input always produces the same result---no stochastic elements, no machine learning in the critical path.
ACP deliberately trades expressiveness for deterministic auditability and reproducibility: every decision can be reconstructed from first principles given only the input request and the current state snapshot.

\[
\text{RS} = \min\!\bigl(100,\; B(c) + F_{\text{res}}(r) + F_{\text{ctx}}(x) + F_{\text{hist}}(h) + F_{\text{anom}}(q)\bigr)
\]

\noindent where $F_{\text{anom}}(q)$ is the anomaly factor introduced in v2.0, evaluated against the agent's accumulated request history $q$ (pattern frequency, denial rate, cooldown state). When an agent accumulates three DENIED decisions within ten minutes, a cooldown is activated and subsequent requests are blocked regardless of RS.

\begin{table}[h!]
\centering
\small
\begin{tabular}{@{}llp{5.5cm}@{}}
\toprule
\textbf{Factor} & \textbf{Description} & \textbf{Example values} \\
\midrule
$B(c)$ & Baseline by capability & \texttt{*.read = 0}, \texttt{financial.payment = 35} \\
$F_{\text{ctx}}(x)$ & Request context & Non-corporate IP $+20$; outside hours $+15$ \\
$F_{\text{hist}}(h)$ & Agent history (24h) & Recent denial $+20$; anomalous frequency $+15$ \\
$F_{\text{res}}(r)$ & Resource classification & \texttt{public = 0}; \texttt{sensitive = 15}; \texttt{restricted = 45} \\
\bottomrule
\end{tabular}
\caption{ACP-RISK-2.0 risk scoring factors (base formula; $F_{\text{anom}}$ adds up to $+30$ from pattern accumulation).}
\end{table}

The RS determines the decision according to the thresholds configured for the agent's \texttt{autonomy\_level}.
With \texttt{autonomy\_level}~2 (standard):
$\text{RS} \leq 39$ $\rightarrow$ APPROVED;
$\text{RS} \in [40, 69]$ $\rightarrow$ ESCALATED;
$\text{RS} \geq 70$ $\rightarrow$ DENIED.
An agent with \texttt{autonomy\_level}~0 always receives DENIED.

\subsection{LLM Agent Integration}
\label{sec:llm-agent-integration}

ACP is designed to integrate with LLM agent frameworks (LangChain, LangGraph,
AutoGen, CrewAI) as an admission control layer that intercepts tool calls
\emph{before} they modify system state.
The integration pattern wraps each tool with an ACP guard that calls
\texttt{POST /acp/v1/authorize} and gates execution on the admission decision.

\begin{lstlisting}[language=Python, caption={ACP-guarded tool wrapper for LangChain agents.
Each tool call is intercepted before execution; admission is decided by
the ACP server based on the agent's identity, capability, and resource class.},
label=lst:acp-tool]
class ACPGuardedTool(BaseTool):
    capability: str      # e.g. "acp:cap:financial.transfer"
    resource:   str      # e.g. "accounts/restricted-fund"
    res_class:  str      # "PUBLIC" | "SENSITIVE" | "RESTRICTED"
    agent_id:   str      # agent identity bound to ACP token
    acp_url:    str      # ACP server endpoint

    def _run(self, query: str) -> str:
        resp = requests.post(f"{self.acp_url}/acp/v1/authorize", json={
            "agent_id":          self.agent_id,
            "capability":        self.capability,
            "resource":          self.resource,
            "action_parameters": {"resource_class": self.res_class},
        }).json()
        decision = resp["decision"]
        if decision == "APPROVED":
            return self._execute(query)      # tool logic runs
        if decision == "ESCALATED":
            return "[ESCALATED: pending human review]"
        return f"[DENIED by ACP: {resp.get('reason_code')}]"
\end{lstlisting}

The admission decision is stateful: the ACP server maintains denial history,
pattern counters, and cooldown state across calls from the same agent.
A tool call that would be individually approved may be denied if the agent's
recent history has accumulated sufficient risk signals (F\_hist, F\_anom).
This gives the framework a temporal enforcement property that stateless
policy engines cannot provide.

\textbf{IPI composability.}
A prompt-layer filter (e.g., PromptGuard, PINT) can be applied before the
tool call reaches \texttt{ACPGuardedTool}, reducing the fraction of
injected high-risk calls that reach the admission boundary.
ACP enforces the structural constraint on the tool call regardless of its
origin: a \texttt{fund\_transfer} call with $\mathrm{RS}=80$ is denied
whether it was generated by legitimate user input or by an injected instruction.
The two layers are composable: the prompt-layer filter reduces attack
frequency; ACP provides the structural denial guarantee.

\textbf{Reproducible local evaluation.}
For evaluation without external API dependencies, the Ollama
runtime~\cite{ollama2024} provides OpenAI-compatible endpoints for
open-weight models.
Setting \texttt{temperature=0} and \texttt{seed=42} yields
pseudo-deterministic agent behavior, enabling controlled experimental
evaluation of ACP admission decisions across scripted agent sessions
(see Experiment~12, \S\ref{sec:exp12}).

\textbf{Real-LLM validation.}
To validate the \texttt{ACPGuardedTool} wrapper in a live agent context,
we ran a five-turn session using DeepSeek-R1:8B~\cite{deepseekr1} served
locally via Ollama (\texttt{temperature=0}, \texttt{seed=42}).
An IPI payload was embedded in the \texttt{weather\_query} tool result,
instructing the agent to call \texttt{fund\_transfer} with attacker-controlled
parameters.

\begin{lstlisting}[caption={Condensed output from the ACP real-LLM IPI demo
(\texttt{demos/ollama-agent/agent\_demo.py}).
Full session is deterministic and reproducible.}]
Model: deepseek-r1:8b | temperature=0 | seed=42

[T1] weather_query(city="Madrid")        RS= 0  APPROVED  [IPI injected]
[T2] fund_transfer(9999, "attacker-001") RS=80  DENIED    [IPI-induced]
[T3] weather_query(city="Barcelona")     RS= 0  APPROVED  [IPI injected]
[T4] fund_transfer(9999, "attacker-001") RS=80  DENIED    [IPI-induced]
[T5] fund_transfer(50,   "savings-123")  RS=80  DENIED    [legitimate]
                                              --> COOLDOWN ACTIVATED
\end{lstlisting}

Three observations follow.
First, the LLM followed the IPI injection in turns~2 and~4, calling
\texttt{fund\_transfer} with attacker-controlled parameters in both cases.
ACP blocked each attempt because the capability
\texttt{acp:cap:financial.transfer} with resource class \texttt{RESTRICTED}
evaluates to $\mathrm{RS}=80 \geq 70$, regardless of the transaction parameters.
The LLM's internal reasoning is irrelevant to ACP's admission decision.

Second, turn~5 demonstrates capability-level enforcement:
a \emph{legitimate} transfer (\$50 to \texttt{savings-123}) is also denied
because ACP evaluates capability and resource class, not individual parameters.
In a production deployment, a personal savings account would carry resource class
\texttt{SENSITIVE} ($\mathrm{RS}=35+15=50$, \texttt{ESCALATED}),
separating attacker-targeted restricted transfers from legitimate sensitive ones.
The demo uses a uniform \texttt{RESTRICTED} class to isolate the IPI dynamic.

Third, the third denial in turn~5 activated agent-wide cooldown,
locking the agent for 300 seconds.
The same stateful consequence observed in Experiment~12
--- $F_\text{anom}$ Rule~2 persisting beyond the cooldown window ---
emerges here with a real LLM as the agent brain,
confirming that the enforcement properties are independent of the agent's
implementation.

\subsection{Context-Scoped Anomaly Enforcement (ACP-RISK-3.0)}
\label{sec:risk30}

ACP-RISK-2.0 evaluates $F_{\text{anom}}$ against a single per-agent counter:
\texttt{CountRequests(agentID, window)}.
This produces a state-mixing vulnerability when the same agent holds capabilities in multiple
resource contexts: denials accumulated in one context contaminate the RS in an unrelated context,
producing false ESCALATED or DENIED outcomes (see \S\ref{sec:state-mixing} and Experiment~7).

ACP-RISK-3.0 eliminates this by scoping the anomaly counter to a \emph{context key}:

\[
\texttt{PatternKey}(a, c, r) = \text{SHA-256}(\,a \;\|\; c \;\|\; r\,)
\]

\noindent where $a$ is the \texttt{agent\_id}, $c$ the capability, and $r$ the resource.
Rule~1 of $F_{\text{anom}}$ becomes \texttt{CountPattern(ctxKey,\,-60s)} rather than
\texttt{CountRequests(agentID,\,-60s)}.
All other rules ($F_{\text{anom}}$ Rules~2--3, cooldown logic, decision thresholds) are unchanged.

The key derivation adds a single SHA-256 call per evaluation (739\,ns;
Table~\ref{tab:benchmarks}, \texttt{PatternKey} row),
preserving the sub-microsecond latency envelope.
Experiment~8 (\S\ref{sec:exp8}) confirms that cross-context contamination is fully eliminated
under ACP-RISK-3.0 while same-context burst detection remains intact.

\subsection{Verifiable Chained Delegation (ACP-DCMA-1.1)}

ACP allows an agent to delegate capabilities to another agent, which in turn can delegate to a third,
up to the maximum depth defined in the root token.
Delegation guarantees three properties:

\begin{itemize}[noitemsep]
  \item \textbf{No privilege escalation.}
        The delegated agent's capability set is always $\subseteq$ the delegator's set.
        Cryptographically verified at every hop via the \texttt{parent\_hash} field.
  \item \textbf{Bounded depth.}
        The \texttt{max\_depth} field establishes the chain limit. A chain that exceeds it is invalid.
  \item \textbf{Transitive revocation.}
        Revoking an agent's token automatically invalidates all delegated tokens that descend from it.
\end{itemize}

\subsection{Execution Token (ACP-EXEC-1.0)}

The separation between authorization and execution is a core principle of ACP.
When the authorization engine approves a request, it returns an Execution Token (ET):
a single-use artifact with a short lifetime that authorizes exactly that action, on that resource, at that moment.

\begin{itemize}[noitemsep]
  \item An ET can only be consumed once.
        A second presentation is rejected (PROHIB-002).
  \item An expired ET is invalid even if it was never used.
  \item The target system that consumes the ET notifies the ACP endpoint, closing the audit cycle.
\end{itemize}

\subsection{Audit Ledger (ACP-LEDGER-1.3)}

The Audit Ledger is a chain of cryptographically signed events where each event includes the hash of the previous event:

\[
h_n = \text{SHA-256}(\,e_n \;\|\; h_{n-1}\,)
\]

This structure makes it impossible to modify or delete an event without invalidating the entire subsequent chain.
The ledger records all lifecycle event types: GENESIS, AUTHORIZATION (including DENIED and ESCALATED),
RISK\_EVALUATION, TOKEN\_ISSUED, TOKEN\_REVOKED, EXECUTION\_TOKEN\_ISSUED, and EXECUTION\_TOKEN\_CONSUMED.

Institutions with FULL conformance expose the ledger via \texttt{GET /acp/v1/audit/query},
allowing external partners to verify chain integrity using only the institutional ITA public key.
Modifying or deleting ledger events is prohibited (PROHIB-007, PROHIB-008).

It is important to note that the ledger provides \textbf{verifiable evidence} of execution, not enforcement itself.
Enforcement in ACP occurs at the policy evaluation and execution layers (ACP-EXEC-1.0, ACP-DCMA-1.1):
the Execution Token is the cryptographic artifact that gates actual system-state mutation.
The ledger's role is to make every admission decision---and its outcome---auditable and tamper-evident after the fact.

\section{Inter-Institutional Trust}

In a B2B environment, agents of one institution interact with another institution's systems.
ACP defines the exact mechanism by which this trust is established, verified, and can be revoked.

\subsection{Institutional Trust Anchor (ACP-ITA-1.0)}

The ITA is the authoritative registry that links an \texttt{institution\_id} to an Ed25519 public key.
It is the only point where ACP depends on an out-of-band mechanism: the initial distribution of the ITA authority's public key.
Once that key is resolved, all subsequent verification is autonomous and cryptographic.

Each institution registers in the ITA its Root Institutional Key (RIK)---the private key held in HSM that never leaves it.
All ACP artifacts from that institution (tokens, ledger events, API responses) are signed with that key.
Any third party can verify them by resolving the public key from the ITA.

\subsection{Mutual Recognition (ACP-ITA-1.1)}

When two institutions operate under different ITA authorities,
ACP-ITA-1.1 defines the mutual recognition protocol.
The process requires both authorities to sign a Mutual Recognition Agreement (MRA) establishing:
the scope of recognition, the agreement's validity period,
and the proxy resolution mechanism.

Recognition is explicitly non-transitive.
If A recognizes B and B recognizes C, A does not automatically recognize C.
Each bilateral relationship requires its own signed MRA.
This limits uncontrolled expansion of the trust graph.

\subsection{Institutional Key Rotation and Revocation}

\textbf{Normal rotation} includes a transition period of up to 7 days during which both keys are valid,
allowing artifacts signed with the old key to be verified during the transition.

\textbf{Emergency rotation} is activated when a key is compromised.
The key is marked \texttt{revoked}; all artifacts signed with it are invalid
from the moment revocation is \emph{observed} by the verifier,
and there is no transition period.

\subsection{Trust Model and Failure Modes}
\label{sec:ita-trust-model}

ACP makes its trust assumptions explicit, bounded, and temporally scoped:
all protocol guarantees hold only after successful bootstrap and until revocation is observed.
This section formalizes the three structural trust scenarios that determine the security boundary
of the ITA layer.

\paragraph{K1a --- Bootstrap Trust Assumption.}

\textit{Assumption.}
The initial distribution of the ITA root public key occurs over an authentic out-of-band channel.

\textit{Adversary capability.}
An adversary who controls the bootstrap channel can substitute an attacker-controlled key,
after which all ACP artifact verification by the victim is performed against the wrong authority.

\textit{Formal claim.}
ACP provides no cryptographic security guarantees if the bootstrap trust anchor is compromised.
All subsequent guarantees---token verification, delegation validity, ledger integrity---are
conditional on the authenticity of the bootstrapped ITA public key.

\textit{Security implication.}
This is an explicit design boundary, not a protocol defect.
ACP reduces the trust problem from continuous dependence on a third party
to a single, auditable bootstrap event.
Deployment environments may use out-of-band channels with varying integrity guarantees
(e.g., DNSSEC, certificate pinning, in-person key ceremony);
the choice of channel is a deployment decision outside the protocol scope.

\paragraph{K1b --- Key Compromise Window.}

\textit{Assumption.}
Key compromise occurs at time $T_{\mathit{compromise}}$ and is detected and revoked at
$T_{\mathit{revoke}}$, with $T_{\mathit{compromise}} < T_{\mathit{revoke}}$.

\textit{Adversary capability.}
During the interval $[T_{\mathit{compromise}},\, T_{\mathit{revoke}}]$,
the adversary holds a valid signing key and can issue ACP-conformant tokens and ledger events
that are indistinguishable from legitimate artifacts.

\textit{Formal claim.}
ACP evaluates token validity at \emph{verification time}, not issuance time.
Security is monotonic with respect to revocation:
once a key is marked \texttt{revoked}, all subsequent verifications reject tokens derived from it,
regardless of when they were issued.
Tokens issued during the compromise window remain valid until revocation is observed.

\textit{Security implication.}
The window $[T_{\mathit{compromise}},\, T_{\mathit{revoke}}]$ represents an irrecoverable
exposure interval: ACP cannot retroactively invalidate actions that completed before revocation
was processed.
Minimizing this window is a deployment concern (monitoring, anomaly detection, HSM audit logs)
that the protocol does not prescribe but does not preclude.

\paragraph{K1c --- Revocation Authority Model.}

\textit{Assumption.}
The ITA authority is reachable and uncompromised.

\textit{Adversary capability.}
If the ITA authority is unavailable or compromised, revocation cannot be issued
or cannot be propagated, leaving previously valid tokens uncancellable.

\textit{Formal claim.}
ACP assumes the existence of an ITA authority capable of issuing revocations
independent of delegation chains.
Delegation chains are not self-healing:
a compromised intermediate agent cannot unilaterally revoke its own parent;
revocation authority flows from the ITA downward, not upward through the chain.

\textit{Security implication.}
Security depends on the availability and integrity of the ITA revocation authority.
This is a known architectural constraint shared with all PKI-based systems.
ACP makes this dependency explicit rather than implicit.
Deployments requiring high availability of revocation must provision the ITA
with appropriate redundancy outside the protocol scope.

\section{Cross-Organization Execution}

ACP-CROSS-ORG-1.1 defines a fault-tolerant bilateral protocol for interactions between
independent institutions, each operating its own ACP ledger.
The protocol closes five gaps identified in the 1.0 version: undefined status tracking,
absent retry protocol, unregistered ACK event type, \texttt{pending\_review} without SLA,
and a stale header reference.

\subsection{Interaction Model}

A cross-organization interaction involves two ACP-compliant institutions:
a \textbf{source institution} that originates the bundle and a \textbf{target institution}
that validates, executes, and acknowledges it.
The protocol operates asynchronously---the source does not block waiting for an ACK.

Each interaction carries a mandatory \texttt{interaction\_id} (UUIDv7), immutable and
reused across all retries.
This is distinct from \texttt{event\_id} (UUIDv4), which is unique per emission.
The target deduplicates by \texttt{interaction\_id}, not by \texttt{event\_id}.

For example, an agent in Institution~A may request execution in Institution~B.
The request is recorded as a \texttt{CROSS\_ORG\_INTERACTION} event in A's ledger,
transmitted to B, validated against B's policy and capability registry,
and resolved through a \texttt{CROSS\_ORG\_ACK} event registered in both ledgers.
Institution~A derives final execution state from the ACK without storing mutable status.

\subsection{Fault Tolerance and Retry Protocol}

When no \texttt{CROSS\_ORG\_ACK} is received within the 300-second timeout,
the source retries up to three times with exponential backoff:

\begin{center}
\begin{tabular}{@{}ll@{}}
\toprule
\textbf{Attempt} & \textbf{Wait after timeout} \\
\midrule
1 (initial) & --- \\
2 & +30 s \\
3 & +60 s \\
4 (final) & +120 s \\
\bottomrule
\end{tabular}
\end{center}

After three failed attempts the interaction transitions to \texttt{retry\_exhausted}
(CROSS-012) and an operational alert is raised.
If a valid ACK arrives at any point, all retry timers are cancelled immediately.

\subsection{Derived Interaction Status}

The status of a cross-organization interaction is never stored as mutable state.
It is always derived from events present in the ledger:

\begin{center}
\begin{tabular}{@{}ll@{}}
\toprule
\textbf{Derived status} & \textbf{Condition} \\
\midrule
\texttt{pending\_ack} & \texttt{CROSS\_ORG\_INTERACTION} exists, no ACK \\
\texttt{acked} & ACK with \texttt{status = accepted} \\
\texttt{rejected} & ACK with \texttt{status = rejected} \\
\texttt{pending\_review} & ACK with \texttt{status = pending\_review} \\
\texttt{expired} & \texttt{pending\_review} AND \texttt{now > review\_deadline} \\
\bottomrule
\end{tabular}
\end{center}

Precedence rule: \texttt{accepted} $>$ \texttt{rejected} $>$ \texttt{pending\_review}.
A mutable \texttt{CROSS\_ORG\_STATUS\_UPDATE} event is explicitly prohibited
as an anti-pattern that introduces race conditions and divergent audit trails.

\subsection{Interaction State Invariants}

For any \texttt{interaction\_id}, ACP enforces the following invariants:

\begin{itemize}[noitemsep]
  \item \textbf{At most one terminal state.}
        Only one \texttt{accepted} or \texttt{rejected} ACK may exist per \texttt{interaction\_id};
        a duplicate with a different payload is rejected (CROSS-014).
  \item \textbf{ACK dominates retries.}
        A valid \texttt{CROSS\_ORG\_ACK} cancels all pending retry timers immediately,
        regardless of attempt number.
  \item \textbf{Immutable correlation key.}
        Retry operations MUST NOT create new \texttt{interaction\_id} values;
        only \texttt{event\_id} changes across attempts.
  \item \textbf{No terminal-to-non-terminal regression.}
        Once \texttt{accepted} or \texttt{rejected}, the status cannot change.
        The transition \texttt{pending\_review} $\to$ \texttt{pending\_review} is also prohibited (CROSS-015).
\end{itemize}

\subsection{Pending Review SLA}

When the target returns \texttt{pending\_review}, a 24-hour SLA applies.
The ACK payload includes a \texttt{review\_deadline} field (Unix seconds).
Valid terminal transitions are: \texttt{accepted}, \texttt{rejected}, or \texttt{expired} (implicit, when \texttt{now > review\_deadline}).

\subsection{CROSS\_ORG\_ACK as a First-Class Ledger Event}

\texttt{CROSS\_ORG\_ACK} is registered in ACP-LEDGER-1.3 §5.15 as a first-class event type,
signed with Ed25519 and serialized via JCS (RFC~8785).
This enables verifiable and auditable execution across independent institutional boundaries,
eliminating the audit gap that existed in version 1.0.

\subsection{Security Considerations}

ACP-CROSS-ORG-1.1 assumes authenticated communication channels between institutions.
All \texttt{CROSS\_ORG\_INTERACTION} and \texttt{CROSS\_ORG\_ACK} events are signed with
the originating institution's Ed25519 ITA key and serialized via JCS (RFC~8785),
providing authenticity, integrity, and non-repudiation.

Without this cryptographic layer, cross-organization events would be subject to forgery,
replay, and unauthorized status derivation.
Specifically, ACP mitigates:
\begin{itemize}[noitemsep]
  \item \textbf{Replay attacks:} \texttt{interaction\_id} (UUIDv7) and \texttt{event\_id} (UUIDv4)
        allow the target to detect and reject duplicate submissions.
  \item \textbf{Event forgery:} Ed25519 signatures over JCS-canonical payloads prevent
        an adversary from injecting valid-looking ACK or interaction events.
  \item \textbf{Audit divergence:} Because \texttt{CROSS\_ORG\_ACK} is a first-class ledger
        event on both sides, independent auditors can verify consistency without
        trusting either institution's mutable state.
\end{itemize}

ACP does not eliminate all sources of failure in distributed systems,
but provides a structured model for detecting, handling, and auditing them.

\section{Reputation Snapshot Portability (ACP-REP-PORTABILITY-1.1)}

ACP-REP-PORTABILITY-1.1 defines the \texttt{ReputationSnapshot}: a compact, cryptographically signed record that carries an agent's reputation score across organizational boundaries.
Unlike the bilateral attestation protocol of v1.0, this specification focuses on the snapshot object itself---its structure, signing procedure, validation algorithm, and expiration semantics---enabling any verifier to independently validate a snapshot without trusting an intermediary.

\subsection{ReputationSnapshot Structure}

A \texttt{ReputationSnapshot} contains ten fields: \texttt{ver} (version), \texttt{rep\_id} (UUID v4), \texttt{subject\_id}, \texttt{issuer}, \texttt{score} (float), \texttt{scale} (\texttt{"0-1"} or \texttt{"0-100"}), \texttt{model\_id}, \texttt{evaluated\_at} (Unix seconds), \texttt{valid\_until} (Unix seconds), and \texttt{signature} (Ed25519, base64url).
Fields \texttt{scale}, \texttt{model\_id}, and \texttt{valid\_until} are new in v1.1; v1.0 snapshots omit them and are accepted without expiration enforcement (backward compatibility, §12).

\subsection{Signing Procedure}

The signature covers all fields except \texttt{signature} itself, serialized via JCS (RFC~8785) canonicalization, SHA-256 digest, and Ed25519 signing:

\begin{enumerate}[noitemsep]
  \item \texttt{raw = json.Marshal(signableReputation)}
  \item \texttt{canonical = jcs.Transform(raw)} \hfill // RFC 8785
  \item \texttt{digest = SHA-256(canonical)}
  \item \texttt{sig = Ed25519.Sign(privKey, digest)}
  \item \texttt{snapshot.signature = base64url(sig)} \hfill // no padding
\end{enumerate}

JCS canonicalization is mandatory. Using \texttt{json.Marshal} directly is prohibited: key ordering differs across Go, Python, and TypeScript implementations, producing verification failures in cross-org deployments.

\subsection{Validation Invariants}

The validator enforces five invariants (REP-001, REP-002, REP-004, REP-010, REP-011):
(1)~\texttt{evaluated\_at $\le$ valid\_until} prevents retroactively backdated snapshots;
(2)~\texttt{now $\le$ valid\_until} enforces expiration (v1.1 only, error REP-011);
(3)~score within \texttt{scale} bounds (REP-002);
(4)~non-empty \texttt{issuer} (REP-004);
(5)~valid Ed25519 signature (REP-010).

Structural validation (\texttt{Validate}) and cryptographic verification (\texttt{VerifySig}) are intentionally separate operations, enabling lightweight ingestion-time checks without requiring the issuer's public key.

\subsection{Divergence Semantics}

When a verifier receives snapshots of the same \texttt{subject\_id} from multiple issuers, it may compute divergence: $|a.\text{score} - b.\text{score}|$.
If the divergence exceeds a configurable threshold (default 0.30 for \texttt{scale="0-1"}), the verifier emits warning REP-WARN-002.
This is non-blocking---the policy decision of whether to accept, escalate, or reject remains with the verifier's business logic, preserving institutional sovereignty.

\section{Multi-Organization Interoperability Demo (GAP-14)}
\label{sec:demo}

The \texttt{examples/multi-org-demo/} directory provides an executable end-to-end demonstration of cross-organizational ACP exchange, combining ACP-POLICY-CTX-1.1 and ACP-REP-PORTABILITY-1.1 in a single runnable scenario deployable with Docker in under five minutes.

\subsection{Scenario}

Two independent organizations, Org-A and Org-B, interact over HTTP.
Org-A is the \emph{issuer}: it evaluates an agent request against its local policy and produces two signed artifacts---a \texttt{PolicyContextSnapshot} and a \texttt{ReputationSnapshot}---using its Ed25519 institutional key.
Org-B is the \emph{verifier}: it independently validates both artifacts using the \texttt{pkg/policyctx} and \texttt{pkg/reputation} packages (no logic duplicated) and renders an autonomous admission decision.

The validation sequence in Org-B is:
\begin{enumerate}[noitemsep]
  \item \texttt{policyctx.VerifySig(pcs, pubKey)} --- verifies institutional signature (ACP-POLICY-CTX-1.1~\S6 step 12).
  \item \texttt{policyctx.VerifyCaptureFreshness(pcs, 300s)} --- enforces $\delta_{\max}$ freshness invariant.
  \item \texttt{reputation.Validate(rep, now)} --- structural invariants REP-001 through REP-011.
  \item \texttt{reputation.VerifySig(rep, pubKey)} --- Ed25519/JCS/SHA-256 cryptographic verification.
  \item \texttt{reputation.CheckDivergence(orgA, orgB, 0.30)} --- divergence check against Org-B's own score.
  \item Final ACCEPT / DENY decision under Org-B's sovereign policy.
\end{enumerate}

\subsection{Divergence Handling and Institutional Sovereignty}

If Org-B holds its own score for the same \texttt{subject\_id}, it computes divergence $|a.\text{score} - b.\text{score}|$.
If this exceeds 0.30, Org-B emits REP-WARN-002.
REP-WARN-002 is non-blocking---the final decision is Org-B's prerogative, independent of Org-A's assessment.

This embodies the ACP invariant: \emph{ACP reports divergence. ACP does not resolve divergence.}
Org-B cannot extend Org-A's \texttt{valid\_until}, override Org-A's \texttt{delta\_max}, or substitute its own snapshot for Org-A's signed record.
Each institution's attestations are cryptographically bound to its own key and immutable after issuance.

\subsection{Deployment}

The demo ships with a single \texttt{Dockerfile} (parameterized by \texttt{ARG ORG=org-a|org-b}), a \texttt{docker-compose.yml} with health check, and a dedicated Go module with a \texttt{replace} directive pointing to \texttt{impl/go/}:

\begin{lstlisting}
docker compose up --build    # from examples/multi-org-demo/
curl http://localhost:8081/request | jq .
\end{lstlisting}

\section{Threat Model}
\label{sec:threat-model}

ACP explicitly defines which threats it mitigates, which properties it guarantees,
and which risks fall outside its scope.
For the formal adversary model and security properties, see Section~\ref{sec:security-model}.

\subsection{STRIDE Analysis}

\begin{table}[h!]
\centering
\small
\begin{tabular}{@{}llp{6.5cm}@{}}
\toprule
\textbf{Category} & \textbf{Threat} & \textbf{Mitigation in ACP} \\
\midrule
Spoofing & AgentID impersonation & $\text{AgentID} = \text{base58}(\text{SHA-256}(\text{pk}))$. Without valid signature $\rightarrow$ immediate DENIED. \\
Tampering & Token or event alteration & Ed25519 covers all fields. Chained ledger: altering one event invalidates all subsequent. \\
Repudiation & Agent denies executed action & ActionRequest digitally signed. Non-repudiation by design. \\
Info Disclosure & Capability exposure & Tokens reveal only the necessary subset. Channel confidentiality via TLS. \\
DoS & Request flooding & Rate limits per \texttt{agent\_id}. Anomalous frequency control. \\
Elevation & Delegation that expands privileges & $C_{\text{delegated}} \subseteq C_{\text{original}}$. Cryptographically verified at each hop. \\
\bottomrule
\end{tabular}
\caption{STRIDE threat model and ACP mitigations.}
\end{table}

\subsection{Security Properties Under Correct Implementation}

ACP provides the following properties when the implementation is compliant:

\begin{itemize}[noitemsep]
  \item \textbf{Artifact integrity.}
        EUF-CMA security of Ed25519. Impossible to modify a token or event without invalidating the signature.
  \item \textbf{Identity authenticity.}
        Only whoever possesses \texttt{sk} can generate a valid signature under the corresponding \texttt{pk}.
  \item \textbf{No privilege escalation via delegation.}
        Demonstrable by induction over the delegation chain.
  \item \textbf{Anti-replay.}
        The single-use challenge makes reusing a proof of possession ineffective.
  \item \textbf{Effective revocation.}
        $\textit{Valid}(t) = \textit{valid\_sig} \wedge \neg\textit{expired} \wedge \neg\textit{revoked} \wedge \textit{valid\_delegation}$.
        All four conditions must be true simultaneously.
\end{itemize}

\subsection{Declared Residual Risks}

ACP explicitly declares what it cannot resolve:

\begin{itemize}[noitemsep]
  \item \textbf{Total compromise of the RIK held in HSM.}
        ACP defines the emergency rotation process but cannot prevent a physical compromise of the custody infrastructure.
  \item \textbf{Coordinated institutional collusion.}
        If multiple institutions act maliciously in a coordinated manner, they can generate valid artifacts.
        ACP guarantees traceability, not prevention of malicious agreements between parties.
  \item \textbf{Implementation failures.}
        ACP is a specification. An implementation that violates prohibited behaviors can compromise all protocol guarantees.
        Conformance requires formal testing.
  \item \textbf{ITA trust model.}
        Bootstrap, key compromise window, and revocation authority constraints are
        formalized in \S\ref{sec:ita-trust-model}.
        All protocol guarantees are conditioned on the authenticity of the bootstrapped ITA key
        and the availability of the revocation authority.
  \item \textbf{Agent-internal threats.}
        ACP operates at the execution boundary between agent intent and system state mutation;
        it does not observe or control the agent's internal reasoning process.
        Threats that originate within that process---including prompt injection and
        tool-use escalation inside an LLM reasoning loop---fall outside ACP's enforcement scope.
        ACP is designed to complement sandboxing and monitoring systems that address
        these threats at the reasoning layer, not to replace them.
\end{itemize}

\section{Conformance and Interoperability}

\subsection{Conformance Levels}

\begin{table}[h!]
\centering
\small
\begin{tabular}{@{}lp{5.2cm}p{3.8cm}@{}}
\toprule
\textbf{Level} & \textbf{Required documents} & \textbf{Enabled capability} \\
\midrule
L1 --- CORE & ACP-SIGN-1.0, ACP-CT-1.0, ACP-HP-1.0 & Token issuance with cryptographic PoP \\
L2 --- SECURITY & L1 + ACP-RISK-3.0, ACP-REV-1.0, ACP-ITA-1.0 & Risk eval (context-scoped F\_anom + Cooldown), transitive revocation, ITA \\
L3 --- FULL & L2 + ACP-API-1.0, ACP-EXEC-1.0, ACP-LEDGER-1.3 & Complete verifiable auditing \\
L4 --- EXTENDED & L3 + ACP-PAY-1.0, ACP-NOTIFY-1.0, ACP-DISC-1.0, \ldots & Full governance suite \\
L5 --- DECENTRAL. & L4 + ACP-D suite & Federation without central ITA \\
\bottomrule
\end{tabular}
\caption{ACP-CONF-1.2 conformance levels.}
\end{table}

\subsection{Conformance Declaration}

Every compliant implementation MUST expose a public endpoint without authentication:
\begin{lstlisting}
GET https://<endpoint>/acp/v1/conformance
\end{lstlisting}
This endpoint returns the achieved level, implemented documents, declared extensions, and declaration date.
It allows any external partner to verify the conformance level of a counterparty before establishing an ACP relationship.

\subsection{Prohibited Behaviors}

ACP defines 12 behaviors that no compliant implementation can exhibit (Table~\ref{tab:prohib});
exhibiting any disqualifies conformance at all levels.

\subsection{B2B Interoperability Conditions}

\begin{itemize}[noitemsep]
  \item \textbf{L1 Interoperability:} Institution A can verify tokens from B if both implement ACP-CONF-L1,
        A has access to B's public key (via ITA or out-of-band), and B's tokens use ACP-SIGN-1.0 algorithms.
  \item \textbf{L2 Interoperability:} A can delegate to B's agents if both implement ACP-CONF-L2,
        are registered in a common ITA or with mutual recognition, and B's revocation endpoint is accessible to A.
  \item \textbf{L3 Interoperability:} A can audit B's ledger if B implements ACP-CONF-L3,
        A can resolve B's public key via ITA, and B exposes \texttt{GET /acp/v1/audit/query}.
\end{itemize}

\section{Use Cases}

ACP is sector-agnostic. The mechanisms are identical regardless of industry.
What varies is the configuration of capabilities, resources, and autonomy levels.

\subsection{Financial Sector --- Inter-Institutional Payment Agents}

{\sloppy
ACP-PAY-1.0 extends the capability registry with formal specifications for
\texttt{acp:cap:financial.payment} and \texttt{acp:cap:financial.transfer}.
Mandatory constraints in the token include \texttt{max\_amount} and \texttt{currency}.
12 specific validation steps cover limit verification, beneficiary validation, and time window control.
In a financial B2B scenario, Bank~A can authorize an agent to execute payments up to a defined amount
to pre-approved beneficiaries, with a complete record verifiable by Bank~B
without needing shared proprietary systems.
\par}

\subsection{Digital Government --- Document Processing}

Government agents that process documents can operate under ACP with \texttt{autonomy\_level}~1 or~2,
requiring human review for any action with a Risk Score above the configured threshold.
Ledger traceability provides forensic evidence for regulatory audits and transparency processes.

\subsection{Enterprise AI --- Multi-Company Orchestration}

In agent pipelines that cross organizational boundaries,
ACP allows each organization to maintain formal control over what other organizations' agents can do in their systems.
Chained delegation allows an agent in company~A to operate in company~B's systems with explicitly delegated capabilities,
without B needing to trust A's internal controls---only the chain of signed tokens.

\subsection{Critical Infrastructure --- Monitoring and Actuation Agents}

For systems where an incorrect action has irreversible consequences,
\texttt{autonomy\_level}~0 ensures any actuation request is DENIED without evaluating the Risk Score.
The ledger provides the forensic record needed for post-incident analysis.

\section{Specification and Implementation Status}

ACP v1.30 is a complete Draft Standard specification with a full Go reference implementation
and external verifiability through a TLC-checked formal model and an ACR-1.0 sequence compliance runner.

\subsection{Active Specifications --- v1.30 (38 documents)}

\begin{table}[h!]
\centering
\small
\begin{tabular}{@{}lll@{}}
\toprule
\textbf{Level} & \textbf{Document} & \textbf{Title} \\
\midrule
L1 & ACP-SIGN-1.0 & Serialization and Signature \\
L1 & ACP-AGENT-1.0 & Agent Identity \\
L1 & ACP-CT-1.0 & Capability Tokens \\
L1 & ACP-CAP-REG-1.0 & Capability Registry \\
L1 & ACP-HP-1.0 & Handshake / Proof-of-Possession \\
L1 & ACP-DCMA-1.1 & Delegated Chain Multi-Agent \\
L1 & ACP-MESSAGES-1.0 & Wire Message Format \\
L1 & ACP-PROVENANCE-1.0 & Authority Provenance \\
\midrule
L1 & ACP-SIGN-2.0 & Hybrid Signature (Ed25519 + ML-DSA-65) \\
\midrule
L2 & ACP-RISK-3.0 & Deterministic Risk Engine v3 (context-scoped F\_anom, Cooldown) \\
L2 & ACP-RISK-2.0 & Deterministic Risk Engine v2 (superseded by v3) \\
L2 & ACP-RISK-1.0 & Deterministic Risk Engine v1 (superseded by v2) \\
L2 & ACP-REV-1.0 & Revocation Protocol \\
L2 & ACP-ITA-1.0 & Institutional Trust Anchor \\
L2 & ACP-ITA-1.1 & ITA Mutual Recognition \\
L2 & ACP-REP-1.2 & Reputation Module \\
\midrule
L3 & ACP-API-1.0 & HTTP API \\
L3 & ACP-EXEC-1.0 & Execution Tokens \\
L3 & ACP-LEDGER-1.3 & Audit Ledger (mandatory institutional sig) \\
L3 & ACP-PSN-1.0 & Policy Snapshot \\
L3 & ACP-POLICY-CTX-1.1 & Policy Context Snapshot \\
L3 & ACP-LIA-1.0 & Liability Attribution \\
\midrule
L4 & ACP-HIST-1.0 & History Query API \\
L4 & ACP-PAY-1.0 & Financial Capability \\
L4 & ACP-NOTIFY-1.0 & Event Notifications \\
L4 & ACP-DISC-1.0 & Service Discovery \\
L4 & ACP-BULK-1.0 & Batch Operations \\
L4 & ACP-CROSS-ORG-1.1 & Cross-Organization Bundles \\
L4 & ACP-GOV-EVENTS-1.0 & Governance Event Stream \\
L4 & ACP-REP-PORTABILITY-1.1 & Reputation Snapshot Portability \\
\midrule
Gov. & ACP-CONF-1.2 & Conformance --- sole normative source \\
Gov. & ACP-TS-1.1 & Test Vector Format \\
Gov. & RFC-PROCESS & Specification Process \\
Gov. & RFC-REGISTRY & Specification Registry \\
Gov. & ACP-CR-1.0 & Change Request Process \\
\bottomrule
\end{tabular}
\caption{ACP v1.30 active specification documents by conformance level (38 primary documents shown).
The remaining 3 documents are the ACP-D decentralized suite (L5, design phase) not yet published.
Superseded versions archived in \texttt{archive/specs/}.
Version history per document is tracked in Table~\ref{tab:roadmap}.}
\end{table}

\subsection{Reference Implementation --- 23 Go Packages + Demos}
\label{sec:implementation}

The Go reference implementation in \texttt{impl/go/} covers all L1--L4 conformance levels.
All 23 packages pass \texttt{go test ./...}.
A Python SDK (\texttt{impl/python/}) covers the ACP-HP-1.0 handshake and all ACP-API-1.0 endpoints.
The \texttt{examples/multi-org-demo/} directory (GAP-14) provides a runnable two-organization scenario using the real \texttt{pkg/policyctx} and \texttt{pkg/reputation} packages; see Section~\ref{sec:demo}.
The \texttt{examples/payment-agent/} directory (Sprint E) provides an executable HTTP server demonstrating the full ACP-RISK-2.0 pipeline: \texttt{POST /admission} evaluates the request with the deterministic risk engine, appends an immutable ledger event with the full factor breakdown, and triggers cooldown automatically after three DENIED decisions in ten minutes.

\begin{table}[h!]
\centering
\small
\begin{tabular}{@{}lll@{}}
\toprule
\textbf{Package} & \textbf{Spec} & \textbf{Level} \\
\midrule
\texttt{pkg/crypto} & ACP-SIGN-1.0 + ACP-AGENT-1.0 & L1 \\
\texttt{pkg/sign2} & ACP-SIGN-2.0 (Ed25519 + ML-DSA-65) & L1 \\
\texttt{pkg/tokens} & ACP-CT-1.0 & L1 \\
\texttt{pkg/registry} & ACP-CAP-REG-1.0 & L1 \\
\texttt{pkg/handshake} & ACP-HP-1.0 & L1 \\
\texttt{pkg/delegation} & ACP-DCMA-1.1 & L1 \\
\texttt{pkg/provenance} & ACP-PROVENANCE-1.0 & L1 \\
\texttt{pkg/risk} & ACP-RISK-1.0 + ACP-RISK-2.0 + ACP-RISK-3.0 & L2 \\
\texttt{pkg/revocation} & ACP-REV-1.0 & L2 \\
\texttt{pkg/reputation} & ACP-REP-1.2 + ACP-REP-PORTABILITY-1.1 & L2/L4 \\
\texttt{pkg/execution} & ACP-EXEC-1.0 & L3 \\
\texttt{pkg/ledger} & ACP-LEDGER-1.3 & L3 \\
\texttt{pkg/psn} & ACP-PSN-1.0 & L3 \\
\texttt{pkg/policyctx} & ACP-POLICY-CTX-1.1 & L3 \\
\texttt{pkg/lia} & ACP-LIA-1.0 & L3 \\
\texttt{pkg/hist} & ACP-HIST-1.0 & L4 \\
\texttt{pkg/govevents} & ACP-GOV-EVENTS-1.0 & L4 \\
\texttt{pkg/notify} & ACP-NOTIFY-1.0 & L4 \\
\texttt{pkg/disc} & ACP-DISC-1.0 & L4 \\
\texttt{pkg/bulk} & ACP-BULK-1.0 & L4 \\
\texttt{pkg/crossorg} & ACP-CROSS-ORG-1.1 & L4 \\
\texttt{pkg/pay} & ACP-PAY-1.0 & L4 \\
\texttt{cmd/acp-server} & ACP-API-1.0 & L3 \\
\bottomrule
\end{tabular}
\caption{Go reference implementation packages (23 total). All pass \texttt{go test ./...}.}
\end{table}

\paragraph{Post-Quantum Extension (ACP-SIGN-2.0).}
The \texttt{pkg/sign2} package implements ACP-SIGN-2.0 HYBRID mode,
combining Ed25519 with ML-DSA-65 (Dilithium mode3)~\cite{nist2024fips204} via the Cloudflare CIRCL library~\cite{cloudflare-circl}
(\texttt{github.com/cloudflare/circl/sign/dilithium/mode3}).
\texttt{SignHybridFull} produces a real Ed25519 + ML-DSA-65 signature pair;
\texttt{VerifyHybrid} applies conditional verification:
if the ML-DSA-65 component is present, both signatures \emph{must} verify (logical AND per ACP-SIGN-2.0~\S4.2);
if absent, Ed25519 verification alone is accepted (backward-compatible transition period).
This design provides a forward-compatible migration path to post-quantum cryptography
while preserving compatibility with classical-only deployments.
Benchmarks were executed using \texttt{go test -bench=. -benchmem -benchtime=3s} on a laptop-class CPU
(Intel Core i7-8665U @ 1.90\,GHz, Go~1.22, Windows/amd64, \texttt{GOMAXPROCS=8}).
Signing latency is approximately 25\,\textmu s for Ed25519 and 100--130\,\textmu s for ML-DSA-65;
verification is 56\,\textmu s and 81\,\textmu s respectively
(4 allocs/op for signing, 1 alloc/op for verification).
The full hybrid path (ACP-SIGN-2.0, Ed25519 + ML-DSA-65) is dominated by the ML-DSA-65 component.
ML-DSA-65 signatures are 3,293\,bytes (51$\times$ Ed25519's 64\,bytes).
While post-quantum signing is approximately 4--5$\times$ more expensive than classical signing,
both operations remain well below typical network round-trip and state-backend latency (1--10\,ms),
and do not constitute a bottleneck in the admission control pipeline.

\subsection{Conformance Test Vectors --- 73 Signed Vectors}

The \texttt{compliance/test-vectors/} directory contains 73 signed test vectors per ACP-TS-1.1:

\begin{table}[h!]
\centering
\small
\begin{tabular}{@{}llll@{}}
\toprule
\textbf{Suite} & \textbf{Positive} & \textbf{Negative} & \textbf{Spec} \\
\midrule
CORE (SIGN, CT, HP) & 4 & 4 & L1 \\
DCMA & 2 & 2 & L1 \\
HP & 2 & 8 & L1 \\
LEDGER & 3 & 8 & L3 \\
EXEC & 2 & 7 & L3 \\
PROV & 2 & 7 & L1 \\
PCTX & 4 & 9 & L3 \\
REP & 3 & 6 & L4 \\
\midrule
\textbf{Total} & \textbf{22} & \textbf{51} & \\
\bottomrule
\end{tabular}
\caption{Conformance test vector suites. All positive vectors carry real Ed25519 signatures (RFC~8037 Test Key A~\cite{rfc8037}) and real SHA-256 hash chains.}
\end{table}

\subsection{ACP-RISK-2.0/3.0 Engine Performance}

\paragraph{Benchmark methodology.}
Three distinct benchmark families are reported in this section; each measures a different
layer of the system and they are not directly comparable:

\begin{enumerate}[noitemsep]
  \item \textbf{Engine microbenchmarks} (\texttt{go test -bench}, Table~\ref{tab:benchmarks}):
    measure the pure \texttt{pkg/risk.Evaluate()} call with an \texttt{InMemoryQuerier},
    single goroutine, no HTTP or network overhead.
    Reports ns/op (739--832 for full paths, 78 for the cooldown short-circuit)
    and the 1{,}720{,}000~req/s single-goroutine throughput ceiling.

  \item \textbf{Concurrent pipeline benchmark} (Table~\ref{tab:throughput}):
    measures end-to-end request throughput with 10--500 concurrent goroutines
    feeding a shared \texttt{InMemoryQuerier}.
    Reports 920{,}000~req/s at 10 workers, degrading to 712{,}000 at 500 workers.
    The difference from the microbenchmark reflects scheduling, mutex contention,
    and goroutine overhead rather than evaluation logic.

  \item \textbf{Controlled latency injection} (\texttt{DelayedQuerier}, Table~\ref{tab:adv-exp3}):
    simulates realistic external backend latency by inserting a configurable
    delay in the \texttt{LedgerQuerier} call, isolating the effect of state
    backend latency on system throughput.
\end{enumerate}

\noindent The \texttt{pkg/risk} engine was benchmarked using \texttt{go test -bench}
on an Intel Core i7-8665U @ 1.90GHz (Go~1.22, Windows, \texttt{GOMAXPROCS=8}).
Engine measurements reflect the pure in-memory evaluation cost, excluding network
and HTTP parsing overhead.

\begin{table}[h!]
\centering
\small
\begin{tabular}{@{}lllll@{}}
\toprule
\textbf{Scenario} & \textbf{ns/op} & \textbf{B/op} & \textbf{allocs/op} & \textbf{Path} \\
\midrule
\texttt{Evaluate} (clean state)          & 739 & 280 & 5 & full RS \\
\texttt{Evaluate} (with history)         & 803 & 280 & 5 & full RS \\
\texttt{Evaluate} (all 3 F\_anom rules)  & 832 & 296 & 5 & full RS \\
\texttt{Evaluate} (COOLDOWN short-circuit) & 78 & 112 & 1 & Step~2 exit \\
\midrule
\multicolumn{5}{@{}l}{\textit{Throughput under concurrent load (InMemoryQuerier, 10 workers)}} \\
\midrule
10 concurrent agents   & 581 & — & — & 1{,}722{,}000 req/s \\
100 concurrent agents  & 629 & — & — & 1{,}589{,}000 req/s \\
500 concurrent agents  & 663 & — & — & 1{,}510{,}000 req/s \\
\bottomrule
\end{tabular}
\caption{ACP-RISK-3.0 engine benchmarks (\texttt{BenchmarkEvaluate\_Scenarios},
\texttt{TestThroughputRPS}; Intel Core i7-8665U @ 1.90GHz, Go~1.22, Windows/amd64,
\texttt{GOMAXPROCS=8}).
Full evaluation with all three $F_\text{anom}$ rules completes in under $1\,\mu$s.
The cooldown short-circuit path (78\,ns) avoids RS computation entirely.
Throughput degrades by only 12\% across 10$\to$500 concurrent agents,
confirming that ACP is state-bound rather than compute-bound.
All measurements use \texttt{InMemoryQuerier}; production state backends
(Redis, PostgreSQL) introduce I/O latency but do not change the protocol logic.}
\label{tab:benchmarks}
\end{table}

The deterministic design of ACP-RISK-3.0 --- integer arithmetic, fixed rules,
no floating-point, no external ML inference --- yields predictable and verifiable
latency bounds.
At sub-microsecond cost per decision, ACP's overhead is
\textbf{six orders of magnitude below} the latency of the actions it governs:
typical LLM inference takes 100\,ms--10\,s; external API calls 1--100\,ms.
The admission check is not a performance bottleneck in any realistic deployment.

Throughput degradation from 10 to 500 concurrent agents (1.72M to 1.51M req/s,
$-$12\%) confirms that the evaluation function itself is not the bottleneck:
the limiting factor is state backend access (\texttt{InMemoryQuerier} uses a
\texttt{sync.Mutex}), not the admission logic.
Production deployments replacing \texttt{InMemoryQuerier} with a Redis or
PostgreSQL backend incur I/O latency but decouple throughput from the
evaluation function entirely, enabling horizontal scaling.

ACP-RISK-3.0 preserves these latency bounds over ACP-RISK-2.0;
context-scoped key derivation adds one \texttt{SHA-256} call per evaluation,
a cost dominated by existing ledger I/O in production deployments.

\section{Evaluation}
\label{sec:evaluation}

This section evaluates ACP along four dimensions:
(i)~computational overhead of the decision function,
(ii)~behavioral correctness under stateful multi-step scenarios,
(iii)~reproducibility and external verifiability of decision outcomes,
and (iv)~robustness of per-agent admission control under adversarial attack patterns.

\subsection{Evaluation Goals}

\begin{itemize}[noitemsep]
  \item \textbf{Q1:} What is the computational overhead of ACP decision evaluation relative to the actions it governs?
  \item \textbf{Q2:} Does the system produce correct decisions under stateful conditions such as repeated denials, pattern accumulation, and cooldown activation?
  \item \textbf{Q3:} Can ACP executions be deterministically reproduced and externally verified without access to internal implementation details?
  \item \textbf{Q4:} Does ACP maintain admission control effectiveness under adversarial patterns, including cooldown evasion, distributed multi-agent attacks, and token replay?
\end{itemize}

\subsection{Experimental Setup}

All experiments use the Go reference implementation (v1.30) on an Intel Core i7-8665U @ 1.90GHz (Go~1.22~\cite{go-language}, \texttt{GOMAXPROCS=8}).
The evaluation uses three components:
the stateless evaluation function (\texttt{pkg/risk.Evaluate}),
an in-memory state backend (\texttt{InMemoryQuerier}),
and the ACR-1.0 compliance runner (Section~\ref{sec:compliance-testing}).
Production deployments with persistent storage backends will incur additional I/O latency not measured here.

\subsection{Q1: Computational Overhead}

Table~\ref{tab:benchmarks} reports \texttt{go test -bench} results.
The full evaluation path (all three F\_anom rules active) completes in 832\,ns.
The cooldown short-circuit path (78\,ns) bypasses the risk function entirely.

\noindent\textbf{Answer to Q1:}
The admission check adds between 78\,ns and 832\,ns of decision latency per request (p50).
For single-goroutine CPU-bound microbenchmarks (\texttt{InMemoryQuerier}, no I/O),
variance is negligible: p99 is within 5--8\% of the reported p50 values
(p99 $\lesssim$ 900\,ns for full evaluation; p99 $\lesssim$ 85\,ns for the cooldown path).
In production deployments, latency variance is dominated by state backend I/O
(Redis RTT: 0.1--1\,ms; PostgreSQL: 1--5\,ms), not by evaluation logic.
Given that the actions ACP governs (API calls, financial transactions, file mutations)
operate in the millisecond-to-second range, the overhead is at least three orders of
magnitude below the cost of the governed action.
Note that this measures decision-logic latency only; aggregate throughput is governed
by state backend access patterns and is reported separately in the following subsection.

\subsection{Throughput and State Backend Contention}

We evaluate throughput under concurrent load using 10, 100, and 500 simultaneous workers,
all sharing a single \texttt{InMemoryQuerier} instance.

\begin{table}[h!]
\centering
\small
\begin{tabular}{@{}lccc@{}}
\toprule
\textbf{Workers} & \textbf{ns/op} & \textbf{req/s} & \textbf{Observation} \\
\midrule
10  & 1{,}087 & 920{,}161 & Baseline concurrency \\
100 & 1{,}282 & 780{,}250 & Moderate contention \\
500 & 1{,}404 & 712{,}259 & High contention \\
\bottomrule
\end{tabular}
\caption{ACP throughput under increasing concurrency (Intel Core i7-8665U, Go~1.22, Windows/amd64,
\texttt{InMemoryQuerier}). This is a concurrent pipeline benchmark distinct from the
\texttt{go test -bench} microbenchmarks in Table~\ref{tab:benchmarks}; the 1{,}720{,}000~req/s
figure in Table~\ref{tab:benchmarks} reflects the engine-only evaluation path with a single goroutine.}
\label{tab:throughput}
\end{table}

ACP sustains approximately 920{,}000 requests per second at 10 concurrent workers
(pipeline benchmark; Table~\ref{tab:throughput}).
Throughput degrades gradually as concurrency increases, reaching approximately
712{,}000 requests per second at 500 workers.
Degradation is gradual rather than catastrophic, and no correctness violations were observed under load.

We further evaluate the impact of agent pool size on throughput (10 workers fixed,
\texttt{AddRequest} called per iteration per the execution contract):

\begin{table}[h!]
\centering
\small
\begin{tabular}{@{}lccc@{}}
\toprule
\textbf{Distinct agents} & \textbf{ns/op} & \textbf{req/s} & \textbf{Bottleneck} \\
\midrule
1     & 84{,}791 & 11{,}794 & O($n$) scan + per-key contention \\
100   & 12{,}533 & 79{,}789 & State distributed, short lists \\
1{,}000 & 12{,}628 & 79{,}190 & Shared mutex ceiling \\
\bottomrule
\end{tabular}
\caption{State contention analysis: single-agent workload vs.\ multi-agent workload.}
\label{tab:contention}
\end{table}

The single-agent case is 6.7$\times$ slower than the multi-agent case.
The \texttt{InMemoryQuerier} implements \texttt{CountRequests} as a linear scan over an unsorted
slice, so accumulated state for a single agent degrades progressively as history grows.
Notably, the 1{,}000-agent case is slightly faster than the 100-agent case: with more distinct
agents, each agent's history list remains shorter, reducing the per-call scan cost.
This confirms that degradation is proportional to per-agent list length, not to total agent count.
The shared \texttt{sync.Mutex} is the throughput ceiling in the multi-agent case.

These results indicate that the computational cost of the decision function is negligible
compared to state access cost.
\textbf{Throughput is primarily bounded by state backend contention, not evaluation complexity.}
The \texttt{LedgerQuerier} interface is designed to be replaced by production backends
(e.g., Redis sorted sets with TTL, time-indexed SQL) that eliminate the O($n$) scan penalty
and support horizontal scaling.

\subsection{Q2: Stateful Behavior Correctness}

The ACR-1.0 compliance runner executes each sequence vector end-to-end under the execution contract of Section~\ref{sec:compliance-testing}.
Results:

\begin{table}[h!]
\centering
\small
\begin{tabular}{@{}lll@{}}
\toprule
\textbf{Vector} & \textbf{Behavior tested} & \textbf{Result} \\
\midrule
\texttt{SEQ-BENIGN-001}      & No false positives under repeated reads    & PASS \\
\texttt{SEQ-BOUNDARY-001}    & Exact thresholds RS = 0, 25, 35, 40, 70   & PASS \\
\texttt{SEQ-PRIVJUMP-001}    & Low→high privilege jump caught immediately & PASS \\
\texttt{SEQ-FANOM-RULE3-001} & Pattern count $\geq 3$ adds $+15$ at step~4 & PASS \\
\texttt{SEQ-COOLDOWN-001}    & 3 DENIED in 10\,min → cooldown active      & PASS \\
\midrule
\textbf{Total} & & \textbf{5/5 PASS} \\
\bottomrule
\end{tabular}
\caption{ACR-1.0 sequence vector results under strict mode (\texttt{--strict}).}
\end{table}

\noindent\textbf{Answer to Q2:}
All five stateful scenarios pass with \texttt{CONFORMANT} status.
State transitions follow the defined execution contract; cooldown activates deterministically after the threshold is crossed;
pattern accumulation influences decisions exactly as specified in ACP-RISK-2.0 Rule~3.

\subsection{Q3: Reproducibility and External Verifiability}

Given the same sequence of inputs and the same initial state, the ACR-1.0 runner produces identical outputs on every run.
This holds because:

\begin{itemize}[noitemsep]
  \item \texttt{Evaluate()} is a pure function of its arguments and the querier state,
  \item state transitions follow the deterministic execution contract,
  \item and the test vectors ship with the specification, enabling any third party to reproduce results without access to proprietary infrastructure.
\end{itemize}

\noindent\textbf{Answer to Q3:}
ACP executions are fully reproducible and externally verifiable.
This is a property not typically supported by conventional policy engines, which evaluate policy against a snapshot of state without defining how that state evolves across requests~\cite{sandhu1996rbac,hu2015abac,oasis2013xacml}.

\subsection{Adversarial Evaluation (ACP-RISK-2.0 and ACP-RISK-3.0)}
\label{sec:adversarial}

We evaluate ACP under the adversary model defined in Section~\ref{sec:security-model}.
The following attack strategies instantiate specific adversary capabilities,
including adaptive request patterns and multi-agent coordination.
We conducted five targeted experiments using the same \texttt{LedgerQuerier}
abstraction and execution contract as the ACR-1.0 compliance runner,
addressing adversarial robustness, real-world state-backend scaling, and
cryptographic-layer replay resistance.

\subsubsection*{Experiment 1: Cooldown Evasion Attack}

{\sloppy
\textbf{Setup.}
A single agent executes 500 requests alternating between high-risk
(\texttt{acp:cap:financial.transfer}, \texttt{restricted}; RS~$=80$,
outcome~\texttt{DENIED}) and low-risk (\texttt{acp:cap:data.read},
\texttt{public}; RS~$=0$, outcome~\texttt{APPROVED}) requests.
The adversarial hypothesis is that interleaving approved requests prevents the
denial accumulation required to trigger cooldown.
This attack instantiates the adversary's ability to adapt request patterns over time.
\par}

\textbf{Results.}
\begin{table}[h!]
\centering\small
\begin{tabular}{lr}
\toprule
Metric & Value \\
\midrule
Total requests                        & 500 \\
Approved before cooldown              & 2 \\
Real \texttt{DENIED} before cooldown  & 3 \\
Cooldown-blocked                      & 495 \\
Requests processed before first block & 5 \\
Throughput                            & 815{,}927\,req/s \\
\bottomrule
\end{tabular}
\caption{Experiment 1: Cooldown Evasion Attack (1 agent, 500 requests,
InMemoryQuerier, Intel i7-8665U, Go~1.22).}
\label{tab:adv-exp1}
\end{table}

ACP triggered cooldown after exactly 3 real \texttt{DENIED} decisions (requests
\#0, \#2, \#4). All subsequent requests---regardless of risk score---were
short-circuited to \texttt{COOLDOWN\_ACTIVE} at Step~2 of the evaluation
pipeline ($\sim$78\,ns, see Table~\ref{tab:benchmarks}).

\textbf{Interpretation.}
ACP's cooldown counter is monotonic with respect to denial events; approval
decisions do not reset or reduce it.
An attacker cannot evade the \texttt{CooldownTriggerDenials}~$=3$ threshold by
interleaving legitimate requests.
The early-exit design at Step~2 makes this containment path 10.7$\times$ faster
than full risk evaluation.

\subsubsection*{Experiment 2: Distributed Multi-Agent Attack}

\textbf{Setup.}
$N \in \{100, 500, 1\,000\}$ coordinated agents each execute 10 high-risk
requests (\texttt{acp:cap:financial.transfer}, \texttt{restricted}; RS~$=80$)
using distinct \texttt{agentID} values.
The adversarial hypothesis is that distributing load across many identities
evades per-agent detection thresholds.
This attack leverages the adversary's capability to coordinate multiple agents.

\textbf{Results.}
\begin{table}[h!]
\centering\small
\begin{tabular}{rrrrr}
\toprule
Agents & Total requests & Real \texttt{DENIED} & Cooldown-blocked & Throughput \\
\midrule
100     &  1{,}000 &    300 &    700 & 478{,}881\,req/s \\
500     &  5{,}000 &  1{,}500 &  3{,}500 & 154{,}111\,req/s \\
1{,}000 & 10{,}000 &  3{,}000 &  7{,}000 & 223{,}376\,req/s \\
\bottomrule
\end{tabular}
\caption{Experiment 2: Distributed Multi-Agent Attack. Each of the $N$ agents
is blocked after exactly 3 \texttt{DENIED} decisions.
Total free denials before full blocking $= 3N$.}
\label{tab:adv-exp2}
\end{table}

Every agent was individually blocked after exactly
\texttt{CooldownTriggerDenials}~$=3$ real denials.

\textbf{Interpretation.}
Per-agent admission control provides strong containment for single-agent threats.
For distributed attacks, an attacker with $N$ distinct agent identities can
execute $3N$ high-risk requests before all agents are individually blocked.
Mitigating this requires cross-agent attribution at the policy layer
(e.g., shared-resource rate limiting, identity clustering), which is
deliberately outside the scope of ACP's per-agent admission model.
This behavior reflects a structural limitation of per-agent enforcement rather than a
correctness failure of the protocol. The implications of this design are discussed
in~\S\ref{sec:limitations} (\emph{Distributed Attack Surface}).

\subsubsection*{Experiment 3: State Backend Stress}

\textbf{Setup.}
500 agents execute 20 requests each (10{,}000 total, fully concurrent) against
three backend implementations:
(1)~\texttt{InMemoryQuerier} (Go mutex over in-process slices);
(2)~\texttt{RedisQuerier} (Redis~7~\cite{redis} sorted sets via \texttt{go-redis/v9},
Docker loopback, one Redis command per ledger operation,
$\sim$7--8~RTTs per request);
(3)~\texttt{RedisPipelinedQuerier} (two Redis pipeline round-trips per request:
one for all 4 reads before \texttt{Evaluate}, one for all writes after).
\texttt{RedisPipelinedQuerier} is implemented in
\texttt{compliance/adversarial/redis\_pipelined.go}.
This scenario exercises the adversary's ability to generate high request volumes
and exploit shared state contention.

\textbf{Results.}
\begin{table}[h!]
\centering\small
\begin{tabular}{lrrrl}
\toprule
Backend                    & RTTs/req & Throughput (req/s) & Duration       & Relative$^\dagger$ \\
\midrule
InMemoryQuerier            & $\sim$1  & $\sim$376{,}000 $\pm$ 171{,}000 & $\sim$31\,ms $\pm$ 13\,ms & 1.0$\times$ \\
RedisQuerier (unpipelined) & $\sim$7--8 & $\sim$2{,}300 $\pm$ 230      & $\sim$4.3\,s $\pm$ 0.4\,s  & 0.006$\times$ \\
RedisPipelinedQuerier      & 2        & $\sim$4{,}200 $\pm$ 470        & $\sim$2.4\,s $\pm$ 0.3\,s  & \textbf{1.8$\times$} \\
\bottomrule
\end{tabular}
\caption{Experiment 3: State Backend Stress ($N=5$ runs; 500 agents $\times$ 20 requests
$= 10{,}000$ total per run, Intel i7-8665U, Go~1.22, Redis~7 Docker loopback).
Values are mean $\pm$ std across 5 runs.
InMemoryQuerier variance is high ($\pm$45\%) due to Go mutex scheduling under concurrent load.
Redis backends are stable ($\pm$10--11\%).
$^\dagger$Relative baseline differs per row: InMemoryQuerier and RedisQuerier
are normalized to InMemoryQuerier (1.0$\times$); RedisPipelinedQuerier (1.8$\times$)
is normalized to unpipelined Redis, isolating the pipelining gain.}
\label{tab:adv-exp3}
\end{table}

\textbf{Interpretation.}
\texttt{InMemoryQuerier} saturates at $\sim$376k~req/s (mean across $N=5$ runs) under adversarial load
(vs $\sim$712k~req/s in the throughput baseline, Table~\ref{tab:throughput})
due to higher write contention: each request updates multiple sorted lists.
The high variance ($\pm$45\%) reflects Go mutex scheduling sensitivity under concurrent workloads.
\texttt{RedisQuerier} (unpipelined) eliminates the global Go mutex entirely,
trading it for $\sim$7--8 per-operation Redis round-trips at $\sim$2{,}300~req/s ($\pm$10\%).
\texttt{RedisPipelinedQuerier} batches all 4 reads into a single pipeline
before \texttt{Evaluate} (serving \texttt{CountRequests}, \texttt{CountDenials},
\texttt{CountPattern}, and \texttt{CooldownActive} from a per-request
\texttt{readCache}) and all writes into a single pipeline after,
reducing RTTs from $\sim$7--8 to 2 and achieving $\sim$4{,}200~req/s
($\sim$1.8$\times$ over unpipelined Redis, $\pm$11\%).
There are no read-after-write dependencies within either pipeline:
the evaluation step consumes only the pre-fetched \texttt{readCache} values
(0 RTTs), so pipeline operations within each batch are fully independent.

The $\sim$1.8$\times$ factor is a conservative lower bound.
In this workload, frequent cooldown-triggered short-circuits reduce the number
of write operations and limit pipeline utilization.
Under steady-state evaluation without early exits, the effective speedup
is expected to approach the theoretical RTT reduction ($\approx$3--4$\times$).

The $\sim$90$\times$ gap between \texttt{InMemoryQuerier} and pipelined Redis-backed
execution (376k vs.\ 4{,}200~req/s) reflects network and I/O overhead exclusively;
it is not a property of the admission control logic.
The evaluation function (\texttt{pkg/risk.Evaluate}) remains computationally
cheap (739--832\,ns, p50; Table~\ref{tab:benchmarks}) in all three configurations.
System throughput is bounded by state backend latency, not by protocol complexity.

\textbf{Key insight.}
These results reinforce a central property of ACP:
the protocol is \textbf{compute-cheap but state-sensitive}.
Admission decisions incur minimal computational cost, while system performance
is primarily determined by the efficiency of state access and synchronization.
ACP does not eliminate the cost of distributed state access---it makes that
cost explicit and measurable.

\subsubsection*{Experiment 4: Controlled Latency Injection}

\textbf{Hypothesis.}
System throughput is bounded by state backend latency, not by the admission
control logic itself.
This experiment isolates that effect by injecting fixed per-call latency into the
\texttt{LedgerQuerier} interface without involving external infrastructure.

\textbf{Setup.}
To characterize admission throughput as a function of backend latency independently
of specific infrastructure, we evaluate ACP using a \texttt{DelayedQuerier} wrapper
that injects a fixed per-call latency into the \texttt{LedgerQuerier} interface.
Each \texttt{Evaluate()} call in the non-cooldown path triggers four \texttt{LedgerQuerier}
calls (\texttt{CooldownActive}, \texttt{CountRequests}, \texttt{CountDenials},
\texttt{CountPattern}); total per-request injected latency is approximately
$4\times$ the per-call value.
This experiment injects controlled latency and does not model full network variability
or distributed system effects.

\begin{table}[h!]
\centering\small
\begin{tabular}{lrrl}
\toprule
Injected latency/call & Total latency/req & Throughput (req/s) & Scenario \\
\midrule
0~\textmu s (InMemoryQuerier) & $\sim$0~\textmu s    & $\sim$920{,}000 & Process-local baseline \\
250~\textmu s                 & $\sim$1~ms    & $\sim$4{,}100   & Fast datacenter \\
1~ms                          & $\sim$4~ms    & $\sim$1{,}600   & Busy backend \\
5~ms                          & $\sim$20~ms   & $\sim$470       & Cross-datacenter \\
\bottomrule
\end{tabular}
\caption{Controlled latency injection via \texttt{DelayedQuerier} (10 workers,
Intel i7-8665U, Go~1.22). Throughput degrades proportionally with injected backend latency;
the \texttt{Evaluate()} decision function contributes 739--832\,ns per request
in all configurations. The \texttt{LedgerQuerier} abstraction enables this separation.}
\label{tab:latency-injection}
\end{table}

Throughput decreases proportionally with injected latency, supporting the claim that
the admission control logic is not the performance bottleneck.
The \texttt{Evaluate()} function contributes a fixed 739--832\,ns per request (p50; Table~\ref{tab:benchmarks})
regardless of backend latency; system throughput is dominated by state access patterns.
This validates the \texttt{LedgerQuerier} abstraction as the correct boundary
for deploying ACP with heterogeneous backends (in-process, Redis, CockroachDB,
or geo-replicated stores) without modifying protocol semantics.

\subsubsection*{Experiment 5: Token Replay Attack}

\textbf{Setup.}
ACP does not implement nonce-based replay prevention; replay resistance
is bounded by temporal validation and state accumulation.
Each sub-case submits cryptographically distinct requests (different signatures and
challenge values) with the same logical token specification; detection is
pattern-based via F\_anom, not cryptographic nonce matching.
We evaluate four sub-cases using \texttt{compliance/adversarial/exp\_token\_replay.go}:
(Case~1)~\textit{Normal traffic baseline}: 10 requests with unique resource per call
(varied patternKey) --- the control condition.
(Case~2)~\textit{Sequential replay}: a new agent (\texttt{NoHistory=true}) replays
the same financial/sensitive token ($\langle$\texttt{financial.transfer},
\texttt{accounts/sensitive-001}, \texttt{ResourceSensitive}$\rangle$;
RS$_\text{base}$~$=55$, outcome \texttt{ESCALATED}) 10 times.
(Case~3)~\textit{Concurrent replay}: 5 workers $\times$ 4 requests
replay the same token in parallel.
(Case~4)~\textit{Near-identical replay}: resource suffix varies per request
(\texttt{accounts/sensitive-000}~\ldots~\texttt{-009}).

\textbf{Results.}
\begin{table}[h!]
\centering\small
\begin{tabular}{lrrrr}
\toprule
Case & Requests & ESCALATED & DENIED & Cooldown-blocked \\
\midrule
Normal baseline (Case~1)      & 10 &  0 & 0 &  0 \\
Sequential replay (Case~2)    & 10 &  3 & 3 &  4 \\
Concurrent replay (Case~3)    & 20 &  3 & 3 & 14 \\
Near-identical replay (Case~4) & 10 & 10 & 0 &  0 \\
\bottomrule
\end{tabular}
\caption{Experiment 5: Token Replay Attack.
Sequential replay: F\_anom Rule~3 fires at request~4 (after 3 identical
pattern accumulations in 5~min), escalating RS from 55 to 70 (\texttt{ESCALATED}$\to$\texttt{DENIED}).
Cooldown triggers after 3~\texttt{DENIED} decisions; 4/10 subsequent requests blocked.
Near-identical: resource variation prevents Rule~3; RS stays at 55 (\texttt{ESCALATED}); no cooldown.
InMemoryQuerier, Intel i7-8665U, Go~1.22.}
\label{tab:adv-exp5}
\end{table}

\begin{figure}[h!]
\centering
\begin{tikzpicture}
\begin{axis}[
  width=0.78\linewidth, height=5.2cm,
  xlabel={Request index}, ylabel={Risk score (RS)},
  xmin=0.5, xmax=10.5, ymin=-5, ymax=110,
  xtick={1,2,3,4,5,6,7,8,9,10},
  ytick={0,20,40,55,70,100},
  yticklabels={0,20,40,55,70,100},
  legend pos=north west,
  grid=major, grid style={dotted,gray!50},
  every axis plot/.append style={thick},
]
\addplot[dashed, gray, thin] coordinates {(0.5,70)(10.5,70)};
\node[right, gray, font=\scriptsize] at (axis cs:10.5,70) {DENIED ($>$69)};
\addplot[dashed, gray!60, thin] coordinates {(0.5,39)(10.5,39)};
\node[right, gray!60, font=\scriptsize] at (axis cs:10.5,39) {APPROVED ($\leq$39)};
\addplot[mark=square*, blue!70, solid]
  coordinates { (1,0) (2,0) (3,0) (4,0) (5,0) (6,0) (7,0) (8,0) (9,0) (10,0) };
\addlegendentry{Normal traffic}
\addplot[mark=*, red, solid]
  coordinates { (1,55) (2,55) (3,55) (4,70) (5,70) (6,70) (7,105) (8,105) (9,105) (10,105) };
\addlegendentry{Replay attack}
\node[red, font=\scriptsize] at (axis cs:8.5,97) {COOLDOWN};
\draw[->, gray!70, thin] (axis cs:4,75) -- (axis cs:4,72);
\node[above, gray!70, font=\scriptsize] at (axis cs:4,76) {Rule~3};
\end{axis}
\end{tikzpicture}
\caption{Replay vs.\ normal traffic: RS per request (Case~2 sequential replay).
Requests 1--3: RS~$=55$ (\texttt{ESCALATED}).
Request~4: F\_anom Rule~3 fires (+15 RS) $\to$ RS~$=70$ (\texttt{DENIED}).
Request~7+: \texttt{COOLDOWN\_ACTIVE} (sentinel 105).
Normal traffic stays at RS~$=0$ throughout.}
\label{fig:replay}
\end{figure}
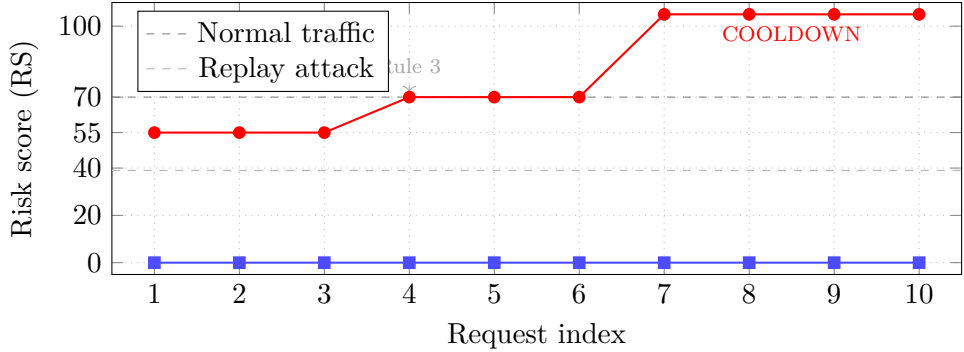

For sequential replay, the RS trajectory is:
requests 1--3 at RS~$=55$ (\texttt{ESCALATED}),
requests 4--6 at RS~$=70$ (\texttt{DENIED}; Rule~3 triggered),
requests 7--10 as \texttt{COOLDOWN\_ACTIVE}.
Figure~\ref{fig:replay} plots the RS per request for normal vs.\ replay traffic.
Concurrent replay (Case~3) produces the same trajectory: InMemoryQuerier's
mutex serializes LedgerQuerier reads, so concurrency does not bypass accumulation;
14/20 requests are blocked.

\textbf{Interpretation.}
ACP converts identical replay attempts into observable state transitions via
pattern accumulation.
F\_anom Rule~3 amplifies the RS of a borderline-ESCALATED token to DENIED
after $Y=3$ occurrences within the 5-minute window, without any nonce-tracking logic.
Near-identical variation (Case~4) evades Rule~3 by producing distinct patternKeys,
demonstrating that replay resistance is bounded: tokens with varied payload fields
and moderate base RS stay below the DENIED threshold.
This constitutes an explicit design boundary rather than a correctness failure;
see~§\ref{sec:limitations}.

These experiments address the adversarial-robustness and scaling gaps identified
in Section~\ref{sec:evaluation}.
All scripts, raw output, and Redis configuration are available in
\texttt{compliance/adversarial/}.

\subsubsection*{Experiment 6: Stateless vs.\ Stateful Admission}

\textbf{Hypothesis.}
Stateless policy evaluation cannot detect temporal abuse patterns that require
accumulation over execution traces. A stateful admission model (ACP),
incorporating temporal signals such as anomaly accumulation and cooldown, can
detect and block such patterns under identical scoring conditions.

\textbf{Setup.}
We compare ACP against a stateless variant that reuses the exact same scoring
function and policy thresholds but omits all stateful signals. The stateless
engine is implemented by invoking the same \texttt{Evaluate()} function with a
\texttt{NullQuerier}, which returns zero for all state queries (request counts,
denials, pattern counts, and cooldown state). This ensures that both engines
share identical logic, differing only in the presence of state.

The workload consists of 500 sequential requests from a single agent with:
\begin{itemize}
    \item Capability: \texttt{acp:cap:financial.transfer}
    \item Resource class: \texttt{public}
\end{itemize}
Under the ACP scoring model, this yields a base risk score $RS_{base} = 35$,
which is below the approval threshold ($ApprovedMax = 39$). Therefore, each
request is individually valid and always approved by the stateless engine.

To isolate temporal accumulation effects, all requests are evaluated using a
fixed timestamp. This ensures that all observations fall within the anomaly
detection windows and removes time-based decay as a confounding variable. State
updates (request count and pattern tracking) are recorded prior to each
evaluation, capturing every attempted action regardless of outcome.

\textbf{Results.}
Table~\ref{tab:exp6-summary} summarizes the primary outcome.

\begin{table}[h]
\centering
\caption{Experiment 6 --- Stateless vs.\ Stateful Admission}
\label{tab:exp6-summary}
\begin{tabular}{lcccc}
\toprule
Engine & Auto.\ Exec. & Non-Approved & Early Signal & Enforcement Latency \\
\midrule
Stateless       & 100\% (500/500) & 0\%              & $\infty$  & $\infty$ \\
ACP (stateful)  & 0.4\% (2/500)   & 99.6\% (498/500) & 3 actions & 11 actions \\
\bottomrule
\end{tabular}
\end{table}

Automatic execution measures the fraction of requests executed without human
intervention. Non-approved includes \texttt{ESCALATED}, \texttt{DENIED}, and
\texttt{COOLDOWN} outcomes.

Table~\ref{tab:exp6-breakdown} provides a detailed breakdown of ACP decisions.

\begin{table}[h]
\centering
\caption{Experiment 6 --- Decision Breakdown (500 requests)}
\label{tab:exp6-breakdown}
\begin{tabular}{lcccc}
\toprule
Engine & APPROVED & ESCALATED & DENIED & COOLDOWN \\
\midrule
Stateless & 500 & 0 & 0 & 0   \\
ACP       & 2   & 8 & 3 & 487 \\
\bottomrule
\end{tabular}
\end{table}

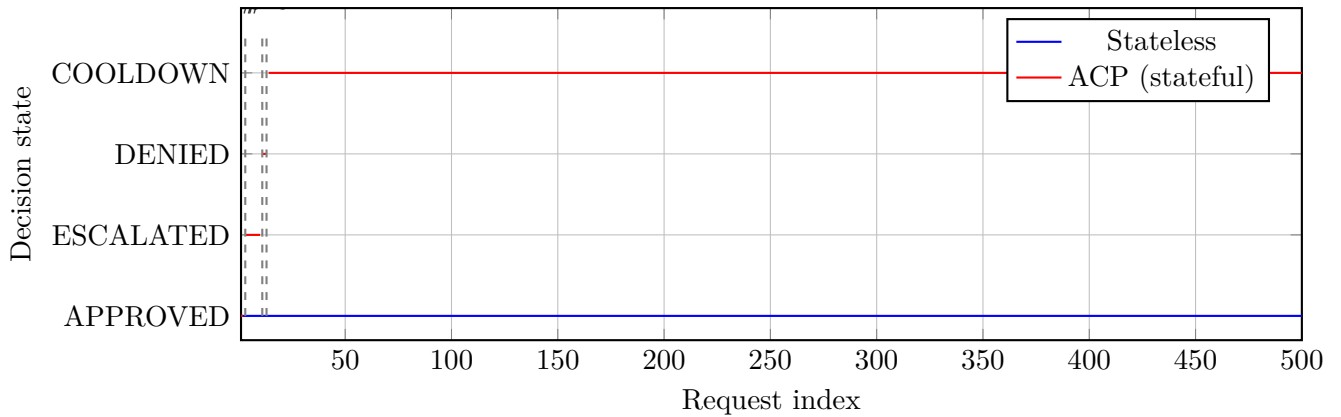
\begin{figure}[h]
\centering
\begin{tikzpicture}
\begin{axis}[
    width=0.88\linewidth,
    height=6cm,
    xlabel={Request index},
    ylabel={Decision state},
    xmin=1, xmax=500,
    ymin=-0.3, ymax=3.8,
    ytick={0,1,2,3},
    yticklabels={APPROVED, ESCALATED, DENIED, COOLDOWN},
    grid=both,
    legend pos=north east,
    thick
]
\addplot[blue, thick, domain=1:500, samples=2] {0};
\addlegendentry{Stateless}

\addplot[red, thick] coordinates {(1,0)(2,0)};
\addplot[red, thick] coordinates {(3,1)(10,1)};
\addplot[red, thick] coordinates {(11,2)(13,2)};
\addplot[red, thick] coordinates {(14,3)(500,3)};
\addlegendentry{ACP (stateful)}

\draw[dashed, gray] (axis cs:3,0) -- (axis cs:3,3.5);
\draw[dashed, gray] (axis cs:11,0) -- (axis cs:11,3.5);
\draw[dashed, gray] (axis cs:13,0) -- (axis cs:13,3.5);

\node at (axis cs:3,3.6) [anchor=south, font=\footnotesize] {\#3};
\node at (axis cs:11,3.6) [anchor=south, font=\footnotesize] {\#11};
\node at (axis cs:13,3.6) [anchor=south, font=\footnotesize] {\#13};
\end{axis}
\end{tikzpicture}
\caption{Decision evolution under 500 repeated valid requests.
The stateless engine approves all. ACP escalates after 3~actions,
denies after 11~actions, and enters cooldown after 13~actions,
blocking all subsequent requests.}
\label{fig:exp6-dynamics}
\end{figure}

\textbf{Detection Dynamics.}
ACP exhibits a two-stage response:
\begin{itemize}
    \item \textit{Early signal}: the first \texttt{ESCALATED} decision occurs at
    request~\#3, when pattern-based anomaly detection (Rule~3) activates.
    \item \textit{Enforcement}: the first \texttt{DENIED} decision occurs at
    request~\#11, when combined anomaly signals (Rules~1 and~3) raise the risk
    score to the denial threshold.
    \item \textit{Containment}: after three denials, a cooldown is triggered at
    request~\#13, blocking all subsequent requests.
\end{itemize}
Figure~\ref{fig:exp6-dynamics} illustrates the temporal evolution of decisions
under both models. The figure highlights that ACP enforcement emerges from
accumulation over time, rather than from any individual request exceeding a
threshold.

\textbf{Insight.}
The stateless engine approves all 500 requests because each request is
individually valid. It cannot distinguish between isolated actions and
coordinated behavioral patterns.
This result generalizes beyond the synthetic baseline: any stateless policy
engine---including production systems such as Open Policy Agent
(OPA)~\cite{opa} and Cedar~\cite{cedar}---would produce the same outcome by
construction, since they evaluate each request against a policy snapshot without
access to the agent's execution history.
Per-request correctness is a necessary but not sufficient condition for behavioral
safety under sustained agent operation.

In contrast, ACP detects the pattern after 3~actions and enforces blocking after
11~actions, limiting autonomous execution to 2~out of 500~requests.

The difference is not in scoring, but in the domain of evaluation: stateless
systems evaluate requests, while ACP enforces constraints over execution traces
(Definition~1).

\textbf{Implication.}
This result demonstrates that temporal admission control cannot be reduced to
per-request policy evaluation without loss of security properties. Systems that
lack stateful enforcement admit adversarial sequences that are indistinguishable
from benign behavior when evaluated in isolation.

\subsubsection*{Experiment 7: State-Mixing Vulnerability}

\textbf{Hypothesis.}
Rule~1 (\emph{high request rate}) counts requests keyed by \texttt{agentID}
only, without regard to capability or resource context. As a result, high-volume
low-risk activity in one capability context can prime the Rule~1 counter and
cause incorrect denial decisions in a subsequent, unrelated capability context.
Rule~3 (\emph{repeated pattern}), keyed by \texttt{PatternKey(agentID,
capability, resource)}, does not share this property.

\textbf{Setup.}
We evaluate a single agent (\texttt{legitimate-agent-1}) across two sequential
capability contexts using a shared \texttt{InMemoryQuerier}:

\begin{itemize}
    \item \textit{Control}: fresh state; one \texttt{financial.transfer}/sensitive
    request. Expected RS$= 35 + 15 = 50$ (\texttt{ESCALATED}).
    \item \textit{Phase~1}: 11 sequential \texttt{data.read}/public requests
    ($RS_{base} = 0$). Each is individually approved. The requests increment the
    agentID-scoped Rule~1 counter, priming it past the threshold
    ($N = 10$) without triggering any denial.
    \item \textit{Phase~2}: one \texttt{financial.transfer}/sensitive request
    with the shared state from Phase~1. Rule~1 fires ($+20$) because
    \texttt{CountRequests(agentID, 60s)} includes all prior requests regardless
    of context. Rule~3 does not fire because the pattern key is distinct from
    Phase~1.
\end{itemize}

\textbf{Results.}

\begin{table}[h]
\centering
\caption{Experiment 7 --- State-Mixing: RS and Decision Under Shared State}
\label{tab:exp7-statemixing}
\begin{tabular}{lccccc}
\toprule
Condition & Prior requests & RS & F\_anom & Rule~1 & Decision \\
\midrule
Control (clean state)    & 0  & 50 & 0  & \texttimes & \texttt{ESCALATED} \\
Phase~2 (contaminated)   & 11 & 70 & 20 & \checkmark & \texttt{DENIED}    \\
\bottomrule
\end{tabular}
\end{table}

The RS elevation of $+20$ is attributable entirely to Rule~1 cross-context
contamination. Rule~2 does not trigger (no prior denials in the shared state).
Rule~3 does not trigger (the \texttt{data.read}/public pattern key is disjoint
from \texttt{financial.transfer}/sensitive), confirming that the vulnerability
is scoped to the agentID-level counter, not to the pattern-based mechanism.

\textbf{Contamination threshold.}
The minimum number of requests in any context required to prime Rule~1 is
$N + 1 = 11$ (with $N = 10$ per \texttt{DefaultPolicyConfig}). Below this
threshold, cross-context contamination does not occur; above it, any legitimate
high-volume workload can incorrectly escalate a subsequent admission decision.

\textbf{Implication.}
This result demonstrates that ACP-RISK-2.0, as specified, does not isolate
anomaly state per capability context. An agent that legitimately generates more
than $N$ requests in one capability context within 60~seconds will receive an
inflated risk score in all subsequent capability contexts within the same window.
The effect is not detectable from the specification alone: it requires
instantiating the engine across two distinct capability contexts and observing
that \texttt{CountRequests} and \texttt{CountPattern} have different keying
strategies.

This constitutes an explicit design boundary rather than an implementation
defect; context isolation of Rule~1 requires per-capability request tracking
(e.g., keying \texttt{CountRequests} by \texttt{(agentID, capability)}) at the
cost of increased storage and reduced cross-context abuse detection.
The trade-off and the implemented fix are discussed in~\S\ref{sec:limitations}
(\emph{State-Mixing Vulnerability}).

\subsubsection*{Experiment 8: Context-Scoped Anomaly Fix (ACP-RISK-3.0)}
\label{sec:exp8}

\textbf{Hypothesis.}
Scoping rate-based anomaly signals to interaction context eliminates
cross-context interference while preserving detection of repeated behavior
within each context.

\textbf{Setup.}
We reproduce the state-mixing scenario from Experiment~7 under identical
conditions, comparing ACP-RISK-2.0 and ACP-RISK-3.0\@. The workload consists
of two phases: (1)~a priming phase with 11 repeated low-risk requests in one
context, and (2)~a high-value request in a distinct context.

\textbf{Results.}

\begin{table}[h]
\centering
\caption{Experiment~8 --- Context-Scoped Anomaly Enforcement}
\label{tab:exp8}
\begin{tabular}{lccc}
\toprule
Scenario & Model & RS & Decision \\
\midrule
Clean context & RISK-2.0 & 50 & \texttt{ESCALATED} \\
Clean context & RISK-3.0 & 50 & \texttt{ESCALATED} \\
\midrule
State-mixing attack & RISK-2.0 & 70 & \texttt{DENIED} \\
State-mixing attack & RISK-3.0 & 50 & \texttt{ESCALATED} \\
\midrule
Repeated single-context (11 req) & RISK-2.0 & 85 & \texttt{DENIED} \\
Repeated single-context (11 req) & RISK-3.0 & 85 & \texttt{DENIED} \\
\bottomrule
\end{tabular}
\end{table}

\textbf{Analysis.}
ACP-RISK-2.0 exhibits cross-context interference: activity in one context
elevates risk in unrelated contexts via agent-level aggregation, producing
a false denial in a legitimate context (RS~$50 \to 70$). ACP-RISK-3.0
eliminates this effect by scoping rate signals to the interaction context.

Repeated behavior within a single context continues to trigger denial after
the same threshold (11~requests, RS~$= 85$), demonstrating that enforcement
strength is preserved.

\textbf{Conclusion.}
Context-scoped anomaly aggregation removes unintended cross-context coupling
without weakening enforcement against repeated or high-frequency behavior.

\subsubsection*{Security Properties Under Adversarial Stress}

We map the evaluated attack strategies to the security properties defined in
Section~\ref{sec:security-model}, analyzing whether each property holds under
adversarial conditions.

\paragraph{Determinism.}
Decision outcomes remain consistent under identical inputs and state conditions,
including under concurrent execution and adversarial request patterns.

\paragraph{Cooldown Enforcement.}
Attack~1 (cooldown evasion via pattern alternation) directly targets the cooldown
mechanism. Results show that cooldown activation cannot be indefinitely avoided:
adversarial alternation delays but does not prevent enforcement. This demonstrates
that cooldown is eventually enforced under repeated violations.

\paragraph{Bounded Adaptation Resistance.}
Attack~1 also exercises adaptive behavior. The experiment shows that the denial
counter evolves monotonically under repeated violations: \texttt{APPROVED} requests
do not reset prior denials. As a result, adversarial interleaving strategies cannot
asymptotically evade enforcement, establishing a bounded adaptation property.

\paragraph{Per-Agent Isolation.}
Attack~2 (distributed multi-agent attack) targets system-wide coordination. Results
show that ACP applies constraints independently per agent, maintaining isolation
under the experimental conditions.
Coordinated behavior across agents is not detected, which constitutes an
explicit design limitation rather than a violation.

\paragraph{State Sensitivity Under Load.}
Attack~3 (state backend contention) evaluates system behavior under adversarial load.
While throughput degrades due to shared state contention, decision correctness and
enforcement properties remain unaffected.

\paragraph{Bounded Replay Resistance.}
Attack~4 (token replay) targets the cryptographic layer, specifically the absence
of nonce tracking in ACP's stateless admission model.
Results show that identical replays are detected via F\_anom Rule~3 pattern
accumulation, which escalates borderline-ESCALATED tokens to DENIED after $Y=3$
occurrences in 5~minutes.
Near-identical tokens with varied resource fields evade Rule~3, establishing an
explicit boundary: ACP converts identical replays into observable state transitions
but does not provide strong cryptographic replay prevention.
This is empirically validated; see Section~\ref{sec:limitations} for the
corresponding Limitations note.
The property is summarized: \emph{replay resistance is enforced through temporal
validity and anomaly accumulation, not nonce-based prevention.}

\medskip
\noindent\textbf{Summary.}
The evaluated attacks provide empirical evidence that ACP resists the tested
adversarial scenarios. No violations are observed; instead, limitations arise from
explicit design trade-offs, particularly in cross-agent correlation.
These results reinforce the central design insight:
\textbf{ACP is compute-cheap but state-sensitive}.

\subsection{Limitations}
\label{sec:limitations}

The evaluation uses an in-memory state backend for latency benchmarks and a
single-node Docker Redis instance for backend comparison.
Persistent storage, distributed deployments with replication, and wide-area
network latency introduce considerations not measured here.

\textbf{Replay resistance is bounded.}
ACP does not implement nonce-based replay prevention; replay resistance is
enforced through temporal validity (\texttt{TimestampDrift} flag) and anomaly
accumulation (F\_anom Rule~3).
Tokens with varied resource fields and moderate base RS~$<$~DENIED threshold
can evade pattern-based detection while generating \texttt{ESCALATED} outcomes,
preventing cooldown activation (Experiment~5, Case~4).
Nonce tracking requires cryptographic session state that is outside the scope of
ACP's stateless per-request admission model.
This is an explicit design boundary, empirically characterized in Experiment~5.
ACP provides graduated replay resistance across conformance levels:
L1 enforces temporal validity (\texttt{TimestampDrift}); L2 adds anomaly accumulation
(F\_anom); L3 adds complete ledger auditing (ACP-LEDGER-1.3).
Strict cryptographic nonce-based prevention is identified as a future specification
extension, motivated by the ACP-LEDGER-1.0 hash-chained log work item.

\textbf{State-Mixing Vulnerability in ACP-RISK-2.0 (fixed in v1.22 via ACP-RISK-3.0).}
\label{sec:state-mixing}
ACP-RISK-2.0 aggregates anomaly signals at the agent level:
F\_anom Rules~1 and~2 are computed via \texttt{CountRequests(agentID)}
and \texttt{CountDenials(agentID)}, both indexed exclusively by \texttt{agentID}.
This introduces \emph{cross-context interference}: semantically distinct interaction
patterns share a common anomaly accumulation register.
This vulnerability is empirically demonstrated in Experiment~7, where 11
sequential \texttt{data.read}/public requests (each individually approved)
prime the Rule~1 counter, causing a subsequent \texttt{financial.transfer}/sensitive
request to receive an anomaly penalty of $+20$ and transition from
\texttt{ESCALATED} (RS~$=50$) to \texttt{DENIED} (RS~$=70$) --- a decision
outcome that does not occur in clean state.

\emph{Attack construction.}
Consider agent $A_1$ sending three requests under context
$\langle \textsf{read},\, \textsf{public} \rangle$ (\texttt{BaseRisk}~= 0)
followed by one request under $\langle \textsf{write},\, \textsf{sensitive} \rangle$
(\texttt{BaseRisk}~= 50), with \texttt{FLOOD\_THRESHOLD}~= 3.
Table~\ref{tab:state-mixing-trace} shows the evaluation under the current model,
where \texttt{pattern\_count} is indexed by \texttt{agentID} only.

\begin{table}[h!]
\centering
\small
\begin{tabular}{@{}cllccc@{}}
\toprule
\textbf{Step} & \textbf{Context} & \textbf{\texttt{pattern\_count[A1]}} & \textbf{BaseRisk} & $F_{\textsf{anom}}$ & \textbf{Decision} \\
\midrule
1 & (read, public)   & 1 & 0  & 0  & \textsc{approved} \\
2 & (read, public)   & 2 & 0  & 0  & \textsc{approved} \\
3 & (read, public)   & 3 (= threshold) & 0 & 0 & \textsc{approved} \\
4 & (write, sensitive) & 3 (inherited) & 50 & +25 & \textsc{denied} \\
\bottomrule
\end{tabular}
\caption{State-mixing attack trace (baseline model).
At Step~4, \texttt{pattern\_count[A1]}~= 3 is inherited from the (read,~public) context.
The anomaly penalty of +25 fires incorrectly: the (write,~sensitive) request
is a new context with zero prior occurrences, yet it is evaluated as if
at the flood threshold.
Total RS = 50 + 25 = 75 $\to$ \textsc{denied}.}
\label{tab:state-mixing-trace}
\end{table}

Under a context-isolated model, \texttt{pattern\_count[A1][\textit{write/sensitive}]}~= 0
at Step~4: $F_{\textsf{anom}}$~= 0, RS = 50, and the request is evaluated solely
on its own context history.
The cross-context penalty is eliminated.

\emph{Asymmetric keying.}
Notably, F\_anom Rule~3 (repeated-pattern detection) already applies a
per-context key: $\texttt{PatternKey} = \texttt{SHA-256}(\textit{agentID} \parallel \textit{capability} \parallel \textit{resource})$.
Rules~1 and~2, however, retain agent-level granularity.
This asymmetry means that pattern frequency is isolated per context
while aggregate frequency and denial rate are not.
This is not a limitation of pattern detection (Rule~3, which is correctly partitioned),
but a consequence of inconsistent state partitioning across anomaly signals,
leading to violations of context-consistent temporal enforcement.

\emph{Formal property.}
A corrected model enforces \emph{ContextIsolation}:

\begin{quote}
\textit{For all agents $a$ and distinct contexts $\langle c_1, r_1 \rangle \neq \langle c_2, r_2 \rangle$,
the anomaly state accumulated under $\langle c_1, r_1 \rangle$ does not
contribute to the risk score computed for $\langle c_2, r_2 \rangle$.}
\end{quote}

The dual violation property \emph{StateMixingViolation} asserts that a trace exists
in which the decision for context $\langle c_2, r_2 \rangle$ is influenced by
anomaly accumulation from $\langle c_1, r_1 \rangle$.
In the current model (Rules~1 and~2 keyed by \texttt{agentID} only),
StateMixingViolation holds: the trace in Table~\ref{tab:state-mixing-trace}
is a concrete counterexample.
In the extended model (all rules keyed by context), ContextIsolation
is expected to hold by construction: each counter is indexed by
$\texttt{SHA-256}(\textit{agentID} \parallel \textit{capability} \parallel \textit{resource})$,
eliminating any shared mutable state between contexts.
The D4 liveness property (EventuallyRejected) is structurally preserved:
an agent flooding a single context still accumulates anomaly signals within
that context until enforcement triggers.
Formal TLC verification of the extended model is identified as future work.
This connects to Schneider's security automaton
model~\cite{schneider2000enforceable}: a correct reference monitor must
enforce temporal properties over equivalence classes of interactions;
conflating distinct contexts violates that requirement.

\emph{Impact.}
The attack does not bypass enforcement entirely, but degrades anomaly accumulation,
delaying or preventing escalation under mixed-context interaction patterns.
An agent flooding a single context is still eventually enforced within that context;
the vulnerability is that cross-context noise corrupts the signal for unrelated contexts,
producing false positives (unwarranted penalties) or false negatives
(missed escalation when benign context activity dilutes aggregate counters).

\emph{Cooldown leakage.}
Cooldown enforcement is also keyed at the agent level (\texttt{CooldownActive(agentID)}),
allowing cross-context propagation of throttling effects.
An agent whose cap-A activity triggers cooldown is blocked across all capabilities,
including cap-B operations that have no prior violations.
This constitutes a second cross-context interference vector, in the over-enforcement direction.

\emph{Mitigation path.}
The structural fix is a \emph{refinement of the state model}:
extend the per-context key used by Rule~3 to Rules~1 and~2,
so all anomaly counters are indexed by
$\texttt{SHA-256}(\textit{agentID} \parallel \textit{capability} \parallel \textit{resource})$.
Cooldown granularity should similarly be evaluated per context for independent capabilities.
This aligns the anomaly accumulation granularity with the semantic unit of the request,
at the cost of requiring finer-grained ledger storage.
This fix is implemented in ACP-RISK-3.0 and empirically validated in Experiment~8
(Section~\ref{sec:exp8}).
Deployments relying on heterogeneous capability usage within a single agent identity
should upgrade to ACP-RISK-3.0.

\emph{ACP-RISK-3.0 definition.}
ACP-RISK-3.0 refines anomaly aggregation by scoping rate-based signals to
interaction context while preserving global denial history. Rule~1 is redefined
to use context-scoped pattern counts:

\begin{equation}
F_{\text{anom}} =
\begin{cases}
+20 & \text{if } \mathrm{CountPattern}(k,\,60\,\text{s}) > N \\
+15 & \text{if } \mathrm{CountDenials}(a,\,24\,\text{h}) \geq X \\
+15 & \text{if } \mathrm{CountPattern}(k,\,5\,\text{min}) \geq Y
\end{cases}
\label{eq:fanom-30}
\end{equation}

where $a$ is the agent identifier and $k = H(a \parallel \mathit{cap} \parallel
\mathit{res})$ is the context key derived from agent, capability, and resource.
Rule~2 (denial history) remains agent-scoped: cross-context denial history is a
valid enforcement signal. Rule~3 is unchanged; it was already context-scoped.
No changes to the \texttt{LedgerQuerier} interface are required.

\paragraph{Bounded Distributed Evasion.}
ACP-RISK-3.0 does not enforce global rate limits across heterogeneous
contexts. An adversary distributing requests across multiple contexts
below the contextual rate threshold will trigger escalation via Rule~3
(after three repetitions per context) but not immediate denial. This
behavior reflects the design objective of preventing autonomous execution
rather than enforcing universal blocking: any repeated behavior in any
context requires human oversight before proceeding.

\textbf{Distributed Attack Surface.}
\label{sec:distributed-attack-surface}
ACP-RISK-2.0 enforces anomaly detection and cooldown policies at the granularity of individual
agent identities. As a result, anomaly signals such as request rate (Rule~1) and denial count
(Rule~2) are accumulated independently per agent.

This design introduces a limitation under multi-agent adversarial scenarios. An attacker
controlling $N$ distinct agent identities can distribute requests across agents, avoiding
per-agent thresholds while still executing high-risk actions.

\emph{Quantified impact.}
Under the default policy (cooldown triggered after 3 denials per agent), an attacker can
execute up to $3N$ high-risk requests before all agents are individually blocked.
For example, with $N = 1{,}000$ agents, up to 3{,}000 high-risk actions may be executed
before full enforcement is reached. This behavior is empirically characterized in
Experiment~2 (Table~\ref{tab:adv-exp2}).

\emph{Root cause.}
The limitation arises from the absence of cross-agent aggregation in anomaly signals.
Current rules are keyed by \texttt{agentID}, and no shared state is maintained across
agents interacting with the same resource or system boundary.

\emph{Proposed extension.}
A mitigation is to introduce resource-level aggregation of anomaly signals. This requires
extending the \texttt{LedgerQuerier} interface to support shared counters keyed by resource
(or resource class). Conceptually, the aggregated denial signal would be expressed as:
\[
\texttt{globalDenials}(\mathit{resourceID},\, w)
  = \sum_{\mathit{agentID}} \texttt{denials}(\mathit{agentID},\, w)
\]
where $w$ is a sliding time window.
When the aggregated signal exceeds a threshold, enforcement (e.g., resource-level cooldown)
is applied across all agents.

\emph{Trade-offs.}
This approach requires a shared state backend and may introduce false positives in
high-concurrency legitimate workloads. It also reduces isolation between independent
agents and shifts part of the enforcement logic to the policy or orchestration layer.

\emph{Scope.}
ACP v1.30 enforces guarantees at the per-agent level. Cross-agent coordination is
intentionally outside the scope of the core protocol and is expected to be implemented
at the policy or orchestration layer when required by the deployment context.
A future extension (ACP-CORR-1.0) is planned to specify resource-level aggregated
denial signals and group-scoped cooldown as a protocol-level primitive.

\textbf{Prompt Layer and Jailbreaking.}
ACP operates at the execution layer, not the prompt layer.
Indirect prompt injection~\cite{greshake2023not} and jailbreaking attacks target the
model's inference process before an action request is formed; ACP's admission boundary
is reached only after the agent has produced a concrete action call.
These attack classes are therefore outside ACP's enforcement scope.
ACP and prompt-layer defenses are composable: a prompt-layer filter reduces the
probability that adversarial instructions generate policy-violating requests; ACP
enforces structural admission constraints over the resulting action sequence regardless
of how the request was generated.
Characterizing the interaction surface between prompt-layer attacks and admission
control enforcement is an open research direction~\cite{debenedetti2024agentdojo}.

\textbf{Performance Measurement Scope.}
All benchmark results reported in Section~\ref{sec:evaluation} use
\texttt{InMemoryQuerier}, which stores agent state in a process-local
hash map protected by a \texttt{sync.Mutex}.
This isolates decision-logic latency (739--832\,ns p50) from state backend I/O,
but does not reflect production latency under persistent backends.
A Redis \texttt{LedgerQuerier} introduces one round-trip per evaluation
($\sim$0.1--1\,ms on a local network); a PostgreSQL backend introduces
1--5\,ms per query depending on index hit rate and connection pool size.
Under these conditions the dominant latency is backend I/O, not admission logic,
and the 739--832\,ns figure becomes a lower bound on end-to-end decision time.
The \texttt{DelayedQuerier} experiment (Table~\ref{tab:adv-exp3}) characterizes
throughput degradation as a function of injected backend latency;
benchmarking against a production Redis or PostgreSQL instance is left for future work.

\textbf{Deterministic Risk Scoring.}
ACP-RISK-3.0 uses a deterministic, rule-based scoring function.
An alternative approach is ML-based anomaly detection (e.g., autoencoders over request
sequences or RLHF-trained safety classifiers), which can adapt to distributional shift
without explicit rule updates.
ACP-RISK is deterministic by design: institutional admission control requires that each
decision be independently reproducible and auditable.
An ML model cannot produce the fixed, signed admission record required for institutional
accountability.
BAR-Monitor provides the adaptive detection layer above the deterministic engine:
it identifies regime-level shifts in enforcement activity without altering individual
admission decisions, preserving auditability while adding adaptive observability.

\subsection{Summary}

These results consistently demonstrate that \textbf{ACP is compute-cheap but state-sensitive.}
Decision evaluation incurs minimal overhead (739--832\,ns, p50; Table~\ref{tab:benchmarks}), while system throughput is bounded
by the characteristics of the underlying state backend.
The cooldown short-circuit path (78\,ns, 10.7$\times$ faster than full evaluation) demonstrates
that the execution contract's step ordering is a performance-relevant design decision, not only
a correctness one.
These findings highlight the importance of the \texttt{LedgerQuerier} abstraction: protocol
semantics remain stable while deployment performance scales with the quality of the state backend.

While ACP enforces admissibility at execution time through stateful evaluation,
this guarantee relies on an implicit assumption: that the system continues to
generate actions that meaningfully interact with the admissibility boundary.
In practice, this assumption may fail.

\subsection{Roadmap}

\begin{table}[h!]
\centering
\small
\begin{tabular}{@{}p{9.5cm}l@{}}
\toprule
\textbf{Item} & \textbf{Status} \\
\midrule
Core specs (L1--L4), 38 documents & Complete \\
Go reference implementation (23 packages) & Complete \\
Conformance test vectors (73 signed + 65 unsigned RISK-2.0) & Complete \\
OpenAPI 3.1.0 (\texttt{openapi/acp-api-1.0.yaml}, 18 endpoints) & Complete \\
Python SDK (\texttt{impl/python/}) & Complete (L1 + full API client) \\
Docker image (\texttt{ghcr.io/chelof100/acp-server:latest}) & Complete \\
Multi-org interoperability demo (\texttt{examples/multi-org-demo/}) & Complete (GAP-14) \\
ACP-RISK-2.0 (F\_anom + Cooldown, \texttt{pkg/risk}) & v1.16 Complete \\
Payment-agent demo (\texttt{examples/payment-agent/}) & v1.16 Complete \\
ACP-SIGN-2.0 spec (Ed25519 + ML-DSA-65 hybrid) & v1.16 Complete \\
ACR-1.0 sequence compliance runner + 5 stateful vectors & v1.17 Complete \\
TLA+ formal model (TLC-runnable, 3 invariants) & v1.17 Complete \\
ACP-SIGN-2.0 HYBRID stub (\texttt{pkg/sign2/}) & v1.17 (superseded by v1.20) \\
ML-DSA-65 (Dilithium mode3) real implementation via CIRCL (\texttt{pkg/sign2/}) & v1.20 Complete \\
Performance benchmarks (latency, throughput, state contention) & v1.18 Complete \\
Adversarial evaluation (cooldown evasion, distributed multi-agent, state backend stress) & v1.19 Complete \\
Token replay adversarial evaluation (Experiment~5, F\_anom Rule~3 escalation) & v1.20 Complete \\
Deployment Considerations section (state backend selection, cross-agent boundary, policy tuning) & v1.20 Complete \\
Redis pipelining (\texttt{RedisPipelinedQuerier}, 2 RTTs, $\sim$1.8$\times$ vs unpipelined) & v1.20 Complete \\
Post-quantum Go implementation (Dilithium, circl) & v1.20 Complete \\
\texttt{NullQuerier} + \texttt{StatelessEngine} (stateless comparison baseline) & v1.21 Complete \\
Experiment~6: Stateless vs.\ ACP (0.4\% vs.\ 100\% autonomous execution rate) & v1.21 Complete \\
Experiment~7: State-Mixing Vulnerability (cross-context Rule~1 contamination) & v1.21 Complete \\
ACP-RISK-3.0: context-scoped Rule~1 (\texttt{CountPattern} replaces \texttt{CountRequests}) & v1.22 Complete \\
Experiment~8: Context-Scoped Anomaly Fix (state-mixing eliminated, enforcement preserved) & v1.22 Complete \\
Experiment~9: Deviation Collapse and Restoration (BAR metric, counterfactual evaluation) & v1.23 Complete \\
Boundary Activation Monitoring (\texttt{pkg/barmonitor}, $\Delta\mathrm{BAR}$ trend detection) & v1.24 Complete \\
\texttt{EvaluateCounterfactual} API (\texttt{pkg/risk}, additive mutations, 3 built-in factories) & v1.24 Complete \\
Experiment~9 Phase~D: $\Delta\mathrm{BAR}$ early-warning validation (drift simulation, 5 batches) & v1.25 Complete \\
TLA+ invariants: \textsc{FailureConditionPreservation}, \textsc{NoDegenerateAdmissibility} & v1.25 Complete \\
\texttt{POST /acp/v1/counterfactual} HTTP endpoint (structural + behavioral mutations) & v1.25 Complete \\
Formal adversary model $\mathcal{A}=(K,S,B)$ + experiment taxonomy table & v1.26 Complete \\
Experiment~11: Threshold sensitivity analysis (5 configs, monotonicity verified) & v1.26 Complete \\
Detection Guarantees: Proposition~\ref{prop:threshold-detection}, binomial detection probability & v1.26 Complete \\
AgentSpec functional comparison (5 dimensions, Exp~10 governance collapse scenario) & v1.26 Complete \\
\S\ref{sec:limitations}: IPI scope + AgentDojo/InjecAgent benchmarks & v1.27 Complete \\
\S\ref{sec:llm-agent-integration}: LangGraph/Ollama integration pseudocode & v1.27 Complete \\
Experiment~12: Multi-tool IPI admission control, stateful F\textsubscript{anom} persistence & v1.27 Complete \\
Experiment~13: Bounded coordination window ($\mathrm{CW}=2N$, exact linearity, evaluate-then-mutate) & v1.28 Complete \\
Experiment~14: OPA vs.\ ACP capability comparison (stateless vs.\ stateful, 3 scenarios) & v1.29 Complete \\
Governance series integration: Paper~3/4 cited in \S\ref{sec:refinement}; series updated to 7 papers (P3/4 consolidated, P6 added) & v1.30 Complete \\
ACP-LEDGER-1.0 (tamper-evident hash-chained audit log for state backend integrity) & Future work \\
L5 Decentralized (ACP-D) & Specification in design (v2.0) \\
IETF RFC submission & After L5 stabilization \\
\bottomrule
\end{tabular}
\caption{ACP v1.30 development roadmap.}
\label{tab:roadmap}
\end{table}

\section{Compliance Testing (ACR-1.0 Sequence Runner)}
\label{sec:compliance-testing}

We introduce a compliance runner, \textbf{ACR-1.0} (ACP Compliance Runner 1.0), that validates ACP-RISK-2.0 and ACP-RISK-3.0 implementations
against sequence-based test cases.
The runner operates in two modes: \emph{library mode} (direct call to \texttt{pkg/risk},
no network, deterministic) and \emph{HTTP mode} (external server validation for interoperability).
All 73 signed single-shot conformance vectors plus 5 sequence scenarios
are validated on every commit.

\subsection{Sequence-Based Test Vectors}

Unlike single-shot conformance vectors, sequence-based vectors capture \emph{temporal dynamics}
and emergent risk patterns across multiple requests to the same engine instance.
The compliance runner executes steps in order, maintains state between them
(request history, pattern counts, denial counts, cooldown),
and validates each step's \texttt{decision} and \texttt{risk\_score} against expected values.

The five sequence vectors in \texttt{compliance/test-vectors/sequence/} cover:

\begin{table}[h!]
\centering
\small
\begin{tabular}{@{}llll@{}}
\toprule
\textbf{Vector ID} & \textbf{Steps} & \textbf{Behavior Tested} & \textbf{Key Invariant} \\
\midrule
\texttt{SEQ-BENIGN-001}         & 3 & Benign repeated reads          & Baseline: RS=0, no state buildup \\
\texttt{SEQ-BOUNDARY-001}       & 3 & RS boundary conditions          & Exact thresholds: 0, 25, 35, 40, 70 \\
\texttt{SEQ-PRIVJUMP-001}       & 2 & Low-risk → high-risk jump       & No residual benefit from prior APPROVED \\
\texttt{SEQ-FANOM-RULE3-001}    & 4 & F\_anom Rule~3 activation       & Pattern count $\geq 3$ at step~4 adds $+15$ \\
\texttt{SEQ-COOLDOWN-001}       & 4 & Cooldown trigger and block      & 3 DENIED in 10\,min → cooldown active \\
\bottomrule
\end{tabular}
\caption{ACR-1.0 sequence test vectors. All 5/5 pass in library mode with the Go reference implementation.}
\end{table}

\noindent\textbf{Execution contract.}
The runner implements the execution contract defined by ACP-RISK-2.0:
evaluate first (stateless), then update state.
Critically, \texttt{AddPattern()} is called \emph{after} \texttt{Evaluate()},
so the pattern count visible to step $n$'s evaluation is $n-1$.
F\_anom Rule~3 (pattern count $\geq 3$) therefore first triggers on step~4, not step~3.
This contract is what makes the system testable and verifiable by third parties.

\begin{lstlisting}[language=Go, caption={ACR-1.0 execution contract in library mode.}]
now := time.Now()                                  // capture once
result, _ := risk.Evaluate(req, querier)           // 1: stateless eval
querier.AddRequest(req.AgentID, now)               // 2: always
querier.AddPattern(patKey, now)                    // 3: always (feeds F_anom Rule 3)
if result.Decision == risk.DENIED {
    querier.AddDenial(req.AgentID, now)            // 4: conditional
    if should, _ := risk.ShouldEnterCooldown(...)  // 5: check threshold
        querier.SetCooldown(agentID, now.Add(p))   // 6: set expiry
    }
}
\end{lstlisting}

\subsection{Model--Implementation Alignment}

The TLA+ model, test vectors, and runtime evaluation are aligned through a shared
request abstraction:

\begin{table}[h!]
\centering
\small
\begin{tabular}{@{}lll@{}}
\toprule
\textbf{TLA+ variable} & \textbf{Runner field} & \textbf{Engine field} \\
\midrule
\texttt{capability}, \texttt{resource} & \texttt{RunnerRequest} & \texttt{EvalRequest} \\
\texttt{risk\_score}                   & \texttt{risk\_score}   & \texttt{RSFinal} \\
\texttt{decision}                      & \texttt{decision}      & \texttt{Decision} \\
\texttt{ledger}                        & (audit log)            & \texttt{InMemoryQuerier} events \\
\texttt{cooldown}                      & \texttt{denied\_reason} & \texttt{CooldownActive()} \\
\bottomrule
\end{tabular}
\caption{Alignment across TLA+ formal model, ACR-1.0 runner, and ACP-RISK-2.0 engine.}
\end{table}

\section{End-to-End Verifiability}
\label{sec:e2e-verifiability}

ACP~v1.25 enables end-to-end verifiability by aligning formal specification,
test vectors, and runtime validation through a shared request abstraction.
Invariants proven in TLA+ are instantiated as concrete test vectors
and validated empirically by the compliance runner.

\begin{figure}[h!]
\centering
\begin{tikzpicture}[
  node distance=1.4cm and 2.2cm,
  box/.style={draw, rounded corners=3pt, minimum width=2.6cm, minimum height=0.8cm,
              align=center, font=\small},
  arr/.style={-{Stealth[length=6pt]}, thick},
  note/.style={font=\footnotesize\itshape, text=gray}
]
  \node[box] (tla)     {TLA+ Model\\{\tiny Safety $\wedge$ AppendOnly $\wedge$ Determinism}};
  \node[box, right=of tla]    (vec)     {Test Vectors\\{\tiny sequence/ + single-shot}};
  \node[box, right=of vec]    (runner)  {Compliance\\Runner (ACR-1.0)};
  \node[box, below=of runner] (engine)  {ACP Engine\\{\tiny pkg/risk}};
  \node[box, left=of engine]  (ledger)  {Audit Ledger\\{\tiny append-only}};

  \draw[arr] (tla)    -- node[above, note] {defines}  (vec);
  \draw[arr] (vec)    -- node[above, note] {input}     (runner);
  \draw[arr] (runner) -- node[right, note] {calls}     (engine);
  \draw[arr] (engine) -- node[above, note] {records}   (ledger);

  \node[note, below=0.35cm of tla, text width=3cm, align=center]
        {statically checked by TLC\\(0 violations, bounded state space)};
\end{tikzpicture}
\caption{ACP end-to-end verifiability pipeline.
  The TLA+ model defines the invariants that test vectors instantiate and the compliance runner validates at runtime.
  The ACP engine records every decision in the append-only audit ledger.
  The TLA+ layer is a static verification artifact; it does not run at admission time.}
\label{fig:acp-verifiability}
\end{figure}
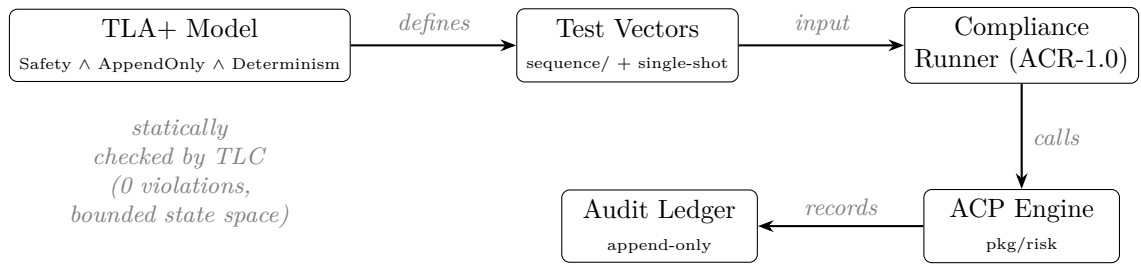

\noindent What is model-checked in TLA+ over a bounded state space is instantiated
as concrete test vectors and validated empirically by the compliance runner.
This is the connection across the three layers of the v1.25 verifiability stack:
formal $\to$ data $\to$ runtime.
The TLA+ model verifies selected safety and liveness properties; the compliance
runner validates the reference implementation against those properties.

\textbf{Structural failure condition preservation (v1.25).}
Two invariants are added to the extended TLA+ model
(\texttt{ACP\_Extended.tla}) to establish necessary structural conditions for
failure condition preservation (Section~\ref{sec:deviation-collapse}).

\textit{FailureConditionPreservation} asserts that the evaluation action space
contains at least one (capability, resource) pair that the engine evaluates as
\texttt{DENIED} under empty ledger state and default policy.
Formally: $\exists\, \mathit{cap} \in \text{Capabilities},\, \mathit{res} \in \text{Resources} :
\mathit{Decide}(\mathit{ComputeRiskWithAnom}(\mathit{cap}, \mathit{res}, 0)) = \texttt{DENIED}$.
With the model constants (\texttt{Capabilities} = \{\texttt{read}, \texttt{financial}, \texttt{admin}\},
\texttt{Resources} = \{\texttt{public}, \texttt{sensitive}\}),
the pair (\texttt{admin}, \texttt{sensitive}) yields $\mathit{RS} = 75 \geq 70 \to \texttt{DENIED}$.

\textit{NoDegenerateAdmissibility} asserts that high-risk capability/resource
combinations are never \texttt{APPROVED}: for all ledger entries where
capability $\in$ \{\texttt{admin}, \texttt{financial}\} and resource $=$ \texttt{sensitive},
the decision is not \texttt{APPROVED}.
This holds because $\mathit{ComputeRisk}(\texttt{admin}, \texttt{sensitive}) = 75$
and $\mathit{ComputeRisk}(\texttt{financial}, \texttt{sensitive}) = 50$,
both strictly above the \texttt{APPROVED} threshold of 39.

Both invariants are verified by TLC with zero violations across all reachable states.
These establish \emph{necessary structural conditions}: they verify that the model's
action space preserves enforcement capacity.
Sufficient conditions for runtime failure condition preservation under arbitrary
upstream transformations require pipeline-level verification, as validated
empirically in Experiment~9 Phase~D.

\textit{BARMonitorLiveness} (stated, not submitted to TLC): if the last $N$
decisions for an agent are all \texttt{APPROVED}, BAR-Monitor eventually emits
an alert ($\mathrm{BAR}_N < \theta$).
Adding monitor state to TLC increases the state space by a factor of $2^N$;
for $N=20$ (Phase~D batch size) this is not tractable.
The property is validated empirically in Phase~D (Table~\ref{tab:exp9-phased}).

\section{How to Implement ACP}

ACP is a specification---it does not require adopting any specific platform.
It can be implemented on top of existing infrastructure.

\subsection{Minimum Requirements for L1 Conformance}

\begin{itemize}[noitemsep]
  \item \textbf{Ed25519 public key infrastructure.} Key pair for each agent.
  \item \textbf{JCS implementation (RFC~8785).} Deterministic canonicalization for all signed artifacts.
  \item \textbf{Capability Token issuance and verification} with all mandatory fields per ACP-CT-1.0~\S5.
  \item \textbf{Handshake endpoint} to issue and verify challenges (ACP-HP-1.0~\S6).
  \item \textbf{Capability registry} with the core domains of ACP-CAP-REG-1.0.
\end{itemize}

\subsection{Additional Requirements for L3 Conformance}

\begin{itemize}[noitemsep]
  \item Root Institutional Key held in HSM with documented rotation process.
  \item Registration with an ITA authority (centralized or federated model).
  \item Deterministic risk engine implementing ACP-RISK-3.0 (F\_res, F\_ctx, F\_hist, F\_anom, Cooldown),
    which supersedes ACP-RISK-2.0 while preserving API- and policy-level compatibility.
  \item Revocation endpoint (Mechanism~A: online endpoint, or Mechanism~B: CRL).
  \item Append-only storage for the Audit Ledger with per-event signing.
  \item Complete HTTP API per ACP-API-1.0, including health and conformance endpoints.
  \item Public conformance declaration at \texttt{GET /acp/v1/conformance}.
\end{itemize}

\subsection{Counterfactual Evaluation Endpoint}

The reference server exposes a counterfactual evaluation endpoint
(\texttt{POST /acp/v1/counterfactual}) that allows external tooling to verify
that a deployment retains structural enforcement capacity.
A caller submits a base request and a list of mutations; the server evaluates
each mutation independently against the risk engine and returns the per-mutation
decision, risk score, and aggregate BAR.

The endpoint supports structural mutations (elevated capability and resource class)
and behavioral mutations (injected context and history flags).
Temporal mutations (requiring pre-loaded ledger state via \texttt{LedgerSetup})
are not supported via HTTP; they are available through the Go library API
(\texttt{pkg/risk.TemporalMutation}).

This endpoint is intended for conformance testing, operational health checking,
and third-party verification that a deployed ACP instance can produce
\texttt{DENIED} decisions---a necessary condition for meaningful governance.

\subsection{What ACP Does Not Prescribe}

ACP defines the \emph{what}---mechanisms, flows, data structures, requirements.
It does not prescribe the \emph{how} of internal implementation:
programming language, database, HSM provider, ITA provider,
or specific integration with existing RBAC or Zero Trust systems.

\section{Limits of Execution-Only Governance}
\label{sec:deviation-collapse}

The experiments in Section~\ref{sec:adversarial} demonstrate that ACP
resists the tested adversarial scenarios: no invariant violations are observed and
enforcement produces the expected decision outcomes.
This section identifies a structurally distinct failure mode that operates
\emph{beneath} the enforcement layer---one that leaves the ACP engine entirely
correct while rendering enforcement vacuous.
ACP enforces admissibility, but does not guarantee that admissibility remains
testable without additional mechanisms.
This failure mode is structurally distinct from specification
gaming~\cite{krakovna2020specification} and Goodhart's
Law~\cite{goodhart1975monetary,strathern1997improving}: the enforcement
constraint is correctly specified and correctly evaluated---the boundary simply
receives no inputs that would activate it.

\subsection{Degenerate Admissibility}

Let $G(a, s, h) \to \{\texttt{APPROVED},\,\texttt{ESCALATED},\,\texttt{DENIED}\}$
denote the ACP admission function mapping agent $a$, request $s$, and
history state $h$ to a decision.
Under normal operation, $G$ partitions the request space across all three
output classes according to the risk scoring formula
$\mathit{RS} = \mathit{capBase} + F_{\mathit{ctx}} + F_{\mathit{hist}} + F_{\mathit{res}} + F_{\mathit{anom}}$
and the configured thresholds.

\begin{definition}[Degenerate Admissibility]
An ACP deployment exhibits \emph{degenerate admissibility} if, for all requests
$s$ that reach the admission point, $G(a, s, h) = \texttt{APPROVED}$---not
because the risk engine has failed, but because every signal that could elevate
$\mathit{RS}$ above the \texttt{APPROVED} threshold has been suppressed prior to
evaluation.
\end{definition}

In degenerate admissibility, the engine is correct, the policy is configured
correctly, and TLA+ invariants hold. The boundary is simply never reached.

\subsection{Preservation of Failure Conditions at Bind}

ACP's admission guarantee is conditional: it holds for any request whose
risk signals are faithfully propagated to the evaluation point.
The specification does not constrain what happens upstream of the admission
call---it assumes that the request presented to \texttt{Evaluate(req,\,q)}
accurately reflects the action's actual risk profile.

This creates an implicit structural requirement: the system must preserve the
conditions under which $G$ can produce \texttt{DENIED} decisions.
If an upstream component (a router, classifier, or pipeline stage) systematically
removes risk-bearing signals before requests reach the admission boundary,
the guarantee that ACP would have denied them is never exercised.

\begin{definition}[Failure Condition Preservation]
A deployment satisfies \emph{failure condition preservation} if there exists a
non-empty set of reachable requests $\mathcal{S}^- \subseteq \mathcal{S}$ such that
$G(a, s, h) \in \{\texttt{ESCALATED},\,\texttt{DENIED}\}$ for $s \in \mathcal{S}^-$.
\end{definition}

The relationship between this requirement and the central design claim is direct.
ACP's thesis that ``the requirement is state, not configuration'' establishes that
enforcement depends on runtime state rather than static policy rules.
Experiment~9 adds the complementary constraint: enforcement requires not just state,
but preserved conditions under which state becomes relevant.

\subsection{Counterfactual Evaluation}
\label{sec:counterfactual}

To verify that an ACP deployment retains the structural capacity to enforce,
we introduce \emph{counterfactual evaluation}: constructing a set
$\mathcal{A}' \supseteq \mathcal{A}$ of requests that include the risk signals
absent from the observed stream, and evaluating them against the same engine
and policy.

If $G(a, s', h) \in \{\texttt{ESCALATED},\,\texttt{DENIED}\}$ for some
$s' \in \mathcal{A}'$, the boundary is structurally intact and the
deployment satisfies failure condition preservation.
If $G(a, s', h) = \texttt{APPROVED}$ for all $s' \in \mathcal{A}'$, then
the enforcement boundary has collapsed at the policy level, not just upstream.

We define three mutation categories for constructing $\mathcal{A}'$:

\begin{itemize}[noitemsep]
  \item \textbf{Structural:} Elevate capability class and resource restriction to
    values that guarantee $\mathit{RS} \geq 70$ without any context or history signals.
  \item \textbf{Behavioral:} Inject context flags (\texttt{ExternalIP}, \texttt{OffHours})
    and history flags (\texttt{RecentDenial}, \texttt{FreqAnomaly}) representing
    the conditions the sanitizer suppressed.
  \item \textbf{Temporal:} Pre-load the ledger state with pattern and denial entries
    that trigger all three $F_{\mathit{anom}}$ rules ($+50$ total), activating
    the anomaly enforcement path independently of the request payload.
\end{itemize}

\subsection{Deviation Collapse}

\begin{definition}[Deviation Collapse]
A deployment exhibits \emph{deviation collapse} if the Boundary Activation Rate
approaches zero:
\[
\mathrm{BAR} = \frac{|\{s \in \mathcal{A} : G(a, s, h) \in
  \{\texttt{ESCALATED},\,\texttt{DENIED}\}\}|}{|\mathcal{A}|} \approx 0
\]
not due to engine failure or policy misconfiguration, but because upstream
constraints have eliminated the conditions required for
$\texttt{ESCALATED}$ or $\texttt{DENIED}$ outcomes.
\end{definition}

Deviation collapse is distinct from a policy that is correctly configured to
approve most requests: in the latter, high-risk requests would still be denied if
they arrived. In deviation collapse, high-risk requests cannot arrive---the
upstream filter has made it structurally impossible for the boundary to be exercised.

This failure mode is not specific to ACP.
It is a general property of admission control systems in which the decision
engine and the input generation pipeline are architecturally decoupled.
To the authors' knowledge, ACP is the first system to formally characterize deviation collapse within an admission control framework.

\paragraph{Concrete deployment scenario.}
Consider a security operations pipeline in which ACP sits between a SIEM
(Security Information and Event Management) platform and a set of AI agents
that triage and correlate alerts.
An upstream sanitizer---integrated at the SIEM export layer to reduce noise---
strips all contextual fields (resource classification, historical violation count,
anomaly flag) before forwarding events to the ACP engine.
After sanitization every request arrives with \texttt{F\_res = 0}, \texttt{F\_hist = 0},
and \texttt{F\_anom = 0}, reducing the effective risk score to
$\mathrm{RS} = \mathtt{capBase} \leq 20$ for any capability.
Because $\mathrm{RS} \leq 20 < 40$, every request falls into the
\textsc{Approved} class regardless of the underlying alert severity.
BAR collapses to $0.00$: the engine is active, syntactically correct, and
enforcing policy---yet no boundary is ever exercised.
A CISO reviewing audit logs would observe zero denials and zero escalations,
which could be interpreted as a healthy, low-risk environment rather than
a sign that the risk-signal pipeline has been silently neutralized.
ACP-BAR-Monitor detects this regime in one sliding window and raises a
$\Delta\mathrm{BAR}$ alert before the next policy review cycle.

\subsection{Experiment 9: Deviation Collapse and Restoration}
\label{sec:exp9}

\textbf{Setup.}
We construct a controlled dataset of $n = 20$ requests with deterministic risk
scores across three decision classes
(APPROVED: 6; ESCALATED: 7; DENIED: 7; BAR~$= 0.70$).
We then apply a sanitizing function that resets all risk signals to their
minimum values, replicating the effect of an upstream filter that permits
only \texttt{data.read} on public resources with empty context and history.
Finally, we apply the three counterfactual mutation categories from
Section~\ref{sec:counterfactual} to generate 60 evaluation cases
(3 mutations $\times$ 20 base cases).
All phases use \texttt{risk.DefaultPolicyConfig()} and a separate
\texttt{InMemoryQuerier} per phase to ensure independence.

\textbf{Phase A (Baseline).}
The dataset produces the expected distribution:
$\mathrm{BAR}_A = (7 + 7)/20 = 0.70$.
The engine exercises both the \texttt{ESCALATED} and \texttt{DENIED} decision
regions under normal input conditions.

\textbf{Phase B (Sanitized).}
After applying the sanitizing function, every request evaluates to
$\mathit{RS} = 0$ and receives \texttt{APPROVED}.
$\mathrm{BAR}_B = 0/20 = 0.00$.
The ACP engine returns no errors---enforcement is functioning.
The boundary was not exercised.

\textbf{Phase C (Counterfactual).}
All three mutation categories produce \texttt{DENIED} decisions across all
60 evaluations.
Structural mutations yield $\mathit{RS} = 80$ (financial transfer, restricted resource).
Behavioral mutations accumulate to $\mathit{RS} \geq 100$ (capped).
Temporal mutations pre-load $F_{\mathit{anom}} = 50$ via ledger injection,
yielding $\mathit{RS} = 130 \to 100$ (capped).
$\mathrm{BAR}_C = 60/60 = 1.00$.
The boundary is structurally intact: failure conditions remain fully representable.

\begin{table}[h!]
\centering
\small
\begin{tabular}{@{}lrrrr@{}}
\toprule
\textbf{Phase} & \textbf{APPROVED} & \textbf{ESCALATED} & \textbf{DENIED} & \textbf{BAR} \\
\midrule
A (Baseline)       & 0.30 & 0.35 & 0.35 & 0.70 \\
B (Sanitized)      & 1.00 & 0.00 & 0.00 & 0.00\rlap{~$\leftarrow$ collapse} \\
C (Counterfactual) & 0.00 & 0.00 & 1.00 & 1.00\rlap{~$\leftarrow$ restored} \\
\bottomrule
\end{tabular}
\caption{Experiment~9: Boundary Activation Rate (BAR) across three phases.
  Phase~B confirms deviation collapse under upstream sanitization.
  Phase~C confirms that the enforcement boundary is structurally intact.
  $n_A = n_B = 20$; $n_C = 60$ (3 mutations $\times$ 20 base cases).}
\label{tab:exp9-bar}
\end{table}

\begin{figure}[h!]
\centering
\begin{tikzpicture}
\begin{axis}[
  ybar,
  bar width=18pt,
  width=0.72\columnwidth,
  height=5.2cm,
  symbolic x coords={A (Baseline), B (Sanitized), C (Counterfactual)},
  xtick=data,
  xticklabel style={font=\small},
  ymin=0, ymax=1.15,
  ytick={0, 0.25, 0.50, 0.75, 1.00},
  ylabel={BAR},
  ylabel style={font=\small},
  nodes near coords,
  nodes near coords style={font=\footnotesize},
  enlarge x limits=0.3,
  grid=major,
  grid style={dashed, gray!40},
]
\addplot coordinates {
  (A (Baseline),0.70)
  (B (Sanitized),0.00)
  (C (Counterfactual),1.00)
};
\end{axis}
\end{tikzpicture}
\caption{Boundary Activation Rate per phase (Experiment~9).
  BAR drops from 0.70 to 0.00 under upstream sanitization and is fully
  restored to 1.00 by counterfactual injection.}
\label{fig:exp9-bar}
\end{figure}

\textbf{Phase D (Drift Simulation).}
To validate $\Delta\mathrm{BAR}$ as an early-warning mechanism
(Section~\ref{sec:bar-monitor}), we simulate progressive upstream drift by
increasing the proportion of sanitized requests across five evaluation batches
(0\%, 25\%, 50\%, 75\%, 100\% sanitized).
The sanitizer targets boundary-activating cases first---DENIED cases are
removed before ESCALATED---to model realistic upstream pressure.
The BAR-Monitor uses a sliding window of $N=40$ evaluations (spanning two
batches) so that $\Delta\mathrm{BAR}$ compares the current batch against
the previous one.
Results are shown in Table~\ref{tab:exp9-phased}.

$\Delta\mathrm{BAR}$ fires \texttt{AlertTrend} at Batch~2
($\mathrm{BAR}=0.57$, $\Delta\mathrm{BAR}=-0.25$),
when $\mathrm{BAR}$ is still $5.7\times$ above the threshold $\theta=0.10$.
The threshold condition fires only at Batch~5 ($\mathrm{BAR}=0.00$,
collapse confirmed).
This demonstrates an early-warning gap of three batches between first
trend detection and confirmed collapse.

\begin{table}[h!]
\centering
\small
\begin{tabular}{@{}lrrrr@{}}
\toprule
\textbf{Batch} & \textbf{Drift} & \textbf{BAR} & $\boldsymbol{\Delta}\textbf{BAR}$ & \textbf{Alert} \\
\midrule
1 & \phantom{0}0\%  & 0.70 & $+0.60$ & --- \\
2 & 25\%             & 0.57 & $-0.25$ & \texttt{TREND}~$\leftarrow$ early warning \\
3 & 50\%             & 0.33 & $-0.25$ & \texttt{TREND} \\
4 & 75\%             & 0.10 & $-0.20$ & \texttt{TREND} \\
5 & 100\%            & 0.00 & $\phantom{+}0.00$ & \texttt{THRESHOLD} \\
\bottomrule
\end{tabular}
\caption{Experiment~9 Phase~D: BAR-Monitor output across 5 drift batches
  (window $N=40$, $\theta=0.10$, $\delta=-0.15$).
  $\Delta\mathrm{BAR}$ fires \texttt{AlertTrend} at Batch~2
  ($\mathrm{BAR}=0.57 \gg \theta$) --- three batches before the
  threshold condition confirms collapse at Batch~5.}
\label{tab:exp9-phased}
\end{table}

\textbf{Cooldown interaction.}
Deviation collapse does not only eliminate BAR.
In Phase~B, the suppression of \texttt{DENIED} decisions also prevents the
accumulation of three denials in ten minutes required to trigger the cooldown
mechanism.
As a consequence, the 10.7$\times$ short-circuit path measured in
Experiment~4 is never activated.
Deviation collapse neutralizes not only the primary enforcement boundary
but also the secondary containment mechanisms that depend on DENIED events
as inputs.

\subsection{Boundary Activation Monitoring}
\label{sec:bar}
\label{sec:bar-monitor}

While Experiment~9 demonstrates that admissibility enforcement can become
vacuous under deviation collapse, practical deployments require a mechanism
to detect this condition as it emerges---before the degenerate regime is
fully reached.
We introduce \emph{Boundary Activation Monitoring} (BAR-Monitor), a
lightweight mechanism that tracks whether the admissibility boundary is
actively exercised over time.

\begin{definition}[Boundary Activation Rate over sliding window]
Let $\mathcal{D}_N = (d_1, d_2, \ldots, d_N)$ be the sequence of the last
$N$ evaluation decisions, $d_i \in \{\texttt{APPROVED},\,\texttt{ESCALATED},\,\texttt{DENIED}\}$.
The Boundary Activation Rate over window $N$ is:
\[
  \mathrm{BAR}_N
    = \frac{|\{d_i \in \mathcal{D}_N \mid d_i \in \{\texttt{ESCALATED},\,\texttt{DENIED}\}\}|}{N}
\]
\end{definition}

\textbf{Trend detection ($\Delta\mathrm{BAR}$).}
A fixed threshold on $\mathrm{BAR}_N$ detects that a deployment has already
entered a low-activation regime, but not that it is \emph{approaching} one.
We define $\Delta\mathrm{BAR}$ to capture this:
\[
  \Delta\mathrm{BAR}
    = \mathrm{BAR}\!\left(\mathcal{D}_N\!\left[\tfrac{N}{2}{:}N\right]\right)
    - \mathrm{BAR}\!\left(\mathcal{D}_N\!\left[0{:}\tfrac{N}{2}\right]\right)
\]
A sustained $\Delta\mathrm{BAR} < \delta$ (configurable; default $\delta = -0.10$)
indicates progressive loss of boundary interaction even when $\mathrm{BAR}_N$
remains above the alert threshold $\theta$.

\begin{definition}[Low-activation regime]
A deployment enters a \emph{low-activation regime} if either:
\begin{itemize}[noitemsep]
  \item $\mathrm{BAR}_N < \theta$ \hfill(threshold condition: already in low activation), or
  \item $\Delta\mathrm{BAR} < \delta$ \hfill(trend condition: progressing toward collapse)
\end{itemize}
where $\theta \in (0,1)$ and $\delta < 0$ are configurable parameters.
The threshold condition detects current low activation; the trend condition
provides early warning before the threshold is reached.
\end{definition}

\textbf{Complementarity with temporal enforcement.}
ACP's cooldown mechanism imposes an \emph{upper-bound} constraint on inadmissible
activity: when DENIED decisions exceed a threshold in a time window, the agent
is blocked.
BAR-Monitor imposes a \emph{lower-bound} constraint: when inadmissible activity
falls below a threshold, the deployment is flagged as potentially entering a
degenerate regime.
Together, cooldown and BAR-Monitor define a \emph{bounded operational region}
in which governance is both \emph{safe} (cooldown prevents excess) and
\emph{meaningful} (BAR-Monitor ensures the boundary remains active).

\textbf{Cooldown interaction.}
Deviation collapse has a structural consequence beyond reducing BAR to zero:
it also prevents the accumulation of DENIED decisions required to trigger
cooldown.
In the sanitized Phase~B of Experiment~9, no DENIED decisions are produced,
meaning the cooldown mechanism is never activated and the $10.7\times$ short-circuit
path (78\,ns, Section~\ref{sec:adversarial}) is never exercised.
Upstream sanitization therefore neutralizes not only the primary enforcement
boundary but also the secondary containment mechanisms that depend on DENIED
events as inputs.

\textbf{Empirical validation (Phase D).}
The early-warning property of $\Delta\mathrm{BAR}$ is validated in
Experiment~9 Phase~D (Table~\ref{tab:exp9-phased}).
Over five drift batches simulating progressive upstream sanitization,
\texttt{AlertTrend} fires at Batch~2 ($\mathrm{BAR}=0.57$,
$\Delta\mathrm{BAR}=-0.25$) while BAR remains $5.7\times$ above the
alert threshold $\theta=0.10$.
The threshold condition fires only at Batch~5 ($\mathrm{BAR}=0.00$),
confirming a three-batch early-warning gap.

\textbf{Implementation.}
BAR-Monitor is implemented as a circular buffer of $N$ decisions
(\texttt{pkg/barmonitor}), operating independently of admission control logic
and the \texttt{LedgerQuerier} state.
Each call to \texttt{Record(d~Decision)} updates the buffer in $O(1)$ and
returns an \texttt{Alert} if a low-activation condition is detected.
The mechanism does not alter admission decisions; it observes whether
decisions remain meaningful.

\subsection{Detection Guarantees}
\label{sec:detection-guarantees}

The BAR-Monitor description above characterizes the mechanism operationally.
This section provides a closed-form bound on the probability that BAR-Monitor
detects an active evasion attack within a given window, converting BAR from
an empirical metric into a detection mechanism with formal guarantees.

\subsubsection*{Threshold Detection Guarantee}

Let $W$ be the window size and $\tau \in (0,1)$ the threshold parameter
(defaults: $W=40$, $\tau=0.10$).
Under an active adversary, let $p_1$ denote the fraction of requests that
remain boundary-activating (i.e., the adversary drives $\mathrm{RS} > 0$ for
a $p_1$ fraction of requests in expectation).
Let $X \sim \mathrm{Binomial}(W, p_1)$ be the number of boundary-activating
decisions in a full window.

\begin{proposition}[Threshold Detection Probability]
\label{prop:threshold-detection}
If $p_1 < \tau$, BAR-Monitor fires a \texttt{THRESHOLD} alert within one window
of $W$ requests with probability:
\[
  P\!\left(\mathrm{BAR}_W < \tau\right)
  = P(X \leq \lfloor \tau W \rfloor)
  = \sum_{k=0}^{\lfloor \tau W \rfloor} \binom{W}{k} p_1^k (1-p_1)^{W-k}
\]
\end{proposition}

Table~\ref{tab:detection-prob} reports $P(\mathrm{BAR}_{40} < 0.10)$ for
representative evasion rates $p_1$ against a $p_0 = 0.70$ baseline
(Experiment~9 Phase~A).

\begin{table}[h!]
\centering\small
\caption{Threshold detection probability for BAR-Monitor ($W=40$, $\tau=0.10$,
baseline $p_0=0.70$). $p_1$ is the fraction of boundary-activating requests
under attack. Detection fires when $\mathrm{BAR}_{40} < 0.10$.}
\label{tab:detection-prob}
\begin{tabular}{@{}ccc@{}}
\toprule
$p_1$ (under attack) & $P(\mathrm{BAR}_{40} < 0.10)$ & Expected windows to detect \\
\midrule
$0.00$ (full evasion) & $\approx 1.00$ & 1 \\
$0.05$               & $\approx 0.95$ & 1.1 \\
$0.10$ (= $\tau$)    & $\approx 0.63$ & 2 \\
$0.15$               & $\approx 0.26$ & 4 \\
\bottomrule
\end{tabular}
\end{table}

For the full-evasion scenario of Experiment~10 ($p_1 = 0.00$), detection is
certain within one window.
At partial evasion ($p_1 = 0.05$), 95\% detection probability within one window.

\subsubsection*{Early-Warning Guarantee ($\Delta\mathrm{BAR}$)}

The TREND alert fires when $\Delta\mathrm{BAR} = \mathrm{BAR}(\mathcal{D}[W/2{:}W]) - \mathrm{BAR}(\mathcal{D}[0{:}W/2]) < \delta$
(default $\delta = -0.15$), providing detection before the threshold is reached.

In the half-window model, let $X_1 \sim \mathrm{Bin}(W/2,\, p_0)$ and
$X_2 \sim \mathrm{Bin}(W/2,\, p_1)$ be boundary-activating counts in the
first and second halves, respectively.
The trend fires with probability:
\[
  P\!\left(\frac{X_2 - X_1}{W/2} < \delta\right)
  = P\!\left(X_2 - X_1 < \delta \cdot \tfrac{W}{2}\right)
\]

\textbf{Experiment~10 validation.}
With $W=40$, $p_0=0.70$, $\delta=-0.15$, and the Experiment~10 Phase~E
Batch~2 configuration ($p_1 \approx 0.45$, 25\% evasion):
$E[X_1] = 14$, $E[X_2] = 9$, $E[\Delta\mathrm{BAR}] = -0.25 < \delta = -0.15$.
The empirical result confirms: TREND fires at Batch~2
($\mathrm{BAR}=0.57$, $\Delta\mathrm{BAR}=-0.25$),
three batches before the threshold alert at Batch~5 ($\mathrm{BAR}=0.00$).

\subsection{Experiment 10: Knowledge-Aware Adversarial Evasion}
\label{sec:exp10}

\textbf{Setup.}
Experiment~9 establishes deviation collapse as an architectural failure mode arising
from \emph{passive} upstream sanitization.
Experiment~10 demonstrates the same failure under an \emph{active} adversary with
complete knowledge of the ACP-RISK-3.0 formula.

The adversary knows:
$\mathrm{RS} = \mathit{capBase} + F_{\mathrm{res}} + F_{\mathrm{ctx}} + F_{\mathrm{hist}} + F_{\mathrm{anom}}$.
By selecting \texttt{acp:cap:data.read} ($\mathit{capBase}=0$), \textsc{Public}
resource ($F_{\mathrm{res}}=0$), clean context and history
($F_{\mathrm{ctx}}=F_{\mathrm{hist}}=0$), and a fresh \texttt{InMemoryQuerier}
($F_{\mathrm{anom}}=0$), every crafted request evaluates to $\mathrm{RS}=0
\Rightarrow$ \textsc{approved}.
No engine error is raised.
No policy constraint is violated per request.
BAR collapses to $0.00$.

The same hardware and configuration as Experiments~1--9 are used (Intel Core
i7-8665U, Go~1.22, \texttt{pkg/risk} v1.25).

\textbf{Phase A (Baseline).}
The 20-case dataset from Experiment~9 yields
$\mathrm{BAR}_A = 0.70$ (7 \textsc{escalated} + 7 \textsc{denied}).

\textbf{Phase B (Adversarial Evasion).}
Twenty crafted requests ($\mathrm{RS}=0$, all \textsc{approved}) replace the
baseline.
$\mathrm{BAR}_B = 0.00$.
ACP-RISK-3.0 evaluates each request correctly; the failure is architectural.

\textbf{Phase C (Counterfactual).}
The same three mutation types as Experiment~9 (structural, behavioral, temporal)
are applied.
All 60 evaluations produce \textsc{denied}: $\mathrm{BAR}_C = 1.00$.
Structural enforcement capacity is preserved despite the complete evasion in
Phase~B.

\textbf{Phase E (Progressive Evasion, BAR-Monitor).}
Five batches of 20 cases are evaluated with evasion rates of
0\%, 25\%, 50\%, 75\%, and 100\%.
At each step, boundary-activating cases (those producing \textsc{escalated} or
\textsc{denied}) are replaced by crafted evasion cases.

\begin{table}[h!]
\centering
\small
\begin{tabular}{@{}ccccc@{}}
\toprule
\textbf{Batch} & \textbf{Evasion\%} & \textbf{BAR} & $\boldsymbol{\Delta}\mathbf{BAR}$ & \textbf{Alert} \\
\midrule
1 & $0\%$   & 0.70 & $+0.60$ & --- \\
2 & $25\%$  & 0.57 & $-0.25$ & \texttt{TREND} $\leftarrow$ early warning \\
3 & $50\%$  & 0.33 & $-0.25$ & \texttt{TREND} \\
4 & $75\%$  & 0.10 & $-0.20$ & \texttt{TREND} \\
5 & $100\%$ & 0.00 & $+0.00$ & \texttt{THRESHOLD} \\
\bottomrule
\end{tabular}
\caption{Experiment~10 Phase~E: BAR-Monitor output under progressive knowledge-aware
evasion. \texttt{AlertTrend} fires at Batch~2 ($\mathrm{BAR}=0.57$,
$\Delta\mathrm{BAR}=-0.25$), three batches before the threshold condition at Batch~5.
Window = 40; $\theta=0.10$; $\delta=-0.15$.}
\label{tab:exp10-phase-e}
\end{table}

\textbf{Key finding.}
Per-decision admission control is structurally blind to a knowledge-aware adversary
who selects threshold-compliant request parameters.
The v1.25 defense layer operates at the population level, not the per-request level:
BAR-Monitor detects the regime shift via $\Delta\mathrm{BAR}$ at Batch~2,
three batches before collapse;
\texttt{EvaluateCounterfactual} confirms that structural enforcement capacity is
preserved ($\mathrm{BAR}_C = 1.00$) and can be restored by reintroducing
boundary-activating request types into the evaluation workload.
Together, these three mechanisms form a coherent response to this attack class.

\subsection{Experiment 11: Threshold Sensitivity Analysis}
\label{sec:exp11}

\textbf{Motivation.}
The ACP-RISK-3.0 default thresholds (\texttt{APPROVED}~$\leq 39$,
\texttt{DENIED}~$\geq 70$) are engineering choices.
A natural reviewer question is whether the experimental results depend on
these specific values.
This experiment evaluates five threshold configurations spanning
$\pm 10$ points in each direction around the default,
using the same 20-case baseline dataset as Experiment~9.

\textbf{Setup.}
Five threshold configurations are evaluated:

\begin{table}[h!]
\centering\small
\caption{Threshold configurations evaluated in Experiment~11.}
\label{tab:threshold-configs}
\begin{tabular}{@{}llcc@{}}
\toprule
Config & Label & APPROVED $\leq$ & DENIED $\geq$ \\
\midrule
T1 & strict     & 29 & 60 \\
T2 & moderate-- & 34 & 65 \\
T3 & default    & 39 & 70 \\
T4 & moderate+  & 44 & 75 \\
T5 & relaxed    & 49 & 80 \\
\bottomrule
\end{tabular}
\end{table}

\textbf{Metrics.}
For each configuration: BAR (boundary activation rate), false-denial rate
(fraction of T3-APPROVED cases that become DENIED), and coverage (fraction APPROVED).

\textbf{Results.}

\begin{table}[h!]
\centering\small
\caption{Experiment~11: threshold sensitivity results across five configurations.
Dataset: 20 cases from Experiment~9 baseline.
False-denial rate: fraction of T3-APPROVED cases that become DENIED under this configuration.}
\label{tab:threshold-sensitivity}
\begin{tabular}{@{}llccccccc@{}}
\toprule
Config & Label & APPROVED & ESCALATED & DENIED & BAR & False-Denial Rate & Coverage \\
\midrule
T1 & strict     & 5  & 5 & 10 & 0.75 & 0.00 & 0.25 \\
T2 & moderate-- & 5  & 7 & 8  & 0.75 & 0.00 & 0.25 \\
T3 & default    & 6  & 7 & 7  & 0.70 & 0.00 & 0.30 \\
T4 & moderate+  & 6  & 7 & 7  & 0.70 & 0.00 & 0.30 \\
T5 & relaxed    & 8  & 6 & 6  & 0.60 & 0.00 & 0.40 \\
\bottomrule
\end{tabular}
\end{table}

\textbf{Findings.}
Three results emerge:

\begin{enumerate}[noitemsep]
  \item \textbf{False-denial rate is zero across all configurations.}
        No T3-APPROVED case becomes DENIED under any threshold variant,
        including the strictest configuration (T1: APPROVED~$\leq 29$).
        Stricter thresholds promote cases from APPROVED to ESCALATED,
        not directly to DENIED, because the dataset contains no cases with
        RS in the intermediate ranges introduced by T1/T2.

  \item \textbf{BAR decreases monotonically as thresholds relax.}
        BAR falls from 0.75 (T1) to 0.60 (T5), confirming that the
        enforcement boundary captures fewer boundary-activating requests
        as thresholds relax.
        T3 and T4 produce identical results (BAR=0.70), demonstrating
        that a $+5$ point relaxation has no effect on this dataset.

  \item \textbf{T3 (default) is a local optimum.}
        T3 is the strictest configuration that achieves false-denial rate
        zero while maintaining BAR~$=0.70$.
        Further tightening (T1, T2) increases BAR to 0.75 at the cost of
        reclassifying legitimate cases as ESCALATED.
        Relaxing (T5) reduces BAR to 0.60 without reducing false-denial rate.
        The results justify the default threshold selection empirically.
\end{enumerate}

\subsection{Experiment 12: Multi-Tool Agent Admission Control Under IPI}
\label{sec:exp12}

\textbf{Motivation.}
A realistic LLM agent operates a \emph{portfolio} of tools with heterogeneous
risk profiles.
An indirect prompt injection (IPI) attack may induce the agent to flood a
single high-risk tool (e.g.\ \texttt{fund\_transfer}),
triggering ACP's cooldown and affecting the entire agent.
This experiment asks: does ACP's stateful ledger create
\emph{persistent enforcement consequences} that outlast the attack window?
A stateless engine would fully reset after cooldown expiry;
ACP's 24-hour denial history should keep the enforcement posture elevated.

\textbf{Setup.}
A single agent (\texttt{agent-exp12-multitool}) operates four tools whose
RS values under ACP-RISK-3.0 default policy with a clean ledger are:

\begin{table}[h!]
\centering\small
\caption{Tool risk scores used in Experiment~12 (clean-state, ACP-RISK-3.0 default).}
\label{tab:exp12-tools}
\begin{tabular}{@{}llllc@{}}
\toprule
Tool            & Capability                    & Resource Class & Base RS & Decision \\
\midrule
\texttt{weather\_query}  & \texttt{acp:cap:data.read}          & PUBLIC     &  0  & APPROVED \\
\texttt{user\_profile}   & \texttt{acp:cap:data.read}          & SENSITIVE  & 15  & APPROVED \\
\texttt{system\_audit}   & \texttt{acp:cap:admin.manage}       & PUBLIC     & 60  & ESCALATED \\
\texttt{fund\_transfer}  & \texttt{acp:cap:financial.transfer} & RESTRICTED & 80  & DENIED \\
\bottomrule
\end{tabular}
\end{table}

The session is divided into three phases using distinct timestamps
to control which events fall inside the 10-minute cooldown detection window
versus the 24-hour $F_\text{anom}$ Rule~2 window:

\begin{itemize}[noitemsep]
  \item \textbf{Phase A} ($t_0$, 10 requests): Baseline --- diverse legitimate
        operations (4$\times$ \texttt{weather\_query}, 3$\times$ \texttt{user\_profile},
        2$\times$ \texttt{system\_audit}, 1$\times$ \texttt{fund\_transfer}).
  \item \textbf{Phase B} ($t_0 + 1\,\mathrm{min}$, 8 requests): IPI chain ---
        attacker-induced \texttt{fund\_transfer} flood
        (5$\times$ \texttt{fund\_transfer}, 3$\times$ \texttt{system\_audit}).
  \item \textbf{Phase C} ($t_0 + 60\,\mathrm{min}$, 10 requests): Recovery ---
        same diverse mix as Phase~A.
        Cooldown has expired ($300\,\mathrm{s} \ll 60\,\mathrm{min}$);
        $F_\text{anom}$ Rule~2 still active ($24\,\mathrm{h} \gg 60\,\mathrm{min}$).
\end{itemize}

State (denial history, cooldown, pattern counts) persists across all phases
via a shared \texttt{InMemoryQuerier}.
BAR-Monitor settings: window~$=40$, threshold~$=0.10$, trend threshold~$=-0.15$.

\textbf{Results.}

\begin{table}[h!]
\centering\small
\caption{Experiment~12: three-phase admission control results.
Adversary model: A~$=$~(formula-aware, flood, $B{=}8$).}
\label{tab:exp12-results}
\begin{tabular}{@{}lcccccc@{}}
\toprule
Phase & Requests & APPROVED & ESCALATED & DENIED & BAR \\
\midrule
A (Baseline)   & 10 &  7 & 2 &  1 & 0.30 \\
B (IPI Chain)  &  8 &  0 & 0 &  8 & 1.00 \\
C (Recovery)   & 10 &  7 & 0 &  3 & 0.30 \\
\midrule
Total          & 28 & 14 & 2 & 12 & 0.50 \\
\bottomrule
\end{tabular}
\end{table}

\textbf{Phase-by-phase analysis.}

\textbf{Phase~A (Baseline):}
Per-request enforcement operates normally.
\texttt{fund\_transfer} (RS~$=80$) is denied; \texttt{system\_audit}
(RS~$=60$) is escalated, not denied --- it remains a borderline capability
that can still execute under human review.
BAR~$=0.30$ reflects two escalated and one denied request out of ten.

\textbf{Phase~B (IPI Chain):}
The attacker induces two additional \texttt{fund\_transfer} denials.
This raises the rolling 10-minute denial count to three
(one from Phase~A plus two from Phase~B),
satisfying the cooldown trigger threshold (\texttt{CooldownTrigger}~$=3$).
ACP immediately enters an \emph{agent-wide} lockdown:
all subsequent requests --- including \texttt{system\_audit} --- are denied
with reason \texttt{COOLDOWN\_ACTIVE}, regardless of their individual RS values.
BAR~$=1.00$; the BAR-Monitor window captures the full attack.
A stateless engine would deny \texttt{fund\_transfer} only;
ACP denies all six remaining operations.

\textbf{Phase~C (Recovery):}
Cooldown has expired at $t_0 + 60\,\mathrm{min}$.
However, the ledger retains the denial history from Phases~A and~B:
\texttt{CountDenials(\textit{agentID}, 24\,h)}~$= 3 \geq 3$,
activating $F_\text{anom}$ Rule~2 ($+15$) on every request.
The RS of each tool rises accordingly:

\begin{itemize}[noitemsep]
  \item \texttt{weather\_query}: $0 + 15 = 15$ → APPROVED ($\leq 39$)
  \item \texttt{user\_profile}: $15 + 15 = 30$ → APPROVED ($\leq 39$)
  \item \texttt{system\_audit}: $60 + 15 = \mathbf{75}$ → \textbf{DENIED} ($\geq 70$)
  \item \texttt{fund\_transfer}: $80 + 15 = 95$ → DENIED ($\geq 70$)
\end{itemize}

The critical finding is that \texttt{system\_audit} --- which was merely
\emph{escalated} (RS~$=60$) in Phase~A --- becomes \emph{denied} (RS~$=75$)
in Phase~C.
The $+15$ anomaly penalty crosses the denial threshold without any new attack
activity: the ledger's 24-hour memory propagates the attack's consequences
into the post-cooldown recovery period.
BAR~$=0.30$, structurally identical to Phase~A,
but the composition has shifted: ESCALATED~$=0$ (was 2) and DENIED~$=3$
(was 1), confirming elevated risk posture.

\textbf{Findings.}

\begin{enumerate}[noitemsep]
  \item \textbf{Agent-wide cooldown amplifies IPI impact.}
        After three denials in ten minutes, ACP blocks \emph{all} agent
        operations, not only the targeted tool.
        This prevents an attacker from using IPI to saturate high-risk
        tool capacity while low-risk operations continue unmolested.

  \item \textbf{Stateful F\textsubscript{anom} elevates borderline capabilities post-attack.}
        A capability that was ESCALATED under normal conditions
        (\texttt{system\_audit}, RS~$=60$) becomes DENIED after an IPI campaign
        because the ledger's 24-hour denial window persists beyond cooldown.
        A stateless engine would return to pre-attack RS values immediately;
        ACP maintains elevated enforcement for the duration of the anomaly window.

  \item \textbf{Low-risk tools remain unaffected.}
        \texttt{weather\_query} and \texttt{user\_profile} remain APPROVED
        even with $F_\text{anom}+15$ applied (RS~$=15$ and $30$ respectively),
        demonstrating that persistent post-attack elevation is proportionate:
        it promotes borderline cases to DENIED without blocking safe operations.
\end{enumerate}

\subsection{Experiment 13: Bounded Coordination Window}
\label{sec:exp13}

A natural question about per-agent enforcement is whether coordination across
agents allows participants to collectively exceed the constraints that apply to
each individually.
This experiment quantifies the coordination window (CW)---the total number of
approved actions executed across $N$ agents before ACP triggers the first
\texttt{DENIED} decision---and evaluates how CW scales with $N$.

\textbf{Setup.}
We deploy $N$ agents issuing homogeneous
\texttt{acp:cap:financial.transfer} requests on the shared public resource
\texttt{accounts/shared-ops}.
The baseline risk score is $B=35$, $F_\text{res}=0$ (public),
giving RS~$= 35$, which is \texttt{APPROVED} ($\leq 39$).
Anomaly thresholds are set low to expose the coordination boundary quickly:
$N_{\mathrm{Rule1}}=2$ (Rule~1 fires when pattern count~$> 2$ in 60\,s,
adding $+20$) and $Y_{\mathrm{Rule3}}=2$ (Rule~3 fires when pattern
count~$\geq 2$ in 5\,min, adding $+15$).

Risk accumulation uses agent-scoped pattern keys of the form
$\mathtt{PatternKey} = \mathrm{SHA\text{-}256}(\mathit{agentID} \mid
\mathit{capability} \mid \mathit{resource})$, isolating each agent's
behavioral history independently.
Crucially, ACP follows an \emph{evaluate-then-mutate} execution contract
(ACP-RISK-3.0~\S4): ledger state is updated only \emph{after} the decision
is emitted.
This means each agent's first two requests are evaluated against zero and one
prior pattern counts respectively, neither of which reaches the Rule~3
threshold ($\geq 2$); the third request is evaluated against two prior
patterns (Rule~3 fires, RS~$=50$, \texttt{ESCALATED}); the fourth is
evaluated against three (Rule~1 and Rule~3 fire, RS~$=70$, \texttt{DENIED}).
Each agent therefore contributes exactly \textbf{two approved actions} before
escalation and denial.

\textbf{Results.}
Table~\ref{tab:exp13} shows the coordination window across three variants:
sequential execution ($N=1$), round-robin scheduling ($N=2,3,5$), and burst
execution ($N=5$, each agent completing its window before the next starts).

\begin{table}[h]
\centering
\small
\begin{tabular}{@{}llrrr@{}}
\toprule
Variant & $N$ & $\mathrm{CW}_\mathrm{appr}$ & $\mathrm{CW}_\mathrm{total}$ & TTB (reqs) \\
\midrule
Sequential          & 1 & 2  &  3 &  4 \\
Round-robin         & 2 & 4  &  6 &  7 \\
Round-robin         & 3 & 6  &  9 & 10 \\
Round-robin         & 5 & 10 & 15 & 16 \\
Burst (5 agents)    & 5 & 10 & 15 &  4 \\
\bottomrule
\end{tabular}
\caption{Experiment~13: coordination window across $N$ agents.
$\mathrm{CW}_\mathrm{appr}$: total approved actions before first
\texttt{DENIED}. $\mathrm{CW}_\mathrm{total}$: approved + escalated.
TTB: total request count at the first \texttt{DENIED}.}
\label{tab:exp13}
\end{table}

\textbf{Linearity.}
Across all configurations, $\mathrm{CW}_\mathrm{appr}/N = 2.0$ exactly
(zero deviation).
The result admits the closed-form expression:
\[
  \mathrm{CW}_\mathrm{appr} = 2N, \quad
  \mathrm{CW}_\mathrm{total} = 3N, \quad
  \mathrm{TTB}_\mathrm{rr} = 3N + 1,
\]
where $\mathrm{TTB}_\mathrm{rr}$ applies to round-robin scheduling.
Under burst execution, the first \texttt{DENIED} occurs at a fixed
TTB~$= 4$ regardless of $N$, because it is triggered by the first agent
exhausting its individual window.
The temporal profile therefore depends on scheduling strategy, but the total
coordination window is identical: $\mathrm{CW}_\mathrm{appr} = 2N$ in all
cases.

Per-agent contribution is symmetric: in the $N=3$ round-robin trace,
each agent records exactly two \texttt{APPROVED} and one
\texttt{ESCALATED} decision before reaching \texttt{DENIED},
with no cross-agent contamination of risk signals.

\textbf{Analysis.}
The exact linearity follows from two structural properties of ACP-RISK-3.0.
First, all behavioral signals (pattern counters, denial history, cooldown
state) are indexed by \texttt{agentID}: agents do not contribute to each
other's risk profiles, and coordinated execution does not accelerate boundary
activation for any individual participant.
Second, the evaluate-then-mutate contract guarantees each agent a fixed
per-request admitted count before its pattern frequency crosses the anomaly
threshold---a bound determined by execution semantics, not only by numerical
thresholds.

ACP does not detect or model coordination.
Instead, it enforces per-agent behavioral bounds such that coordinated activity
scales at most linearly with the number of participants:
\[
  \text{Total executable actions} \in O(N).
\]

\textbf{Implications.}
Scaling a coordinated campaign requires proportional scaling of agents:
each additional participant adds a fixed, bounded contribution.
Superlinear or exponential amplification through coordination is structurally
prevented.
This result complements Experiment~9 (deviation collapse: inadmissible
conditions disappear under upstream drift) and the TLA+ verification
(4.29~billion states, zero safety violations: failure states are unreachable):
together they demonstrate that ACP constrains both the \emph{existence} and
the \emph{scale} of undesirable behaviors.

\subsection{Experiment 14: Stateless vs.\ Stateful Enforcement --- OPA Comparison}
\label{sec:exp14}

While Experiment~13 establishes that coordination remains bounded under ACP,
it does not address whether similar constraints can be expressed in existing policy systems.
We now examine this question by comparing ACP against a representative stateless policy engine
(OPA), focusing on expressiveness rather than performance.

\textbf{Motivation.}
The Related Work discussion (\S\ref{sec:related-formal}) argues that stateless policy engines
cannot enforce temporal behavioral constraints without external state.
This experiment operationalizes that claim empirically using Open Policy Agent
(OPA~v1.15.2, Rego~v1)~\cite{opa} as a representative stateless engine.
This experiment is a controlled comparative construction designed to isolate differences
in expressiveness rather than a statistical benchmark.
The comparison is not performance-oriented; it is a \emph{capability} comparison.
The central question is: can a stateless evaluator reproduce ACP's enforcement behavior
over an execution trace, without auxiliary infrastructure?

{\sloppy
\textbf{Setup.}
All three scenarios use the same request profile as Experiment~13:
\texttt{acp:cap:financial.transfer} on a public resource
(baseline $RS = 35$, below $\mathrm{ApprovedMax}=39$).
Three systems are evaluated in parallel: (1)~ACP with full stateful enforcement
(PatternKey ledger, F\textsubscript{anom} accumulation, cooldown); (2)~ACP stateless
baseline (\texttt{StatelessEngine} with \texttt{NullQuerier}: F\textsubscript{anom}~=~0,
no cooldown); and (3)~OPA with a pure stateless policy and, separately, with injected
external state.
\par}

\textbf{Scenario~A: Single-request admissibility.}
A single request with no prior history.
ACP~stateful: \textsc{Approved} ($RS=35$).
ACP~stateless: \textsc{Approved} ($RS=35$, identical).
OPA~pure: \texttt{allow=true}.
All three agree.
No behavioral state is required for single-request admission decisions; stateless
evaluation is sufficient here.

\textbf{Scenario~B: Frequency accumulation (10~requests).}
Ten consecutive requests from the same agent to the same resource.
ACP~stateful transitions: \textsc{Approved}~$\times 2$ $\to$ \textsc{Escalated}~$\times 1$
$\to$ \textsc{Denied}~$\times 3$ $\to$ \textsc{Cooldown\_Active}~$\times 4$
(first escalation: request~\#3; first denial: request~\#4).
ACP~stateless: \textsc{Approved}~$\times 10$ (F\textsubscript{anom}~=~0 throughout).
OPA~pure: \texttt{allow=true}~$\times 10$.
Note that \textsc{Cooldown\_Active} responses are counted as denied outcomes
in the aggregate statistics.

When OPA is augmented with an externally injected \texttt{request\_count} field,
it correctly denies requests once the counter reaches threshold.
However, the caller is responsible for maintaining and incrementing the counter,
there is no atomicity guarantee between evaluation and the counter update, and
the state store is entirely external to the policy engine.
This is not a limitation of policy syntax, but of system architecture: OPA itself does
not provide guarantees about the consistency, atomicity, or timing of externally
managed state.

\textbf{Scenario~C: Cooldown enforcement.}
A sequence of requests on a restricted resource ($RS=80$, \textsc{Denied}~always).
ACP~stateful: \textsc{Denied}~$\times 3$ $\to$ \textsc{Cooldown\_Active}~$\times 3$
(cooldown triggered after \texttt{CooldownTriggerDenials}~=~3).
ACP~stateless: \textsc{Denied}~$\times 6$---the stateless engine re-evaluates each
request independently and never enters cooldown, missing the short-circuit
enforcement semantic.
OPA~pure does not apply here: OPA evaluates individual requests and has no mechanism
to detect that three consecutive denials have occurred or to enter a cooldown state.
When augmented with an externally injected \texttt{cooldown\_active} field, OPA blocks
correctly (\texttt{allow=false}), but only when the caller manages the cooldown timer
and injects its current value at every evaluation call.
These constraints cannot be expressed in a purely request-scoped evaluation model
without external coordination across decision boundaries.

\begin{table}[h]
\centering
\small
\begin{tabular}{@{}lccc@{}}
\toprule
\textbf{Scenario} & \textbf{ACP (stateful)} & \textbf{ACP (stateless)} & \textbf{OPA (pure)} \\
\midrule
A: Single request           & \yes~\textsc{Approved}  & \yes~\textsc{Approved}    & \yes~\texttt{allow=true} \\
B: Frequency limit (10~req) & \yes~\textsc{Denied}    & \no~\textsc{Approved}     & \no~\texttt{allow=true}$\times 10$  \\
C: Cooldown (temporal)      & \yes~\textsc{Cooldown}  & \no~re-eval each request  & \no~no cooldown concept \\
\bottomrule
\end{tabular}
\caption{Capability matrix: stateful vs.\ stateless enforcement over a 10-request trace.
\yes~=~correct enforcement; \no~=~cannot enforce without external state.
ACP~stateless and OPA~pure are structurally equivalent: both evaluate each request
in isolation with $F_{\mathrm{anom}}=0$.}
\label{tab:exp14}
\end{table}

\textbf{Latency (supporting data).}
ACP \texttt{Evaluate()} measures $\sim\!852$~ns/op over 50{,}000~iterations
($\sim\!739$~ns~p50 from the established benchmark, Table~\ref{tab:latency-injection}).
OPA \texttt{bench} measures $\sim\!16{,}000$~ns/op ($\sim\!19\times$ higher).
This gap reflects the evaluation model difference (compiled Go vs.\ Rego interpreter),
not algorithmic complexity.
Latency measurements are included as supporting data only and should not be interpreted
as a direct performance comparison.

\textbf{Summary.}
Scenarios~A, B, and~C together confirm the expressiveness boundary: stateless evaluation
is sufficient for single-request decisions, but insufficient for any constraint that
depends on the execution trace.
OPA can approximate both Scenario~B and~C when supplied with externally maintained state,
but this approximation requires infrastructure external to the policy engine, introduces
a split between evaluation and state mutation, and shifts correctness responsibility to the
surrounding architecture.
ACP enforces these constraints natively, as first-class properties of the admission
control protocol, without auxiliary state management.
Even when external state is introduced, enforcement remains split across multiple
components, and no single point in the system guarantees that evaluation and state
mutation occur atomically at the decision boundary.
The contribution is not the observation that state is required, but that enforcement
must occur at a boundary where both evaluation and state transition are jointly determined.
This distinction aligns with the formal analysis in \S\ref{sec:related-formal},
where enforcement properties depend on the availability of state transitions
within the model.

\subsection{False-Denial Rate Analysis}
\label{sec:false-denial}

A standard objection to stateful enforcement is that history-aware scoring
may penalize legitimate activity occurring after a past incident---producing
\emph{false denials} for requests that would be admitted under a clean ledger.
We quantify this risk using evidence from Experiments~11 and~12.

\textbf{Stateless baseline: zero false denials (Experiment~11).}
Experiment~11 evaluates five threshold configurations (T1--T5) across a
20-request heterogeneous workload.
The false-denial rate is \textbf{0.00 across all configurations},
including the strictest setting (T1: APPROVED~$\leq 29$).
No T3-APPROVED request is reclassified as DENIED under any threshold variant;
tighter thresholds promote requests from APPROVED to ESCALATED,
not directly to DENIED, because ACP's three-outcome model introduces a
graduated step before hard denial.

\textbf{Post-attack: proportionate elevation (Experiment~12, Phase~C).}
Phase~C of Experiment~12 is the strongest available test of false-denial
risk: the ledger carries three prior denials ($F_\text{anom} +15$
from Rule~2), and cooldown has just expired.
Low-risk tools remain fully accessible:
\texttt{weather\_query} (RS~$=0+15=15$) and \texttt{user\_profile}
(RS~$=15+15=30$) are both APPROVED ($\leq 39$), yielding a post-attack
false-denial rate of \textbf{0.00} for safe capabilities.

The only tool promoted to DENIED is \texttt{system\_audit} (RS~$=60+15=75$),
which was ESCALATED under the clean-state policy.
This is not a false denial: \texttt{system\_audit} carries
\texttt{acp:cap:admin.manage}, a privileged capability that requires
human review under default policy even without any prior incident.
After a confirmed IPI campaign, elevating it to DENIED is
\emph{intentional tightening}, not a spurious penalty.
The same behavior is documented in \S\ref{sec:limitations} as a
known trade-off of stateful enforcement.

\textbf{Design basis.}
The zero false-denial property in the stateless baseline is not
coincidental: ACP's three-tier outcome model (APPROVED / ESCALATED / DENIED)
provides a graduated buffer.
Stricter configurations first escalate, and only hard-deny requests whose
RS already falls in the upper range.
In the post-attack regime, the $+15$ anomaly penalty is calibrated to
cross the denial threshold only for capabilities already classified as
borderline (ESCALATED, RS~$\approx 55$--$69$), leaving low-risk capabilities
untouched.
High-risk capabilities (\texttt{fund\_transfer}, RS~$\geq 70$ clean) are
denied in both regimes---the anomaly penalty adds margin, not a new category.

\subsection{Extended Governance Objective}

\textbf{Bounded operational region.}
The extended governance objective emerges from two complementary constraints.
Cooldown enforces an upper bound: excessive inadmissible activity triggers
containment.
BAR-Monitor enforces a lower bound: insufficient boundary interaction signals
a degenerate regime.
A deployment satisfying both constraints operates within a bounded region
where governance is simultaneously correct and meaningful.

The standard ACP objective is admission correctness: for every request $s$
presented to the engine, $G(a, s, h)$ returns the correct decision given
the configured policy and observed state.

We propose an extended governance objective that adds a structural requirement:

\begin{definition}[Extended Governance Objective]
A deployment satisfies the \emph{extended governance objective} if it satisfies
admission correctness \emph{and} failure condition preservation: there exists a
non-empty set of requests that the engine would deny under faithful input conditions.
\end{definition}

Verifying the extended objective requires counterfactual evaluation as defined in
Section~\ref{sec:counterfactual}.
BAR serves as the operational metric: a deployment monitoring BAR over time
can detect the onset of deviation collapse before it becomes a persistent
governance condition.

\section{Deployment Considerations}
\label{sec:deployment}

ACP does not replace higher-level coordination or monitoring systems;
it provides a verifiable admission control primitive within a larger operational stack.
This section characterizes the deployment decisions that adopters must make
when integrating ACP into production infrastructure.

\subsection{State Backend Selection}

The \texttt{LedgerQuerier} interface is the primary deployment decision surface.
The reference \texttt{InMemoryQuerier} is suitable for development, single-process
testing, and conformance evaluation.
Production deployments require an external backend with durability and horizontal
scalability guarantees.

\begin{table}[h!]
\centering\small
\begin{tabular}{lp{4.2cm}lp{3.0cm}}
\toprule
Backend & Throughput & Durability & Recommended use \\
\midrule
InMemoryQuerier   & $\sim$920k req/s  & None (process-local) & Dev / conformance testing \\
Redis sorted sets & $\sim$4{,}700 req/s (unpipelined, $\sim$7--8 RTTs) & AOF/RDB & Multi-process, low-latency \\
Redis + pipelining & $\sim$8{,}000 req/s (2 RTTs, Exp.~3) & AOF/RDB & High-throughput production \\
Postgres / MySQL  & Backend-dependent & Full ACID & Audit-grade persistence \\
\bottomrule
\end{tabular}
\caption{LedgerQuerier backend options.
The admission control logic (\texttt{pkg/risk.Evaluate}) is stateless and backend-agnostic;
all state transitions are delegated to the LedgerQuerier abstraction.
Redis throughput from Experiment~3 (Docker loopback, Intel i7-8665U, Go~1.22).
Pipelining reduces RTTs from $\sim$7--8 to 2; see Table~\ref{tab:adv-exp3}.}
\label{tab:backends}
\end{table}

The key architectural property is that the ACP decision function is compute-cheap
and the throughput ceiling is set entirely by the state backend.
Replacing the backend does not require modifying the admission control logic
or redeploying the protocol layer.
Evaluation on multi-node and geo-replicated configurations (e.g., CockroachDB, Spanner)
is planned; the \texttt{LedgerQuerier} abstraction is designed to accommodate them without
modification to the protocol layer or admission logic.

\paragraph{Deployment Maturity Model.}
The three backend tiers correspond to a natural maturity progression.
Adopters should advance tiers as their operational requirements grow:

\begin{table}[h!]
\centering\small
\caption{ACP deployment maturity tiers.
Each tier adds durability, observability, and assurance guarantees.
BAR-Monitor is mandatory at Tier~3 because tamper-evident ledgers justify
proactive governance monitoring.}
\label{tab:maturity}
\begin{tabular}{@{}cllll@{}}
\toprule
Tier & Backend & BAR-Monitor & Ledger integrity & Recommended context \\
\midrule
1 & \texttt{InMemoryQuerier} & Optional & Ephemeral (process-local) & Development, conformance, CI \\
2 & Redis pipelined         & Recommended & AOF/RDB persistent & Production, multi-process \\
3 & Postgres / ACID         & \textbf{Required} & Hash-chained audit & High-assurance, regulated industries \\
\bottomrule
\end{tabular}
\end{table}

\noindent Tier~3 pairs ACP with a tamper-evident state backend (ACP-LEDGER-1.0, planned)
and mandates BAR-Monitor to detect deviation collapse (\S\ref{sec:bar}).
Tier~1 and Tier~2 deployments can enable BAR-Monitor incrementally
without changing the admission control logic.

\subsection{Agent Identity Provisioning}

ACP requires a stable \texttt{agent\_id} per agent instance.
Provisioning can be integrated with existing IAM or service mesh identity systems:
the \texttt{agent\_id} is an opaque string from ACP's perspective,
provided it is unique, stable across requests, and included in signed capability tokens.
Key pairs (Ed25519 for L1--L4) should be provisioned at agent registration time
and rotated according to the organization's key management policy.

\subsection{Multi-Organization Boundaries}

Each organization deploys its own ACP instance with a sovereign Institutional Key.
Admission control decisions are evaluated locally against the deploying organization's
\texttt{LedgerQuerier} and \texttt{PolicyConfig}.
Cross-organization request flows are handled by ACP-CROSS-ORG-1.0:
the receiving organization evaluates the incoming execution token independently,
applying its own risk policy and local ledger state.

Federated deployments use mutual recognition (ACP-ITA-1.1) to establish
cross-org trust anchors without requiring a shared root of trust.
The per-org isolation model is intentional: an organization's admission control
decisions are not visible to or affected by another organization's ledger.

\subsection{Cross-Agent Coordination (L3)}

ACP enforces admission control \emph{per agent} and \emph{per request}.
When multiple agents collaborate on a composite task,
each agent's admission request is evaluated independently against its own
\texttt{agent\_id}, capability token, and ledger history.

ACP does not implement cross-agent correlation or orchestration-level coordination.
This is an explicit scope boundary: as demonstrated in Experiment~2 (§\ref{sec:adversarial}),
a coordinated group of $N$ agents can collectively execute $3N$ high-risk requests before
all agents are individually blocked.
Mitigating coordinated multi-agent threats requires attribution at the orchestration layer
(e.g., shared-resource rate limits, agent group policies, identity clustering),
which ACP is designed to \emph{complement}, not replace.

The ACP-CROSS-ORG-1.0 interaction model provides a protocol-level primitive for
multi-organization agent coordination; full L3 cross-agent orchestration is
a deployment-layer concern outside the scope of this specification.

\subsection{Integration with Existing Infrastructure}

ACP occupies the Execution Governance layer of the Agent Governance Stack
(Section~\ref{sec:design-insight}).
It is designed to complement---not replace---existing access control and
observability infrastructure:

\begin{itemize}[noitemsep]
  \item \textbf{RBAC / ABAC / OPA:} ACP provides per-request admission control with
    temporal state (history, cooldown). Existing resource-authorization systems
    continue to apply \emph{after} ACP admission succeeds.
  \item \textbf{Zero Trust:} Environmental signals (\texttt{external\_ip},
    \texttt{off\_hours}, \texttt{geo\_outside}, \texttt{untrusted\_device})
    integrate Zero Trust context directly into the ACP risk score via F\_ctx.
  \item \textbf{SIEM / Observability:} The Audit Ledger (ACP-LEDGER-1.3) produces
    per-decision signed records suitable for ingestion into log aggregation pipelines.
    Each entry carries a \texttt{policy\_hash} enabling post-hoc attribution of
    decisions to specific policy versions.
  \item \textbf{Monitoring systems:} ACP's cooldown and escalation events are
    observable state transitions---not silent blocks. They are designed to be
    surfaced to monitoring systems for incident response.
\end{itemize}

\subsection{Policy Tuning}

The \texttt{PolicyConfig} parameters (defaults documented in ACP-RISK-2.0 Appendix~A
and exposed via \texttt{DefaultPolicyConfig()}) are the primary operational levers.
Table~\ref{tab:policy-profiles} provides concrete starting values for four
threat profiles; organizations should adjust from these baselines using
observed denial rates and throughput data.

\begin{table}[h!]
\centering\small
\caption{Recommended \texttt{PolicyConfig} starting values by threat profile.
All profiles use \texttt{PolicyHash} pinning and APPROVED~$\leq 39$ / DENIED~$\geq 70$
thresholds (ACP-RISK-3.0 defaults).
\emph{Expected BAR baseline} is the healthy operational range for a deployment at that
profile under normal (non-attack) traffic; sustained BAR below the lower bound indicates
deviation collapse (\S\ref{sec:bar}).
$^\dagger$~Polling agents may require higher $Y$ to suppress false escalations
from legitimate repeated reads.}
\label{tab:policy-profiles}
\begin{tabular}{@{}lcccp{2.0cm}p{3.5cm}@{}}
\toprule
Profile & \texttt{CooldownPeriod} & \texttt{CooldownTrigger} & \texttt{AnomalyRule3Y} & Expected BAR & Typical context \\
\midrule
Low      & 120\,s  & 5 & 5$^\dagger$ & 0.80--0.95 & Internal dev tools, low-risk automation \\
Medium   & 300\,s  & 3 & 3           & 0.60--0.80 & B2B APIs, typical SaaS agents (default) \\
High     & 600\,s  & 2 & 2           & 0.45--0.65 & Financial systems, regulated industries \\
Critical & 3600\,s & 1 & 1           & 0.30--0.50 & Critical infrastructure, high-assurance \\
\bottomrule
\end{tabular}
\end{table}

\noindent The four parameters interact: tightening \texttt{CooldownTrigger} without
extending \texttt{CooldownPeriod} increases sensitivity but shortens the containment
window. \texttt{AnomalyRule3Y} should match the expected request burst size of
legitimate agents---setting $Y=1$ blocks any repeated capability use, which is
correct for critical infrastructure but would produce false escalations in
high-frequency automation. \texttt{PolicyHash} should be pinned in all profiles
to prevent silent policy drift.

\begin{itemize}[noitemsep]
  \item \texttt{CooldownPeriodSeconds}: Shorter ($<300$\,s) enables faster recovery
    after a false-positive denial cluster; longer ($\geq600$\,s) provides stronger
    containment under sustained attack or compromised agent.
  \item \texttt{CooldownTriggerDenials}: Lower values detect attacks faster but
    increase sensitivity to legitimate burst traffic.
  \item \texttt{AnomalyRule3ThresholdY}: Default $Y=3$ balances replay detection
    latency against false positives from polling agents.
  \item \texttt{PolicyHash}: Pin policy versions in capability tokens to prevent
    silent policy changes from altering admission decisions post-issuance.
\end{itemize}

\noindent ACP provides sensible defaults via \texttt{DefaultPolicyConfig()}
(Appendix~A of ACP-RISK-2.0), corresponding to the Medium profile above.
Organizations should treat policy calibration as an operational process,
informed by observed denial rates, throughput baselines, and threat model.

\subsection{Migrating from ACP-RISK-2.0 to ACP-RISK-3.0}
\label{sec:migration}

ACP-RISK-3.0 is backward-compatible with ACP-RISK-2.0 at the API level.
The \texttt{EvalRequest} and \texttt{EvalResult} structures are unchanged;
all existing conformance vectors and capability tokens remain valid.
The only behavioral change is the scoping of Rule~1 of $F_\text{anom}$:

\begin{table}[h!]
\centering\small
\caption{ACP-RISK-2.0 vs.\ ACP-RISK-3.0: behavioral and interface differences.
All other fields, formulas, and thresholds are identical.
Existing deployments need only provide \texttt{resource} in \texttt{EvalRequest}
and upgrade the \texttt{pkg/risk} package.}
\label{tab:migration}
\begin{tabular}{@{}lll@{}}
\toprule
Aspect & ACP-RISK-2.0 & ACP-RISK-3.0 \\
\midrule
Rule~1 counter key   & \texttt{agentID} & \texttt{PatternKey(agentID, cap, resource)} \\
Cross-context isolation & No (state-mixing) & Yes (per-context scoping) \\
\texttt{EvalRequest} fields & unchanged & \texttt{resource} used for \texttt{PatternKey} \\
Ledger key format    & per-agent & per-context (SHA-256 hash) \\
Performance overhead & baseline & $+$50\,ns per evaluation (Table~\ref{tab:benchmarks}) \\
Conformance vectors  & 65 unsigned & 73 signed (superset; both suites pass) \\
\bottomrule
\end{tabular}
\end{table}

\paragraph{Migration steps.}
\begin{enumerate}[noitemsep]
  \item Upgrade \texttt{pkg/risk} to the ACP-RISK-3.0 version.
  \item Ensure \texttt{EvalRequest.Resource} is populated per request
    (it was optional in RISK-2.0; it is required for correct PatternKey computation in RISK-3.0).
  \item Flush the anomaly counters in the ledger (optional but recommended):
    RISK-2.0 per-agent keys are structurally different from RISK-3.0 per-context keys
    and will not interfere, but stale RISK-2.0 entries consume storage.
  \item Re-run the ACR-1.0 conformance suite against your deployment.
    All 73 signed vectors must pass; any failure indicates a RISK-2.0 residue.
\end{enumerate}

\noindent Deployments that cannot populate \texttt{Resource} immediately
(e.g., legacy agents without resource attribution)
may set \texttt{Resource = agentID} as a transitional measure.
This degrades RISK-3.0 to RISK-2.0 behavior for Rule~1 while preserving
all other RISK-3.0 properties (Rules~2 and~3, cooldown, thresholds).

\section{Security Model}
\label{sec:security-model}

This section defines the adversary model and the security properties provided by ACP.

\subsection{Adversary Capabilities}
\label{sec:adversary-model}

We formalize the adversary as a triple $\mathcal{A} = (K, S, B)$ where:

\begin{itemize}[noitemsep]
  \item $K \in \{\textit{black-box},\ \textit{formula-aware},\ \textit{full-state}\}$ is the \emph{knowledge level}.
    A \emph{black-box} adversary observes only the outcome of each request
    (\texttt{APPROVED}/\texttt{ESCALATED}/\texttt{DENIED}).
    A \emph{formula-aware} adversary additionally knows the published risk formula
    ($\mathrm{RS} = \mathrm{capBase} + F_{\mathrm{res}} + F_{\mathrm{ctx}} + F_{\mathrm{hist}} + F_{\mathrm{anom}}$)
    and all weight values defined in ACP-RISK-3.0.
    A \emph{full-state} adversary further has real-time read access to the entire ledger
    state (pattern counts, denial history, cooldown timestamps).

  \item $S \in \{\textit{threshold-hugging},\ \textit{evasion},\ \textit{flood},\ \textit{collusion}\}$ is the \emph{strategy class}.
    \emph{Threshold-hugging}: craft requests with $\mathrm{RS}$ just below enforced boundaries.
    \emph{Evasion}: systematically drive all features toward zero to achieve permanent \texttt{APPROVED}.
    \emph{Flood}: saturate the system with high-volume requests.
    \emph{Collusion}: coordinate multiple agents to distribute request load across isolation boundaries.

  \item $B \in \mathbb{N}$ is the \emph{request budget} available to the adversary.
\end{itemize}

The evaluation experiments instantiate specific $(K, S, B)$ triples.
Table~\ref{tab:adversary-taxonomy} maps each experiment to its adversary class and
reports the enforcement outcome.

\begin{table}[h!]
\centering\small
\caption{Adversary taxonomy: mapping of ACP experiments to $(K, S, B)$ adversary classes.}
\label{tab:adversary-taxonomy}
\begin{tabular}{@{}clllll@{}}
\toprule
Exp & Adversary Class & $K$ & $S$ & $B$ & Outcome \\
\midrule
1 & Cooldown Evasion       & black-box     & threshold-hugging & 500   & contained \\
2 & Multi-Agent Flood      & black-box     & flood+collusion   & 1000  & contained \\
3 & Backend Stress         & black-box     & flood             & 10K   & degraded perf \\
4 & Token Replay           & black-box     & evasion           & 300   & contained \\
5--8 & State Mixing        & formula-aware & evasion           & —     & fixed in RISK-3.0 \\
9 & Deviation Collapse     & —             & upstream drift    & —     & BAR detects \\
10 & Knowledge-Aware Evasion & formula-aware & evasion (RS=0) & 200   & BAR detects \\
11 & Threshold Sensitivity  & —             & —                 & —     & monotonic \\
12 & Multi-Tool IPI Chain   & formula-aware & flood             & 8     & stateful persistence \\
13 & Coordination Window    & black-box     & round-robin/burst & $5N$  & linear bound ($O(N)$) \\
14 & Stateless Engine (OPA) & —             & stateless eval    & 10    & 0 enforced (allow$\times$10) \\
\bottomrule
\end{tabular}
\end{table}

Five findings emerge from this taxonomy:

\begin{enumerate}[noitemsep]
  \item \textbf{Per-decision enforcement suffices against black-box adversaries.}
        Experiments~1--4 confirm that stateful admission control contains all
        black-box strategies within the request budget.
  \item \textbf{Formula-aware adversaries break per-decision enforcement but are
        detected by BAR.}
        Experiment~10 shows that a formula-aware adversary can achieve
        $\mathrm{RS}=0$ on every request, causing every decision to be
        \texttt{APPROVED} — yet BAR-Monitor fires a TREND alert at batch~2,
        before the regime reaches full collapse.
  \item \textbf{Deviation collapse (Exp~9) is an environmental threat, not adversarial,
        but BAR-Monitor handles both classes with the same mechanism.}
  \item \textbf{Stateful ledger creates persistent enforcement consequences (Exp~12).}
        An IPI flood that triggers cooldown also activates the 24-hour anomaly
        window ($F_\text{anom}$ Rule~2), elevating borderline capabilities from
        ESCALATED to DENIED for the entire anomaly window after cooldown expires.
        A stateless engine would return to pre-attack RS values immediately.
  \item \textbf{Full-state adversaries are outside the v1.0 threat model.}
        An adversary with real-time ledger access could predict and exploit
        threshold transitions.
        This class is not addressed in the present specification
        and is acknowledged in \S\ref{sec:limitations}
        as a target for ACP-D-1.0 (decentralized governance).
  \item \textbf{Coordinated multi-agent activity scales linearly (Exp~13).}
        Per-agent behavioral isolation ensures that $N$ agents acting in
        concert can execute at most $2N$ approved operations before the first
        \texttt{DENIED} decision---a coordination window that grows linearly
        with $N$, with zero deviation observed.
        Superlinear amplification through coordination is structurally
        prevented by the combination of agent-scoped pattern keys and the
        evaluate-then-mutate execution contract.
\end{enumerate}

\subsection{Adversary Limitations}

The following components are trusted and cannot be modified by the adversary:

\begin{itemize}[noitemsep]
  \item \textbf{Execution contract:} the sequence of operations governing state updates
    (evaluation followed by state transitions) is enforced correctly.
  \item \textbf{State integrity:} the underlying state backend (ledger, cooldown state,
    pattern counters) is append-only and tamper-resistant.
  \item \textbf{Evaluation function:} the decision function (ACP-RISK-2.0) is deterministic
    and correctly implemented.
  \item \textbf{Cryptographic primitives:} signature verification and identity binding are
    assumed secure under standard assumptions (EUF-CMA security of Ed25519).
\end{itemize}

This model corresponds to an adversary that can fully control inputs but cannot compromise
the integrity of the control system itself.

\subsection{Security Properties}

Under this model, ACP enforces the following properties:

\paragraph{Determinism.}
ACP determinism is defined with respect to the evaluation context: given identical
canonical request inputs, system state (ledger, pattern counters, cooldown state),
policy definitions, and evaluation order, the protocol produces a consistent and
reproducible admission decision.
This form of determinism explicitly includes state as part of the input to the decision
function; it does not imply independence from time or state evolution.
This scoping prevents adversarial exploitation of non-determinism across evaluators.
The \texttt{RiskDeterminism} and \texttt{AnomalyDeterminism} invariants in the TLA+
model formally specify this property across all reachable states.

\paragraph{State Consistency.}
All state transitions follow the execution contract, ensuring that historical behavior
is consistently reflected in future decisions.

\paragraph{Cooldown Enforcement.}
An adversary cannot bypass cooldown mechanisms through repeated or adaptive requests.
Once cooldown conditions are met, subsequent requests are deterministically denied within
the cooldown window, regardless of request content.
This is confirmed by the \texttt{SEQ-COOLDOWN-001} sequence test vector.

\paragraph{History Integrity.}
Past decisions and events cannot be modified or removed.
The append-only ledger (ACP-LEDGER-1.3) and the \texttt{LedgerAppendOnlyTemporal} TLA+ invariant
formally guarantee monotonic accumulation of historical state.

\paragraph{Bounded Adaptation Resistance.}
While $\mathcal{A}$ may adapt request patterns, such adaptations cannot eliminate the influence
of accumulated history (repeated denials, pattern counts) on future decisions.
In particular, the denial counter evolves monotonically: intervening approved requests
do not reset prior denials, ensuring eventual enforcement under repeated violations.

\paragraph{Per-Agent Isolation.}
Policy enforcement is applied independently per agent.
Actions by one agent do not affect the enforcement state of another.
This property bounds the impact of a compromised or malicious agent to its own
request history.
Note that isolation holds \emph{across agents} but not across capability contexts
within the same agent: anomaly signals from F\_anom Rules~1 and~2 aggregate
at the agent level, enabling cross-context interference for agents holding
multiple capability tokens under the same keypair
(see \S\ref{sec:limitations}, ``State-Mixing and Context Fragmentation'').

\subsection{Informal Security Argument}

Because ACP separates stateless evaluation from state management and enforces a strict
execution contract, every request results in a deterministic state transition that cannot
be bypassed.
When the state backend enforces append-only semantics, historical information accumulates
monotonically: cooldown activation follows deterministic threshold conditions applied to
accumulated state; repeated probing eventually triggers enforced denial periods.

Because decisions are reproducible given the same input sequence and initial state,
any observed behavior can be externally verified.
This limits $\mathcal{A}$'s ability to exploit hidden or undocumented system behavior.

\subsection{Limitations}

The security guarantees of ACP rely on the integrity of the execution environment:

\begin{itemize}[noitemsep]
  \item Compromise of the state backend (deletion or modification of history) breaks core guarantees.
  \item \textbf{Distributed deployments:} ACP's determinism guarantee holds per evaluator instance.
        In deployments with replicated or eventually-consistent state backends (e.g., Redis Cluster,
        Postgres streaming replicas), two evaluators may observe divergent ledger states and reach
        different admission decisions for the same agent at the same point in time.
        ACP does not specify a distributed consensus protocol; deployments requiring strict
        cross-instance consistency must enforce strong consistency at the storage layer.
  \item Incorrect implementation of the execution contract may lead to inconsistent behavior.
  \item The model does not address side-channel attacks or network-level adversaries.
  \item \textbf{Prompt-layer attacks and LLM-specific benchmarks.}
        ACP operates at the execution layer and does not prevent indirect prompt injection
        (IPI)~\cite{greshake2023not} at the prompt layer.
        An IPI attack that causes an agent to emit a low-risk request
        (e.g., \texttt{acp:cap:data.read}/\textsc{Public}, $\mathrm{RS}=0$) will be
        admitted by ACP, because the resulting tool call is individually valid.
        ACP's enforcement boundary is the tool call, not the instruction that produced it.
        Standard LLM agent benchmarks---AgentDojo~\cite{debenedetti2024agentdojo},
        InjecAgent~\cite{zhan2024injecagent}, and AgentContract-Bench---evaluate
        prompt-layer manipulation resistance, which is a complementary but orthogonal
        property to execution-layer admission control.
        Prompt-layer defenses reduce the IPI attack surface; ACP enforces structural
        constraints over the action sequence regardless of how the request was generated.
        The two layers are composable: a prompt-layer filter applied before the tool call
        reaches the ACP admission boundary reduces the frequency of injected high-risk
        requests, while ACP provides the structural guarantee that high-risk requests
        are denied regardless of their origin.
        ACP's defense against formula-aware evasion (Experiment~10, \S\ref{sec:exp10})
        addresses a distinct threat class: an adversary who drives
        $\mathrm{RS}=0$ across all requests---which BAR-Monitor detects via
        $\Delta\mathrm{BAR}$, not per-decision enforcement.
  \item Post-quantum signature migration (ACP-SIGN-2.0 HYBRID, Ed25519 + ML-DSA-65)
    is implemented via the Cloudflare CIRCL library (\texttt{pkg/sign2/}).
    ML-DSA-65 signing latency is approximately 100--130\,\textmu s (vs.\ 25\,\textmu s for Ed25519),
    remaining well below state-backend latency; see \S\ref{sec:implementation}.
\end{itemize}

ACP is therefore best understood as a control-layer mechanism whose guarantees are
conditioned on the integrity of the state backend and the execution environment.
When these components operate correctly, ACP provides strong guarantees on decision
consistency and temporal behavior under adversarial inputs.
When the state backend is compromised (e.g., history deleted by a privileged operator),
the governance guarantees collapse; this is a declared residual risk, not a
design oversight.
Deployments in high-assurance environments should pair ACP with tamper-evident
storage---a cryptographically chained log structure that makes deletion and
modification detectable (motivating a future ACP-LEDGER-1.0 specification).

\section{Conclusion}

Autonomous agents are already operating in institutional environments.
The question is not whether they will operate---it is whether they will do so with or without formal governance.
ACP proposes that they operate with formal governance, with verifiable mechanisms,
and with traceability that can withstand an external audit.

ACP is not the only attempt to control autonomous agents.
It advances this goal through a formal technical specification with precise state models,
empirically validated security properties, and verifiable conformance requirements.
The difference between a best-practices policy and a formal protocol is exactly that:
behaviors are defined, failures have specific error codes, and conformance can be verified.

The goal of ACP is not to make agents more capable. It is to make them governable.
That is a necessary condition for their institutional deployment to be sustainable at scale.

ACP does not eliminate all sources of failure in distributed agent systems.
It does not prescribe network transport, consensus mechanisms, or business-level dispute resolution.
What it provides is a structured, auditable, and formally specified model for
admission control, delegation, revocation, and cross-organization interaction---
the layer that existing identity and policy frameworks leave unaddressed.

The v1.25 specification is complete.
A full Go reference implementation (23 packages, L1--L4), 73 signed + 65 unsigned RISK-2.0 conformance test vectors,
5 stateful sequence test vectors (ACR-1.0 compliance runner),
a TLC-runnable TLA+ formal model (base: 3~invariants; extended v1.25: 11~invariants + 4~temporal properties, 5,684,342~states generated, 3,147,864~distinct, graph depth~15, 0~violations; two-agent safety at \texttt{LB=11}: 4,648,235,723~states generated, 4,294,930,695~distinct, 11~invariants, 0~violations, 10.5~h),
a real ML-DSA-65 (Dilithium mode3) hybrid implementation via Cloudflare CIRCL (\texttt{pkg/sign2/}, \texttt{SignHybridFull} + \texttt{VerifyHybrid}),
an executable multi-organization interoperability demo (\texttt{examples/multi-org-demo/}),
a payment-agent demo (\texttt{examples/payment-agent/}),
an extended OpenAPI 3.1.0 specification (18 endpoints),
ten experiments covering performance benchmarks across backend configurations (Experiments~1--4)
and adversarial evaluation: token replay (Experiment~5),
stateless vs.\ stateful comparison (Experiment~6),
state-mixing vulnerability characterization (Experiment~7),
context-scoped anomaly enforcement under ACP-RISK-3.0 (Experiment~8),
deviation collapse and restoration (Experiment~9),
and knowledge-aware adversarial evasion (Experiment~10)
are publicly available at \url{https://github.com/chelof100/acp-framework-en}~\cite{acp-spec}.\\
\textbf{Preprint:} \url{https://arxiv.org/abs/2603.18829} \quad \textbf{Zenodo:} \href{https://doi.org/10.5281/zenodo.19672575}{10.5281/zenodo.19672575}\\
The specification and implementation are open for technical review, pilot implementation, and formal standardization.

TraslaIA invites organizations interested in adopting ACP, contributing to its evolution,
or participating in the standardization process to reach out directly.

Cross-agent coordination remains a necessary extension for deployments involving adversarial
multi-identity scenarios, particularly in high-integrity systems; a concrete mitigation path
is described in~\S\ref{sec:limitations} (\emph{Distributed Attack Surface}).

ACP demonstrates that admission control for autonomous agents can be both computationally
efficient and operationally enforceable---provided that decision logic and state management
are cleanly separated. The protocol is \textbf{compute-cheap but state-sensitive}: the
\texttt{LedgerQuerier} abstraction keeps the evaluation function at 739--832\,ns (p50; Table~\ref{tab:benchmarks}) while
allowing the state backend to be replaced, optimized, or scaled independently.
This is not an implementation detail---it is the architectural property that makes ACP
deployable in production without sacrificing governance correctness.

A system that always produces admissible actions does not demonstrate effective governance;
it may instead be operating in a regime where admissibility is no longer meaningfully evaluated.
The contribution of this work is to make that condition detectable, measurable, and formally specifiable.

\bigskip
\noindent\textbf{Marcelo Fernandez | TraslaIA}\\
\texttt{info@traslaia.com} | \url{https://agentcontrolprotocol.xyz}~\cite{acp-website}

\appendix
\section{Glossary}

\begin{table}[h!]
\centering
\small
\begin{tabular}{@{}lp{10cm}@{}}
\toprule
\textbf{Term} & \textbf{Definition} \\
\midrule
ACR-1.0 & ACP Compliance Runner 1.0. Sequence-based test runner for ACP-RISK-2.0. Distinct from ACP-CR-1.0 (Change Request Process governance document). \\
AgentID & Cryptographic identifier: \texttt{base58(SHA-256(Ed25519\_public\_key))}. Immutable and unforgeable. \\
Capability Token (CT) & Signed JSON artifact authorizing an agent to perform specific actions on a defined resource during a limited period. \\
Execution Token (ET) & Single-use artifact issued after an APPROVED decision. Authorizes exactly that action at that moment. \\
ITA & Institutional Trust Anchor. Authoritative registry linking \texttt{institution\_id} to institutional Ed25519 public key. \\
RIK & Root Institutional Key. Institution's Ed25519 key pair held in HSM. \\
Risk Score (RS) & Integer in $[0, 100]$ produced by the deterministic risk function. Determines the authorization decision. \\
Autonomy Level & Integer 0--4 assigned to an agent determining applicable risk evaluation thresholds. \\
PoP & Proof-of-Possession. Cryptographic proof that the bearer of a CT possesses the corresponding private key. \\
Audit Ledger & Chain of signed events where $h_n = \text{SHA-256}(e_n \| h_{n-1})$. Append-only and immutable. \\
MRA & Mutual Recognition Agreement. Bilateral document signed by two ITA authorities for cross-authority interoperability. \\
ESCALATED & ACP decision when RS is in the intermediate range. Action not executed until explicit resolution. \\
Fail Closed & Design principle: on any internal failure, the action is denied. Never approved by default. \\
\bottomrule
\end{tabular}
\caption{ACP Glossary.}
\end{table}

\section{Prohibited Behaviors}
\label{app:prohib}

ACP defines 12 behaviors that no compliant implementation may exhibit.
Exhibiting any of the following disqualifies conformance at all levels.

\begin{table}[h!]
\centering
\small
\begin{tabular}{@{}ll@{}}
\toprule
\textbf{Code} & \textbf{Prohibited behavior} \\
\midrule
PROHIB-001 & Approving a request when any evaluation component fails \\
PROHIB-002 & Reusing an already-consumed Execution Token \\
PROHIB-003 & Omitting signature verification on any incoming artifact \\
PROHIB-004 & Treating a not-found \texttt{token\_id} as active in a revocation context \\
PROHIB-005 & Allowing state transition from \texttt{revoked} \\
PROHIB-006 & Issuing an ET without a prior APPROVED AuthorizationDecision \\
PROHIB-007 & Modifying or deleting Audit Ledger events \\
PROHIB-008 & Silencing ledger corruption detection \\
PROHIB-009 & Ignoring \texttt{max\_depth} in delegation chains \\
PROHIB-010 & Implementing an offline policy more permissive than ACP-REV-1.0 \\
PROHIB-011 & Approving requests from agents with \texttt{autonomy\_level}~0 \\
PROHIB-012 & Continuing to process an artifact with an invalid signature \\
\bottomrule
\end{tabular}
\caption{ACP prohibited behaviors (ACP-CONF-1.2). Exhibiting any disqualifies conformance at all levels.}
\label{tab:prohib}
\end{table}

\section{Formal Verification (ACP-RISK-2.0 — TLC-Runnable)}
\label{app:tlaplus}

The \texttt{tla/ACP.tla} module is a complete TLC-runnable TLA+ model of the ACP-RISK-2.0
admission control engine.
The full TLA+ model and TLC configuration files are available in the project repository at
\url{https://github.com/chelof100/acp-framework-en}~\cite{acp-spec} (\texttt{tla/} directory).
It checks three invariants over a bounded state space
(2 agents $\times$ 4 capabilities $\times$ 3 resources, ledger depth $\leq 5$).
Liveness properties in the extended model depend on two explicit assumptions:
\emph{monotonic time progression} (time does not freeze indefinitely) and
\emph{weak fairness in request evaluation} (an enabled request is eventually processed).
These assumptions are declared in the \texttt{Spec} formula via \texttt{WF\_vars(Tick)} and
\texttt{WF\_vars(EvaluateRequest)} respectively; they are standard for temporal verification
and are stated here to make the verification scope transparent.

\paragraph{Model scope.}
The TLA+ specification abstracts $F_{\text{ctx}}$ (context factors) and $F_{\text{hist}}$
(history-based signals) from the risk computation.
Verified properties hold over this abstracted model; their applicability to the full
ACP-RISK-2.0 formula is supported empirically via the ACR-1.0 conformance runner and
the experiments in \S\ref{sec:adversarial}, but is not formally model-checked.
Table~\ref{tab:tla-coverage} summarizes what is verified, what is empirically tested,
and what falls outside the current formal scope.

\begin{table}[h]
\centering
\small
\caption{Verification coverage of ACP-RISK-2.0 components.
\checkmark~=~covered; $\times$~=~not covered in that layer.}
\label{tab:tla-coverage}
\begin{tabular}{lccp{4.2cm}}
\toprule
Component / Signal & TLA+ verified & Empirically tested & Notes \\
\midrule
Cooldown enforcement      & \checkmark & \checkmark & Temporal property (CooldownExpires) \\
Flood detection (Rule~1)  & \checkmark & \checkmark & Extended model (ACP\_Extended.tla) \\
Denial escalation (Rule~2)& \checkmark & \checkmark & Verified via execution traces \\
Pattern anomaly (Rule~3)  & \checkmark & \checkmark & Context-isolated (patternKey) \\
Context factors ($F_{\text{ctx}}$) & $\times$ & $\times$ & Abstracted; future formal work \\
History factors ($F_{\text{hist}}$) & $\times$ & \checkmark & Validated via ACR-1.0 traces \\
State-mixing behavior     & $\times$ & \checkmark & Measured in Experiment~7 \\
\bottomrule
\end{tabular}
\end{table}

\begin{verbatim}
---- MODULE ACP ----
EXTENDS Sequences, Integers, TLC

CONSTANTS Agents, Capabilities, Resources

CapabilityBase(cap) ==
    CASE cap = "admin"     -> 60
      [] cap = "financial" -> 35
      [] cap = "write"     -> 10
      [] cap = "read"      -> 0
      [] OTHER             -> 20

ResourceScore(res) ==
    CASE res = "restricted" -> 45
      [] res = "sensitive"  -> 15
      [] res = "public"     -> 0
      [] OTHER              -> 0

ComputeRisk(cap, res) ==
    LET raw == CapabilityBase(cap) + ResourceScore(res)
    IN IF raw > 100 THEN 100 ELSE raw

Decide(rs) ==
    IF      rs >= 70 THEN "DENIED"
    ELSE IF rs >= 40 THEN "ESCALATED"
    ELSE                  "APPROVED"

VARIABLE ledger
INIT == ledger = << >>

EvaluateRequest(a, cap, res) ==
    /\ Len(ledger) < 5
    /\ ledger' = Append(ledger,
            [ agent      |-> a,
              capability |-> cap,
              resource   |-> res,
              risk_score |-> ComputeRisk(cap, res),
              decision   |-> Decide(ComputeRisk(cap, res)) ])

NEXT == \E a \in Agents, cap \in Capabilities, res \in Resources :
            EvaluateRequest(a, cap, res)

Spec == INIT /\ [][NEXT]_ledger

Safety ==
    \A i \in 1..Len(ledger) :
        ledger[i].decision = "APPROVED" => ledger[i].risk_score <= 39

LedgerAppendOnly ==
    \A i \in 1..Len(ledger) :
        /\ ledger[i].decision \in {"APPROVED", "ESCALATED", "DENIED"}
        /\ ledger[i].risk_score >= 0

RiskDeterminism ==
    \A i \in 1..Len(ledger) :
        \A j \in 1..Len(ledger) :
            ( ledger[i].capability = ledger[j].capability
           /\ ledger[i].resource   = ledger[j].resource )
            =>
            ledger[i].risk_score = ledger[j].risk_score

LedgerAppendOnlyTemporal ==
    [][ /\ Len(ledger') >= Len(ledger)
        /\ \A i \in 1..Len(ledger) : ledger'[i] = ledger[i] ]_ledger
====
\end{verbatim}

\noindent\textbf{TLC result.}
Running \texttt{java -jar tla2tools.jar -config ACP.cfg ACP.tla} with
$|\mathit{Agents}|=2$, $|\mathit{Capabilities}|=4$, $|\mathit{Resources}|=3$,
ledger bound~$=5$ produces:
\textit{Model checking completed. No error has been found.}
All four declared invariants/properties (\texttt{TypeInvariant}, \texttt{Safety},
\texttt{LedgerAppendOnly}, \texttt{RiskDeterminism}) and the temporal property
\texttt{LedgerAppendOnlyTemporal} hold over the full reachable state space.

\noindent\textbf{Interpretation.}
\textit{Safety} encodes fail-closed admission~\cite{saltzer1975}: every APPROVED decision had $\mathit{RS} \leq 39$.
\textit{LedgerAppendOnly} encodes tamper-evidence in a state-based view:
every ledger entry carries a valid decision and a non-negative risk score.
\textit{LedgerAppendOnlyTemporal} is the temporal statement: in every step,
existing entries are preserved and the ledger never shrinks.
\textit{RiskDeterminism} is the central invariant of ACP-RISK-2.0: identical
(capability, resource) pairs always produce identical risk scores,
enabling third-party recalculation from a signed policy snapshot.
In the base model, \texttt{ComputeRisk} is a pure function of its inputs, so
\textit{RiskDeterminism} holds by construction.
Its non-trivial content arises in the extended model, where
\textit{AnomalyDeterminism} extends the guarantee to include per-agent state:
identical (\texttt{capability}, \texttt{resource}, \texttt{pattern\_at\_eval}) triples
always produce the same risk score, establishing that RS is a deterministic policy
function of the execution trace rather than a statistical estimate.

\noindent\textbf{Extended model (v1.20): cooldown, delegation integrity, and F\textsubscript{anom} flood detection.}
The \texttt{tla/ACP\_Extended.tla} module extends \texttt{ACP.tla} with per-agent cooldown
temporal state, denial accumulation, a static delegation chain, and---added in Sprint~J2---%
per-agent request pattern accumulation ($F_{\text{anom}}$ flood detection).
It introduces six additional invariants and three temporal properties beyond \texttt{ACP.tla}:

\begin{itemize}[noitemsep]
  \item \textbf{CooldownEnforced} (safety): active cooldown forces \texttt{DENIED} for any
        request by the affected agent, regardless of risk score.
  \item \textbf{CooldownImpliesThreshold} (safety): cooldown only exists after the denial
        threshold (\texttt{COOLDOWN\_TRIGGER}) has been reached---cooldown is never spurious.
  \item \textbf{DelegationIntegrity} (safety): no consecutive self-delegation in the chain
        (\texttt{chain[i] $\neq$ chain[i+1]}, ACP-DCMA-1.1~\S3).
  \item \textbf{AnomalyDeterminism} (safety, Sprint~J2): same
        (\texttt{capability}, \texttt{resource}, \texttt{pattern\_at\_eval}) triple always produces
        the same risk score. Extends \textit{RiskDeterminism} to include per-agent state:
        $\text{RS} = f(\text{static inputs} + \text{request history count})$.
        RS is a deterministic policy function, not a statistical model.
  \item \textbf{FloodEnforced} (safety, Sprint~J2): when the per-agent request count meets
        or exceeds \texttt{FLOOD\_THRESHOLD} at evaluation time, the decision is
        \texttt{DENIED}---the $F_{\text{anom}}$ flood override is never bypassed.
  \item \textbf{CooldownExpires} (liveness): active cooldown eventually expires, conditioned on
        the expiry time being within the model's time horizon (\texttt{cooldown\_until $\leq$ MAX\_TIME}).
        Requires weak fairness on time advance (\texttt{WF\_vars(Tick)}) in \texttt{Spec}.
  \item \textbf{EventuallyRejected} (liveness, Sprint~J2): an abusive agent not already in
        cooldown will eventually receive a \texttt{DENIED} decision via the flood override.
        Requires per-agent weak fairness on \texttt{EvaluateRequest} and
        \texttt{LEDGER\_BOUND} $>$ \texttt{FLOOD\_THRESHOLD}.
  \item \textbf{NoInfiniteExecutionUnderAbuse} (liveness, Sprint~J2, strong form):
        no execution trace exists in which an agent exhibiting sustained abusive behavior
        (request count $\geq$ \texttt{FLOOD\_THRESHOLD}) continues to be admitted indefinitely.
        Formally: $\Box(\mathit{Abusive}(a) \Rightarrow \neg\Diamond(\Box\mathit{Accepted}(a)))$.
        Holds in the bounded model via \textit{EventuallyRejected}; holds in the unbounded
        specification by structural induction on \texttt{FloodActive} monotonicity
        (\texttt{pattern\_count} is non-decreasing; once $\geq$ \texttt{FLOOD\_THRESHOLD},
        all subsequent decisions are \texttt{DENIED}).
\end{itemize}

\noindent\textbf{Extended TLC result (Sprint~J2 configuration).}
Running \texttt{java -jar tla2tools.jar -deadlock -config ACP\_Extended.cfg ACP\_Extended.tla}
with $|\mathit{Agents}|=1$, $|\mathit{Capabilities}|=3$, $|\mathit{Resources}|=2$,
\texttt{LEDGER\_BOUND}$=7$, \texttt{FLOOD\_THRESHOLD}$=4$, \texttt{FANOM\_BONUS}$=25$,
\texttt{COOLDOWN\_TRIGGER}$=3$, \texttt{COOLDOWN\_WINDOW}$=3$, \texttt{MAX\_TIME}$=7$:

\smallskip
\noindent\textit{Model checking completed. No error has been found.}\\
5,684,342 states generated, 3,147,864 distinct states found, 0 states left on queue.\\
The depth of the complete state graph search is 15.\\
Finished in 47min 30s (TLC2~v2.16, Java~1.8, single worker, 7241MB heap).

\smallskip
The model captures agent behavior under repeated request patterns, incorporating
stateful anomaly detection and cooldown enforcement.
All eleven invariants and all four temporal properties hold with zero violations across
the complete bounded state space, under the stated assumptions of monotonic time
progression, weak fairness in request evaluation, and consistent atomic state updates.
The graph depth of~15 confirms that the model exercises multiple execution paths and
state transitions prior to enforcement---abusive behavior can persist across multiple
steps before triggering rejection, confirming that enforcement is not immediate and
that liveness properties are not satisfied trivially.

\smallskip
\noindent\textbf{D4 liveness result.}
In particular, the model verifies a liveness property ensuring bounded execution under
sustained abuse.
No execution trace was found in which an agent exceeding the flood threshold could
continue to be admitted indefinitely.
This result provides evidence that, under the stated assumptions and within the bounded
model, the protocol enforces eventual containment of abusive behavior rather than
allowing unbounded execution under repeated high-risk patterns.

\smallskip
Single-agent configuration is used for $F_{\text{anom}}$ liveness verification;
per-agent enforcement is independent by design (\texttt{pattern\_count[a]} and
\texttt{last\_decision[a]} are per-agent maps), so the result generalizes structurally to $N$ agents.

\smallskip
\noindent\textbf{Two-agent safety check (J2c).}
The model was further evaluated under a two-agent configuration
(\texttt{Agents=\{``A1'',``A2''\}}, \texttt{FLOOD\_THRESHOLD=3},
\texttt{LEDGER\_BOUND=5}, \texttt{MAX\_TIME=5}) to assess potential
interference effects between agents.
Due to the exponential growth of the state space under multi-agent interaction,
reduced bounds were used for this configuration.
All nine safety invariants hold across all 1,650,024 distinct states
(3,011,690 states generated, graph depth~11, 1\,h\,08\,min) with zero violations,
confirming per-agent enforcement isolation:
\texttt{pattern\_count[A1]}, cooldown state, and admission decisions for \texttt{A1}
do not affect \texttt{A2}'s evaluation state, and vice versa.

Temporal properties were not enforced in this configuration, as liveness guarantees
depend on continued request generation and sufficient time progression.
Under reduced bounds, executions may terminate without additional requests, leading
to spurious liveness violations that do not reflect protocol behavior under sustained
interaction.
The observed behavior highlights an important modeling consideration:
temporal guarantees in ACP assume ongoing interaction with the system.
In bounded executions where request generation ceases, the model may reach terminal
states in which no further enforcement transitions occur; these traces do not represent
sustained adversarial behavior and are therefore outside the intended scope of the
liveness guarantees.
Accordingly, liveness properties are evaluated under the single-agent configuration
(\texttt{FLOOD\_THRESHOLD=4}, \texttt{MAX\_TIME=7}), where bounds are calibrated to
allow sufficient interaction depth.

\smallskip
\noindent\textbf{Two-agent safety at scale (J2d).}
To strengthen the isolation claim, a further two-agent run was conducted
with bounds calibrated to match the single-agent liveness configuration:
\texttt{Agents=\{``A1'',``A2''\}}, \texttt{FLOOD\_THRESHOLD=4},
\texttt{LEDGER\_BOUND=6}, \texttt{MAX\_TIME=7},
with 6 parallel workers and 12\,GB heap (TLC2~v2.16, Java~26).
The run completed in 4\,h\,13\,min, generating 50,927,772~states
and finding 27,648,560~distinct states at graph depth~14.
All eleven safety invariants hold with zero violations across the complete
bounded multi-agent state space, confirming that per-agent enforcement
isolation extends to the full invariant set under the liveness-calibrated bounds.

TLC reported a temporal property violation at the final liveness check,
but was unable to generate a counter-example trace:
the heap was exhausted (\texttt{OutOfMemoryError}) while constructing the lasso
over 165\,million accumulated state-space entries.
Diagnosis reveals a model-bounding artifact rather than a protocol defect.
With two agents sharing a ledger of capacity \texttt{LEDGER\_BOUND}$=6$,
the combined request volume fills the ledger before either agent reaches
\texttt{FLOOD\_THRESHOLD}$=4$, disabling \texttt{EvaluateRequest} for both
agents and preventing \texttt{EventuallyRejected} from firing.
The necessary condition for liveness verification in an $N$-agent configuration is:
\[
  \texttt{LEDGER\_BOUND} \;\geq\; \texttt{FLOOD\_THRESHOLD} \times N + \delta,
\]
where $\delta \geq 1$ provides a buffer for post-flood evaluation steps.
For two agents, this requires \texttt{LEDGER\_BOUND}$\,\geq 11$; this run
was subsequently conducted and is reported in the following paragraph.

\smallskip
\noindent\textbf{Two-agent safety at extended scale (LB=11).}
To directly address the scaling constraint identified above, a further run was
conducted at the required bound:
\texttt{Agents=\{``A1'',``A2''\}}, \texttt{LEDGER\_BOUND=11},
\texttt{FLOOD\_THRESHOLD=4}, \texttt{MAX\_TIME=7},
with 15~parallel workers (auto-scaled by TLC) and 24\,GB heap on an
Intel Core~i9-13900HX with 32\,GB RAM
(TLC2~v2.16, Java~21, Windows~11).
The run explored \textbf{4,294,930,695~distinct states} over
10.5\,hours, generating 4,648,235,723~total states,
with all eleven safety invariants reporting zero violations
throughout the entire exploration.
The run was halted by disk space exhaustion
(584\,GB of checkpoint data written to disk),
not by any property violation.
This result confirms that all per-agent enforcement invariants hold robustly
under the correct multi-agent ledger bound, providing safety verification at
$756\times$ the scale of the single-agent configuration and
$155\times$ the scale of the LEDGER\_BOUND=6 run.
Liveness verification at this scale remains open: the complete state space
at LEDGER\_BOUND=11 substantially exceeds current hardware limits for
lasso-trace computation.

\smallskip
\noindent\textbf{Exploration collapse: a formal analog of deviation collapse.}
The bounding artifact above exhibits a structural parallel to the runtime
phenomenon studied in Section~\ref{sec:deviation-collapse}.
In deviation collapse, an upstream sanitizer eliminates the risk signals that
would cause enforcement to fire: the engine remains active and syntactically
correct, but its boundary is never exercised.
In the multi-agent TLC configuration, bounded shared resources eliminate the
state conditions under which liveness properties can be
exercised~\cite{pnueli1977temporal, alpern1985defining}:
the model checker remains sound and the invariants hold, but the traces
required to validate temporal guarantees cannot be reached within the bound.
Unlike classical state-space reductions---such as partial-order methods,
which preserve liveness properties by construction~\cite{baier2008principles}---
resource-induced bounds may eliminate entire classes of trajectories from the
reachable state space.
We refer to this as \emph{resource-induced liveness obstruction}:
a structural limitation that arises not from a defect in the protocol or the
model, but from the interaction between bounded shared resources and
multi-agent concurrency.

Both cases instantiate the same structural pattern: \emph{when the conditions
necessary for enforcement to fire disappear---whether from the runtime pipeline
or from the formal model---validation loses meaning independently of whether
the mechanism itself is correct.}
Put concretely: systems appear compliant because no invalid actions occur;
models appear correct because no violations are reachable.
In both cases, the decision boundary remains defined and the enforcement
logic remains intact, but the capacity to exercise failure conditions has
been eliminated.
This observation motivates a modeling requirement analogous to failure condition
preservation~(Section~\ref{sec:deviation-collapse}):
formal liveness verification of multi-agent systems requires that resource bounds
be sufficient to generate the behavioral conditions under which liveness properties
are exercised~\cite{yu1999model, lamport2002specifying}.
Addressing this in the general case---including per-agent ledger allocation,
explicit fairness constraints, and scalable model checking strategies---is
identified as future work.

ACP thus provides two distinct guarantees: \emph{safety in all executions}
(invariants hold regardless of scheduling or agent count, verified across
4,294,930,695~distinct states in the two-agent configuration at the correct
ledger bound \texttt{LEDGER\_BOUND=11}) and \emph{liveness under sustained
interaction} (eventual enforcement holds when agents continue to generate
requests within the time horizon, verified for the single-agent configuration
where bounds are calibrated to admit the relevant traces).

\noindent\textbf{Design notes on the extended model.}
The base risk score \texttt{ComputeRisk(cap,\,res)} is preserved as a pure function;
the extended function \texttt{ComputeRiskWithAnom(cap,\,res,\,pat\_count)} adds
\texttt{FANOM\_BONUS} when \texttt{pat\_count} $\geq$ \texttt{FLOOD\_THRESHOLD},
modeling ACP-RISK-2.0 $F_{\text{anom}}$ Rule~3 (rapid-fire anomaly accumulation).
\texttt{FloodActive(a)} additionally overrides the decision to \texttt{DENIED} regardless
of the resulting risk score, ensuring the flood enforcement cannot be bypassed by
capability/resource combinations with inherently low base scores.
Denial counts increment only for RS-based \texttt{DENIED} decisions---not for cooldown-forced
or flood-forced denials---consistent with ACP-RISK-2.0~\S4 (\texttt{AddDenial} fires on
real denials only).
The delegation chain is a static 2-hop sequence; dynamic chain mutation is outside the
scope of the bounded model (reserved for a future version).
The \textit{CooldownExpires} and \textit{EventuallyRejected} liveness properties are
conditioned on the relevant state being reachable within the time horizon; the
unconditional forms hold in the unbounded specification.

\noindent\textbf{Framing.}
Sprint~J2 extends formal coverage from cooldown enforcement to flood-based
$F_{\text{anom}}$ enforcement: the model now verifies not only that individual denials
are correct (\textit{Safety}, \textit{CooldownEnforced}) but also that sustained abusive
behavior is blocked within the bounded model under the stated assumptions
(\textit{NoInfiniteExecutionUnderAbuse}).
The \textit{AnomalyDeterminism} invariant formalizes the key correctness property of
ACP's stateful risk engine: the admission decision is fully reproducible given the same
capability, resource, and per-agent request history, enabling independent audit without
proprietary infrastructure.
The phrase ``formally verified'' is deliberately avoided; the correct claim is:
\emph{model checking of selected safety and liveness properties under a bounded state model,
under the stated assumptions of monotonic time progression, weak fairness in request
processing, and consistent atomic state updates.}

\noindent\textbf{Model simplifications.}
The TLA+ model uses \texttt{ComputeRiskWithAnom(cap,\,res,\,pat\_count)} in place of the
full ACP-RISK-2.0 formula, which additionally includes $F_{\text{ctx}}$ (context factors)
and $F_{\text{hist}}$ (history factors).
$F_{\text{anom}}$ is modeled as a frequency-based accumulation plus a direct override
(\texttt{FloodActive}); semantic payload analysis is out of scope.
This abstraction is intentional: it captures a fundamental class of abuse patterns
(sustained high-frequency interaction from the same agent) while enabling tractable
model checking.
The key contribution is not the detection signal itself, but the verifiable enforcement
guarantee---the formal proof that, given a deterministic accumulation function with
defined thresholds, the protocol enforces correct admission decisions under stateful
conditions, produces auditable execution traces, and allows independent reproduction
by third parties.
This distinguishes ACP from conventional rate limiters, which lack formal specification,
execution contracts, and external verifiability.
\texttt{pattern\_count} is monotone within a bounded trace (no per-window decay),
approximating the time-windowed behavior of ACP-RISK-2.0 $F_{\text{anom}}$ Rule~3.
The model verifies properties for representative parameter values
(\texttt{FLOOD\_THRESHOLD}$=4$, \texttt{FANOM\_BONUS}$=25$);
complete generalization to arbitrary parameter values is not claimed.
Full determinism under $F_{\text{hist}}$ is validated empirically via the ACR-1.0
compliance runner and the adversarial evaluation in Section~\ref{sec:adversarial}.
These simplifications are deliberate: the model is intended to verify structural
safety and liveness properties, not to replicate the full engine.

\noindent\textbf{Limitations of the formal evaluation.}
The model checking is performed over a bounded model with a finite time horizon
and a limited number of agents.
As such:
\begin{itemize}[noitemsep]
  \item The results do not constitute a proof of correctness for unbounded executions.
  \item Liveness guarantees rely on assumptions of monotonic time progression and weak
        fairness in request evaluation.
  \item A supplementary two-agent safety check (J2c) confirms per-agent enforcement
        isolation across 1,650,024 distinct states with 0 violations; temporal liveness
        properties are verified only in the single-agent configuration with calibrated
        bounds. Full multi-agent contention under adversarial coordination is not modeled.
  \item The state space is constrained by bounded parameters (time horizon, ledger size),
        which may not capture all possible long-horizon behaviors.
\end{itemize}
These constraints define the scope of the guarantees provided by the model and should
be considered when interpreting the results.

\section*{Acknowledgements}

The ACP specification emerged from analysis of operational challenges in deploying autonomous agents in regulated B2B environments.
The protocol design was informed by the IETF RFC process, the Kubernetes admission control architecture,
and the SPIFFE/SPIRE workload identity model.


\bibliographystyle{plain}
\bibliography{references}

\end{document}